\documentclass[printer,longauth,usenatbib]{aa}

\usepackage{natbib}
\usepackage{times}
\usepackage[T1]{fontenc}
\usepackage{graphicx}
\usepackage{tabularx}
\usepackage[skip=1ex]{caption}
\usepackage{amsmath}

\usepackage{pdflscape}

\usepackage[utf8]{inputenc}
\usepackage{tabularx}

\def\deg{\hbox{$^\circ$}}
\usepackage[normalem]{ulem}

\begin{document}

\title{Scaling slowly rotating asteroids \break 
       with stellar occultations}

\authorrunning{Marciniak et al.}
\titlerunning{Scaling slowly rotating asteroids by stellar occultations}

\author{A. Marciniak \inst{\ref{inst:UAM}}
  \and J. \v{D}urech \inst{\ref{inst:Prague}}
  \and A. Choukroun \inst{\ref{inst:UAM}}
  \and J. Hanu\v{s} \inst{\ref{inst:Prague}}
  \and W. Og{\l}oza \inst{\ref{inst:Suh}}
  \and R. Szak{\'a}ts \inst{\ref{inst:Konkoly},\ref{inst:Excellence}}
  \and L. Moln{\'a}r \inst{\ref{inst:Konkoly},\ref{inst:Excellence},\ref{inst:Lendulet},\ref{inst:ELTE}}
  \and A.~P{\'a}l \inst{\ref{inst:Konkoly},\ref{inst:Excellence},\ref{inst:ELU}} 
  \and F.~Monteiro \inst{\ref{inst:OASI}} 
  \and E.~Frappa \inst{\ref{inst:EURASTER}}
  \and W.~Beisker \inst{\ref{inst:IOTA-ES}}
  \and H.~Pavlov \inst{\ref{inst:IOTA-ES}}
  \and J.~Moore \inst{\ref{inst:IOTA}}
  \and R.~Adomavi\v{c}ien{\.e} \inst{\ref{inst:Vilnius}}
  \and R.~Aikawa  \inst{\ref{inst:JOIN}}
  \and S.~Andersson \inst{\ref{inst:IOTA-ES}}
  \and P.~Antonini \inst{\ref{inst:Geneva}}
  \and Y.~Argentin \inst{\ref{inst:IOTA-ES}}
  \and A.~Asai \inst{\ref{inst:JOIN}}
  \and P.~Assoignon \inst{\ref{inst:IOTA-ES}}
  \and J.~Barton \inst{\ref{inst:IOTA}}
  \and P.~Baruffetti \inst{\ref{inst:IOTA-ES}}
  \and K.~L.~Bath \inst{\ref{inst:IOTA-ES}}
  \and R.~Behrend \inst{\ref{inst:Geneva}}  
  \and L.~Benedyktowicz \inst{\ref{inst:IOTA-ES}}
  \and L.~Bernasconi \inst{\ref{inst:Engar}} 
  \and G.~Biguet \inst{\ref{inst:T60}}
  \and M.~Billiani \inst{\ref{inst:IOTA-ES}}
  \and D.~B{\l}a{\.z}ewicz \inst{\ref{inst:IOTA-ES}}
  \and R.~Boninsegna \inst{\ref{inst:IOTA-ES}}
  \and M.~Borkowski \inst{\ref{inst:IOTA-ES}}
  \and J.~Bosch \inst{\ref{inst:Collonges}} 
  \and S.~Brazill \inst{\ref{inst:IOTA}}
  \and M.~Bronikowska \inst{\ref{inst:UAM}}
  \and A.~Bruno \inst{\ref{inst:T60}}
  \and M.~Butkiewicz - B\k{a}k \inst{\ref{inst:UAM}}
  \and J.~Caron \inst{\ref{inst:Geneva}}
  \and G.~Casalnuovo \inst{\ref{inst:IOTA-ES}}
  \and J.~J. Castellani \inst{\ref{inst:IOTA-ES}}
  \and P.~Ceravolo \inst{\ref{inst:IOTA}}
  \and M.~Conjat \inst{\ref{inst:IOTA-ES}}
  \and P.~Delincak \inst{\ref{inst:IOTA-ES}}
  \and J.~Delpau \inst{\ref{inst:IOTA-ES}}
  \and C.~Demeautis \inst{\ref{inst:Geneva}}
  \and A.~Demirkol \inst{\ref{inst:Torun}}
  \and M.~Dr{\'o}{\.z}d{\.z} \inst{\ref{inst:Suh}}
  \and R.~Duffard \inst{\ref{inst:Granada}}
  \and C.~Durandet \inst{\ref{inst:T60}}
  \and D.~Eisfeldt \inst{\ref{inst:IOTA}}
  \and M.~Evangelista \inst{\ref{inst:OASI}}
  \and S.~Fauvaud \inst{\ref{inst:Bardon},\ref{inst:T60}}
  \and M.~Fauvaud \inst{\ref{inst:Bardon},\ref{inst:T60}}
  \and M.~Ferrais \inst{\ref{inst:Arecibo}}
  \and M.~Filipek \inst{\ref{inst:IOTA-ES}}
  \and P.~Fini \inst{\ref{inst:IOTA-ES}}
  \and K.~Fukui  \inst{\ref{inst:JOIN}}
  \and B.~G\"{a}hrken \inst{\ref{inst:IOTA-ES}}
  \and S.~Geier \inst{\ref{inst:IAC},\ref{inst:GRANTECAN}}
  \and T.~George \inst{\ref{inst:IOTA}}
  \and B.~Goffin \inst{\ref{inst:IOTA-ES}}
  \and J.~Golonka \inst{\ref{inst:Torun}}
  \and T.~Goto  \inst{\ref{inst:JOIN}}
  \and J.~Grice \inst{\ref{inst:Open}}
  \and K.~Guhl \inst{\ref{inst:IOTA-ES}}
  \and K.~Hal{\'i}\v{r} \inst{\ref{inst:IOTA-ES}}
  \and W.~Hanna \inst{\ref{inst:TTOA}}
  \and M.~Harman \inst{\ref{inst:IOTA-ES}}
  \and A.~Hashimoto \inst{\ref{inst:JOIN}}
  \and W.~Hasubick \inst{\ref{inst:IOTA-ES}}
  \and D.~Higgins \inst{\ref{inst:Hill}}
  \and M.~Higuchi \inst{\ref{inst:JOIN}}
  \and T.~Hirose \inst{\ref{inst:JOIN}}
  \and R.~Hirsch \inst{\ref{inst:UAM}}
  \and O.~Hofschulz \inst{\ref{inst:IOTA-ES}}
  \and T.~Horaguchi  \inst{\ref{inst:JOIN}}
  \and J.~Horbowicz \inst{\ref{inst:UAM}}
  \and M.~Ida  \inst{\ref{inst:JOIN}}
  \and B.~Ign{\'a}cz \inst{\ref{inst:Konkoly},\ref{inst:Excellence}}
  \and M.~Ishida  \inst{\ref{inst:JOIN}}
  \and K.~Isobe  \inst{\ref{inst:JOIN}}
  \and E.~Jehin \inst{\ref{inst:Liege}}
  \and B.~Joachimczyk \inst{\ref{inst:Torun}}
  \and A.~Jones \inst{\ref{inst:BAA}}
  \and J.~Juan \inst{\ref{inst:IOTA-ES}}
  \and K.~Kami{\'n}ski \inst{\ref{inst:UAM}}
  \and M.~K.~Kami{\'n}ska \inst{\ref{inst:UAM}}
  \and P.~Kankiewicz \inst{\ref{inst:Kielce}}
  \and H.~Kasebe  \inst{\ref{inst:JOIN}}
  \and B.~Kattentidt \inst{\ref{inst:IOTA-ES}}
  \and D.-H.~Kim \inst{\ref{inst:Chungbuk},\ref{inst:KASI}}
  \and M.-J.~Kim \inst{\ref{inst:KASI}}
  \and K.~Kitazaki  \inst{\ref{inst:JOIN}}
  \and A.~Klotz \inst{\ref{inst:IOTA-ES}}
  \and M.~Komraus \inst{\ref{inst:Torun}}
  \and I.~Konstanciak \inst{\ref{inst:UAM}}
  \and R.~K{\"o}nyves - T{\'o}th \inst{\ref{inst:Konkoly},\ref{inst:Excellence}}
  \and K.~Kouno  \inst{\ref{inst:JOIN}}
  \and E.~Kowald \inst{\ref{inst:IOTA-ES}}
  \and J.~Krajewski \inst{\ref{inst:UAM}}
  \and G.~Krannich \inst{\ref{inst:IOTA-ES}}
  \and A.~Kreutzer  \inst{\ref{inst:IOTA-ES}}
  \and A.~Kryszczy{\'n}ska \inst{\ref{inst:UAM}}
  \and J.~Kub{\'a}nek \inst{\ref{inst:IOTA-ES}}
  \and V.~Kudak \inst{\ref{inst:Uzhhorod}}
  \and F.~Kugel \inst{\ref{inst:Geneva}}
  \and R.~Kukita  \inst{\ref{inst:JOIN}}
  \and P.~Kulczak \inst{\ref{inst:UAM}}
  \and D.~Lazzaro \inst{\ref{inst:OASI}}
  \and J.~Licandro \inst{\ref{inst:IAC},\ref{inst:ULL}}
  \and F.~Livet \inst{\ref{inst:T60}}
  \and P.~Maley  \inst{\ref{inst:IOTA},\ref{inst:Houston}}
  \and N.~Manago  \inst{\ref{inst:JOIN}}
  \and J.~M{\'a}nek \inst{\ref{inst:IOTA-ES}}
  \and A.~Manna \inst{\ref{inst:IOTA-ES}}
  \and H.~Matsushita  \inst{\ref{inst:JOIN}}
  \and S.~Meister \inst{\ref{inst:IOTA-ES}}
  \and W.~Mesquita \inst{\ref{inst:OASI}}
  \and S.~Messner  \inst{\ref{inst:IOTA}}
  \and J.~Michelet \inst{\ref{inst:T60}}
  \and J.~Michimani \inst{\ref{inst:OASI}}
  \and I.~Mieczkowska \inst{\ref{inst:UAM}}
  \and N.~Morales \inst{\ref{inst:Granada}}
  \and M.~Motyli{\'n}ski \inst{\ref{inst:Torun}}
  \and M.~Murawiecka \inst{\ref{inst:UAM}}
  \and J.~Newman  \inst{\ref{inst:TTOA}}
  \and V.~Nikitin  \inst{\ref{inst:IOTA}}
  \and M.~Nishimura  \inst{\ref{inst:JOIN}}
  \and J.~Oey \inst{\ref{inst:Blue}}
  \and D.~Oszkiewicz \inst{\ref{inst:UAM}}
  \and M.~Owada   \inst{\ref{inst:JOIN}}
  \and E.~Pak\v{s}tien{\.e} \inst{\ref{inst:Vilnius}}
  \and M.~Paw{\l}owski \inst{\ref{inst:UAM}}
  \and W.~Pereira \inst{\ref{inst:OASI}}
  \and V.~Perig \inst{\ref{inst:Uzhhorod}}
  \and J.~Per{\l}a \inst{\ref{inst:UAM}}
  \and F.~Pilcher \inst{\ref{inst:Organ}}
  \and E.~Podlewska-Gaca \inst{\ref{inst:UAM}}
  \and J.~Pol{\'a}k \inst{\ref{inst:IOTA-ES}}
  \and T.~Polakis \inst{\ref{inst:Tempe}}
  \and M.~Poli{\'n}ska \inst{\ref{inst:UAM}}
  \and A.~Popowicz \inst{\ref{inst:Gliwice}}
  \and F.~Richard \inst{\ref{inst:T60}}
  \and J.~J~ Rives \inst{\ref{inst:T60}}
  \and T.~Rodrigues \inst{\ref{inst:OASI}}
  \and {\L}.~Rogi{\'n}ski \inst{\ref{inst:UAM}}
  \and E.~Rond{\'o}n \inst{\ref{inst:OASI}}
  \and M.~Rottenborn \inst{\ref{inst:IOTA-ES}}
  \and R.~Sch\"{a}fer \inst{\ref{inst:IOTA-ES}}
  \and C.~Schnabel \inst{\ref{inst:IOTA-ES}}
  \and O.~Schreurs \inst{\ref{inst:IOTA-ES}}
  \and A.~Selva \inst{\ref{inst:IOTA-ES}}
  \and M.~Simon \inst{\ref{inst:IOTA-ES}}
  \and B.~Skiff \inst{\ref{inst:Lowell}}
  \and M.~Skrutskie \inst{\ref{inst:Virginia}}
  \and J.~Skrzypek \inst{\ref{inst:UAM}}
  \and K.~Sobkowiak \inst{\ref{inst:UAM}}
  \and E.~Sonbas \inst{\ref{inst:Adi}}
  \and S.~Sposetti \inst{\ref{inst:IOTA-ES}}
  \and P.~Stuart \inst{\ref{inst:IOTA}}
  \and K.~Szyszka \inst{\ref{inst:Torun}}
  \and K.~Terakubo  \inst{\ref{inst:JOIN}}
  \and W.~Thomas \inst{\ref{inst:IOTA}}
  \and P.~Trela \inst{\ref{inst:UAM}}
  \and S.~Uchiyama  \inst{\ref{inst:JOIN}}
  \and M.~Urbanik \inst{\ref{inst:IOTA-ES}}
  \and G.~Vaudescal \inst{\ref{inst:IOTA-ES}}
  \and R.~Venable \inst{\ref{inst:IOTA}}
  \and Ha.~Watanabe  \inst{\ref{inst:JOIN}}
  \and Hi.~Watanabe  \inst{\ref{inst:JOIN}}
  \and M.~Winiarski \inst{\ref{inst:Suh}}$^\dagger$
  \and R.~Wr{\'o}blewski \inst{\ref{inst:UAM}}
  \and H.~Yamamura  \inst{\ref{inst:JOIN}}
  \and M.~Yamashita  \inst{\ref{inst:JOIN}}
  \and H.~Yoshihara  \inst{\ref{inst:JOIN}}
  \and M.~Zawilski \inst{\ref{inst:IOTA-ES}}
  \and P.~Zelen{\'y} \inst{\ref{inst:ValMez},\ref{inst:IOTA-ES}}
  \and M.~{\.Z}ejmo \inst{\ref{inst:Kepler}}
  \and K.~{\.Z}ukowski \inst{\ref{inst:UAM}}
  \and S.~{\.Z}ywica \inst{\ref{inst:Torun}}
}

 \institute{
   Astronomical Observatory Institute, Faculty of Physics, Adam Mickiewicz University,
  S{\l}oneczna 36, 60-286 Pozna{\'n}, Poland. E-mail: am@amu.edu.pl \label{inst:UAM}
  \and Astronomical Institute, Faculty of Mathematics and Physics, Charles University, V Hole\v{s}ovi\v{c}k{\'a}ch 2,
  180 00 Prague 8, Czech Republic \label{inst:Prague}
  \and Mt. Suhora Observatory, Pedagogical University, Podchor\k{a}{\.z}ych 2, 30-084, Cracow, Poland \label{inst:Suh}
  \and Konkoly Observatory, Research Centre for Astronomy and Earth Sciences, E\"{o}tv\"{o}s Lor{\'a}nd Research Network (ELKH),
       H-1121 Budapest, Konkoly Thege Mikl{\'o}s {\'u}t 15-17, Hungary \label{inst:Konkoly}
  \and CSFK, MTA Centre of Excellence, Budapest, Konkoly Thege Mikl{\'o}s {\'u}t 15-17, H-1121, Hungary \label{inst:Excellence}
  \and MTA CSFK Lend{\"u}let Near-Field Cosmology Research Group \label{inst:Lendulet}
  \and ELTE E{\"o}tv{\"o}s Lor{\'a}nd University, Institute of Physics, 1117, P\'azm\'any P\'eter s\'et\'any 1/A, Budapest, Hungary \label{inst:ELTE}
  \and Astronomy Department, E\"otv\"os Lor\'and University, P\'azm\'any P. s. 1/A, H-1171 Budapest, Hungary \label{inst:ELU}
  \and Observat{\'o}rio Nacional, R. Gen. Jos{\'e} Cristino, 77 - S{\~a}o Crist{\'o}v{\~a}o, 20921-400, Rio de Janeiro - RJ, Brazil \label{inst:OASI}
  \and EURASTER, 8 rue du Tonnelier, 46100 Faycelles, France \label{inst:EURASTER}
  \and International Occultation Timing Association/European Section, Am Brombeerhag 13, 30459, Hannover, Germany \label{inst:IOTA-ES}
  \and International Occultation Timing Association (IOTA), PO Box 7152, WA, 98042, USA \label{inst:IOTA}
  \and Institute of Theoretical Physics and Astronomy, Vilnius University, Saul{\.e}tekio al. 3, 10257 Vilnius, Lithuania \label{inst:Vilnius}
  \and Japanese Occultation Information Network \label{inst:JOIN}
  \and Geneva Observatory, CH-1290 Sauverny, Switzerland \label{inst:Geneva}
  \and Les Engarouines Observatory, F-84570 Mallemort-du-Comtat, France \label{inst:Engar}
  \and Association T60, Observatoire Midi-Pyr{\'e}n{\'e}es, 14, avenue Edouard Belin, 31400 Toulouse, France \label{inst:T60}
  \and Collonges Observatory, F-74160 Collonges, France \label{inst:Collonges}
  \and Institute of Astronomy, Faculty of Physics, Astronomy and Informatics, Nicolaus Copernicus University in Toru{\'n}, 
  ul.~Grudzi\k{a}dzka~5, 87-100 Toru{\'n}, Poland \label{inst:Torun}
   \and Departamento de Sistema Solar, Instituto de Astrof{\'i}sica de Andaluc{\'i}a (CSIC),
  Glorieta de la Astronom{\'i}a s/n, 18008 Granada, Spain \label{inst:Granada}
  \and Observatoire du Bois de Bardon, 16110 Taponnat, France \label{inst:Bardon}
  \and Arecibo Observatory, University of Central Florida, HC-3 Box 53995, Arecibo, PR 00612, USA \label{inst:Arecibo}
  \and Instituto de Astrof{\'i}sica de Canarias, C/ V{\'i}a Lactea, s/n, 38205 La Laguna, Tenerife, Spain \label{inst:IAC}
  \and Gran Telescopio Canarias (GRANTECAN), Cuesta de San Jos{\'e} s/n, E-38712, Bre{\~n}a Baja, La Palma, Spain \label{inst:GRANTECAN}
  \and Open University, School of Physical Sciences, The Open University, MK7 6AA, UK \label{inst:Open}
  \and Trans-Tasman Occultation Alliance (TTOA), Wellington, PO Box 3181, New Zealand \label{inst:TTOA}
  \and Hunters Hill Observatory, 7 Mawalan Street, Ngunnawal, ACT 2913, Australia \label{inst:Hill}
  \and Space sciences, Technologies and Astrophysics Research Institute, Universit{\'e} de Li{\`e}ge, All{\'e}e du 6 Ao{\^u}t 17,
  4000 Li{\`e}ge, Belgium \label{inst:Liege}
  \and British Astronomical Association, Burlington House, Piccadilly, Mayfair, W1J 0DU, London, UK \label{inst:BAA}
  \and Institute of Physics, Jan Kochanowski University, ul. Uniwersytecka 7, 25-406 Kielce \label{inst:Kielce}
  \and Chungbuk National University, 1, Chungdae-ro, Seowon-gu, Cheongju-si, Chungcheongbuk-do, Republic of Korea \label{inst:Chungbuk}
  \and Korea Astronomy and Space Science Institute, 776 Daedeok-daero, Yuseong-gu, Daejeon 34055, Korea \label{inst:KASI}
  \and Laboratory of Space Researches, Uzhhorod National University, Daleka st. 2a, 88000, Uzhhorod, Ukraine \label{inst:Uzhhorod}
  \and Departamento de Astrof{\'i}sica, Universidad de La Laguna - ULL, Tenerife, Spain \label{inst:ULL}
  \and NASA Johnson Space Center Astronomical Society, Houston, TX USA \label{inst:Houston}
  \and Blue Mountains Observatory, Leura, Australia \label{inst:Blue}
  \and Organ Mesa Observatory, 4438 Organ Mesa Loop, Las Cruces, New Mexico 88011 USA \label{inst:Organ}
  \and Command Module Observatory, 121 W. Alameda Dr., Tempe, AZ 85282 USA \label{inst:Tempe}
  \and Silesian University of Technology, Department of Electronics, Electrical Engineering and Microelectronics, Akademicka 16,
  44-100 Gliwice, Poland \label{inst:Gliwice}
  \and Lowell Observatory, 1400 West Mars Hill Road, Flagstaff, Arizona, 86001 USA \label{inst:Lowell}
  \and Department of Astronomy, University of Virginia, Charlottesville, VA, 22904 USA \label{inst:Virginia}
  \and Department of Physics, Adiyaman University, 02040 Adiyaman, Turkey \label{inst:Adi}
  \and Observatory, Vset{\'i}nsk{\'a} 78, Vala\v{s}sk{\'e} Mezi\v{r}{\'i}\v{c}{\'i}, Czechia \label{inst:ValMez}
  \and Kepler Institute of Astronomy, University of Zielona G{\'o}ra, Lubuska 2, 65-265 Zielona G{\'o}ra, Poland \label{inst:Kepler}
}

\date{Received 20 February 2023 / Accepted ...}

\abstract
{%Context 
 As evidenced by recent survey results, the majority of asteroids are slow rotators (spin periods longer than 12 hours), 
 but lack spin and shape models because of selection bias.
 This bias is skewing our overall understanding of the spins, shapes, and sizes of  asteroids, as well as of their other properties. 
 Also, diameter determinations for large (>60km) and medium-sized asteroids (between 30 and 60 km) often vary by over 30\% for multiple reasons.}
{%Aims
 Our long-term project is focused on a few tens of slow rotators with periods of up to 60 hours. We aim
 to obtain their full light curves and reconstruct their spins and shapes. We also precisely scale the models, 
 typically with an accuracy of a few percent.}
{%Methods
 We used wide sets of dense light curves for spin and shape reconstructions via light-curve inversion. 
 Precisely scaling them with thermal data was not possible here because of poor infrared datasets: large bodies tend to saturate in WISE 
 mission detectors. Therefore, we recently also launched a special campaign among stellar occultation observers, both in order to scale these models and to verify the shape solutions, often allowing us to break the mirror pole ambiguity.}
{%Results
 The presented scheme resulted in shape models for 16 slow rotators, most of them for the first time.
 Fitting them to chords from stellar occultation timings resolved previous inconsistencies in size determinations. 
 For around half of the targets, this fitting also allowed us to identify a clearly preferred pole solution from the pair 
 of two mirror pole solutions, thus removing the ambiguity inherent to light-curve inversion.
 We also address the influence of the uncertainty of the shape models on the derived diameters.  
}
{%Conclusions
 Overall, our project has already provided reliable models for around 50 slow rotators.
 Such well-determined and scaled asteroid shapes will, for example, constitute a solid basis for precise density determinations 
 when coupled with mass information. Spin and shape models in general continue to fill the gaps caused by various biases. 
}


\maketitle

\section{Introduction} 
\label{sec:intro}

 One of the ultimate aims of asteroid research is to obtain density determinations that would enable studies of their internal
 structure (micro- and macroporosity), and mineralogical composition in order to link meteorites with their parent bodies.
 Asteroids larger than 100 km in diameter are considered primitive and largely unchanged since their formation
 \citep{Morbidelli2009}, while smaller ones are seen as remnants of collisional disruptions and reaccumulations,
 with substantial macroporosity \citep{Carry2012}.
 
 Sizes of asteroids, a necessary prerequisite to deriving conclusions as to internal structure and density where mass is
 available, are not easy to determine. 
 Accurate size determinations are of vital importance, as their relative error is tripled in density determinations, 
 while to clearly distinguish between the different mineralogies of asteroid interiors, 
 the density needs to be known with 20\% precision \citep{Carry2012}.
 In the majority of cases, asteroid sizes are calculated from a known value of absolute
 magnitude, and an assumed geometric albedo \citep{Harris1997}, which leads to large uncertainties in the derived diameter, namely at the
 level of 20\%-30\% for main-belt asteroids, and reaching over 50\% in some cases \citep{Tanga2013}\footnote{https://mp3c.oca.eu/}. 
 The solution is to use infrared measurements: 
 in the visible light, a small target with a highly reflective surface is equally as bright as a large target with a dark surface. 
 However, in the infrared, the two objects appear substantially different. A visually dark object, due to its smaller albedo, absorbs most of the 
 solar radiation, and is heated to higher temperatures than its visually brighter counterpart \citep{Harris2002}. Therefore, thermal data combined with absolute 
 brightness in the visible light enable us to put tight constraints on the object size, typically reducing the uncertainty to 5\%--10\% 
 \citep{Delbo2015}, but only if spin and shape model is available. Without a model, the accuracy of this method 
 is greatly reduced. Also, this method is limited to objects with rich thermal datasets from multiple space observatories. 
 For example, large objects of around 100 km in diameter tend to be saturated in W3 and/or W4 bands (11.1 and 22.64 $\mu$m respectively) 
 of the WISE mission \citep{Wright2010}, and so data from WISE cannot be used in their thermophysical modelling  \citep{Delbo2015}. 
 The diameters of these targets must therefore be determined with other methods.
  
 A somewhat complementary method that enables large asteroid surfaces to be imaged and their shapes and sizes to be determined is
 high-resolution imaging with the state-of-the-art adaptive optics systems used with modern deconvolution techniques. A recent large programme 
 at the VLT using the SPHERE/ZIMPOL instrument provided detailed shapes and accurate sizes for around 40 main-belt asteroids representing 
 all of the main taxonomic classes \citep[see e.g.][]{Vernazza2021}. This technique, although spectacular, is nevertheless limited to the relatively 
 nearby and large asteroids.

 Another powerful technique for measuring sizes of asteroids is the stellar occultation 
 timing analysis \citep{Herald2020}, which is almost completely independent of the their size or distance. Asteroids in their on-sky paths sometimes cross background stars, casting a shadow on the Earth surface. 
 Observations of these rare events from a few separate sites within the predicted shadow path ---thanks to precise timings--- enable
 almost direct asteroid size measurements in kilometers. The target silhouette can also be outlined by  
 the occultation, which can be used to verify the shape models \citep{Durech2011}. Such events sometimes lead to discoveries of asteroid satellites or rings
 \citep{Gault2022, Ortiz2017}, enabling mass determinations. 
 An important advantage of this technique is the possibility to break the symmetry between two mirror pole solutions, 
 which is inherent to photometric light-curve inversions of asteroids orbiting close to the ecliptic plane \citep{Kaas2006}.
 
 A desirable situation would be to have at least three occultation chords for each event \citep{Durech2011}. A dense 
 network of ground-based observatories is therefore required, such as the network built within the Lucky Star project\footnote{http://lesia.obspm.fr/lucky-star/}, 
 the RECON network\footnote{http://tnorecon.net/} \citep{Buie2016}, or the wide and well-organised regional networks of the International Occultation Timing 
 Association\footnote{https://occultations.org/}. With Gaia catalogues available \citep{GAIA_EDR3}, the accuracy of occultation 
 predictions has substantially improved \citep{Tanga2022}, and is expected to improve even more with the release
 of the full Gaia catalogue. Now there is a unique opportunity for successful observations, since both stellar positions 
 and asteroid orbits are determined with unprecedented accuracy, resulting in most reliable occultation predictions to date. 
 There is a time-window in which to take this opportunity, because over the following  decades both quantities will
 start to deteriorate because of uncertainties in stellar proper motions and perturbations of asteroid orbits.

 The vast majority of asteroid shape models published today are scale-free, approximate shape representations, because
 the focus is mainly on determining their spin parameters \citep[see e.g.][]{Hanus2018, ATLAS}.
 These shape models, given their angular appearance with sharp edges and large planar areas, tend to be problematic when used 
 in further applications such as thermophysical modelling or fitting their silhouettes to stellar occultations 
 \citep{Hanus2016}, with the main outcome of both being diameter determinations. 
 
 In parallel, the targets on which most of these studies have focused do not necessarily represent all of the various populations of main-belt asteroids \citep{M2015}. Most of the spin and shape
 asteroid models available today are for relatively fast-rotating targets, while the TESS mission,  for example,
 revealed that slow rotators (bodies rotating with periods of longer than 12 hours) strongly dominate in all asteroid
 populations at various sizes (e.g. around 5700 objects out of 9900 in the TESS DR1 sample\footnote{https://archive.konkoly.hu/pub/tssys/dr1/}, 
 \citep{Pal2020}). 
 
 Motivated by the trends described above, our survey \citep[see e.g.][]{M2018} 
 is designed to reconstruct precise spin and three-dimensional shape models for the most challenging, slowly rotating asteroids, and to determine their sizes with the best available methods (thermophysical modelling, and/or stellar occultation fitting). 
 Thanks to this survey, we now have the most accurate shape models and size values to date   for over 30 such asteroids \citep{M2021}.
 In this work, we present a further 16 slow rotators, most of which previously lacked shape models or even 
 good-quality light curves. Unlike our previous targets \citep[published in][]{M2018, M2019,M2021}, these 16 asteroids 
 could not be reliably scaled using thermal data because of the poor quality of the infrared datasets existing for them. Here, we therefore focus on 
 scaling them using stellar occultations.

 Section 2 describes our two observing campaigns, which target these bodies: one to obtain their light curves in multiple 
 apparitions, and another to observe them in stellar occultations. 
 Sections 3 and 4 present the methods we used for shape reconstruction and to fit the models to stellar occultation chords.
 Section 5 presents asteroid spin and shape models, verified and scaled using occultation fitting, 
 which in many cases enabled  us to identify a preferred pole solution. 
 A final section summarises our findings and presents future plans, with possible applications of our results.

 \section{Observing campaigns}
 \subsection{Photometric campaign for dense light curves}
 
 Our wide photometric campaign, which is described in detail in \citet{M2015}, is motivated by the lack of a good-quality representation of the main-belt asteroid population as a whole by spin and shape models. 
 As expected, models have mainly been constructed for targets for which this process is relatively straightforward, while others have been left behind. 
 Therefore, we focused on asteroids with long periods and low amplitudes in order to counterbalance this unequal distribution. 
 Later, slow rotators were shown to be important targets for thermal studies, as it has been speculated 
 that they might allow us to study deeper surface regolith layers \citep[see][for discussions]{HarrisDrube2016, HarrisDrube2020, M2021}. 
 Furthermore, as evident from the results from Kepler \citep{Molnar2018, Kalup2021} and confirmed by TESS missions \citep{Pal2020}, the main focus 
 in asteroid physical studies should be on slow rotators.
 
 The light-curve campaign for slow rotators relies on a wide, relatively uniformly distributed, and efficiently coordinated network of small telescopes of up to 1m in diameter (Table \ref{obs} in the Appendix presents the observing runs, sites, 
 and participating observers). This allows full coverage of the light curves of targets with rotation periods reaching 60 hours 
 within a reasonable time-frame. Data collection took a long time, partly because of the long periods 
 of our targets, but also because of the paucity of previous data, which meant it was necessary to observe each target in five or more apparitions in order to obtain a unique spin and shape model using the convex inversion method by \citet{Kaas2001}. 
 Wherever possible, our light curves are supplemented with data from the Kepler and TESS missions, and from previous 
 ground-based observations, resulting in dense light curves (see Table \ref{obs} in the Appendix for references), mostly through the 
 ALCDEF database\footnote{https://alcdef.org}.

\subsection{Stellar occultation campaign}

 In October 2020, we launched an auxiliary campaign for slow rotators. Although some of them have previously been observed 
 in stellar occultations, the results were mostly single positive chord observations, which are unusable for precise scaling \citep{Herald2019}. 
 Therefore, in cooperation with the European Section of the International Occultation Timing Association (IOTA/ES), we started 
 an occultation campaign for `Neglected Asteroids'\footnote{https://www.iota-es.de/neglected\_asteroids.html}.
 Later, this campaign gained its separate tag in the Occult Watcher Cloud\footnote{https://cloud.occultwatcher.net/},
 a tool widely used for occultation planning and coordination.
 The aim is to register multi-chord stellar occultations for slow rotators from our target list. 
 In contrast to the light-curve campaign, where hundreds of hours on-target are needed, occultation events require observations of only  
 one or two minutes, as they last only a few seconds. This means that good coordination and timing are even more important than in the 
 light-curve campaign. The occultation predictions were made 
 using the JPL Horizons database for asteroid orbits\footnote{https://ssd.jpl.nasa.gov/horizons/},   
 and Gaia EDR3\footnote{https://www.cosmos.esa.int/web/gaia/earlydr3} for star positions. Although these events are rare, 
 the campaign has already resulted in a dozen or so successful occultation observations, with the number of positive chords per event reaching nine 
 (see e.g. the events from the years 2021 and 2022 in Figures \ref{439occ} and \ref{566occ}). 

 Archival occultation data were downloaded from the Planetary Data System (PDS) archive\footnote{http://sbn.psi.edu/pds/resource/occ.html} 
 \citep{Herald2019}, and more recent ones were obtained from the archive of the Occult 
 programme\footnote{http://www.lunar-occultations.com/iota/occult4.htm}, or directly from the regional IOTA 
 coordinators, who perform data evaluation and vetting in order to achieve good-quality and consistent occultation results. 
 Most of the occultation observers are amateur astronomers, which makes this campaign 
 a good example of a successful professional--amateur, or `pro--am', collaboration.
 Table \ref{occult_obs} in the Appendix provides the observer names and site locations 
 for each occultation event used here.

\section{Convex inversion with shape regularisation}

 The majority of asteroid models presented in this work were constructed
 using the common convex inversion method of \citet{KaasTorppa2001,
 Kaas2001}. When light-curve data are sufficiently abundant to uniquely determine the
 sidereal rotation period and the pole direction, the shape model
 reconstructed from the Gaussian image is usually such that its rotation
 axis $z$ is very close to the principal axis of the inertia tensor with
 the largest moment of inertia. Shapes that fit the data but strongly
 violate this condition have to be rejected as physically unacceptable
 solutions of the inverse problem. However, in some cases the data are
 so abundant that it is clear that, although the solution is correct, the
 shape is not physically acceptable. By `squeezing' the shape along the
 rotation axis, the inertia tensor can be changed in such a way that its
 greatest axis is close to the rotation axis. However, this can
 only be done after the best-fit model is found, and we must also check whether or not this
 shrinking along the rotation axis affects the light-curve fit
 significantly.
      
\begin{figure*}
\centering
\includegraphics[width=0.7\textwidth]{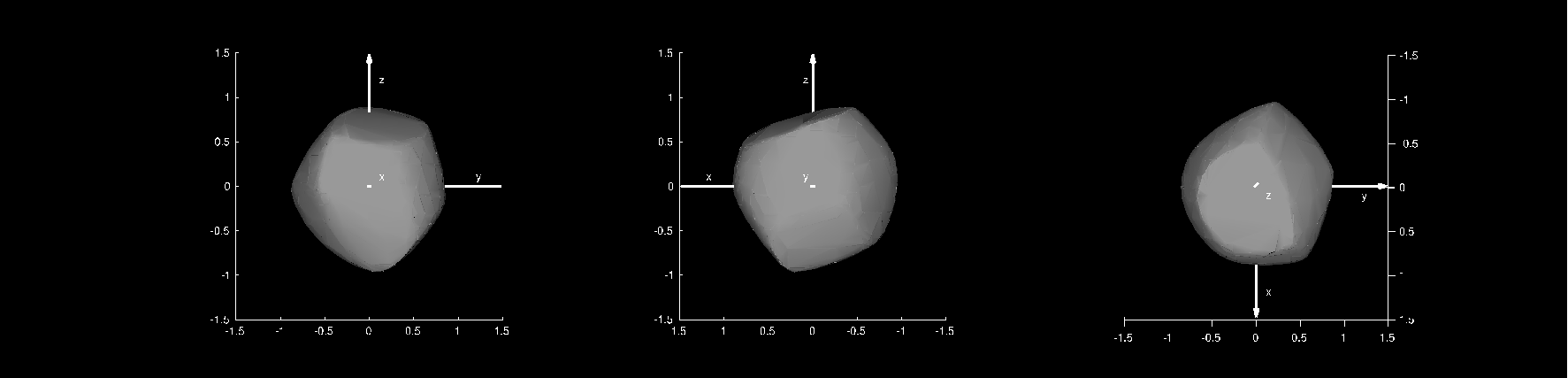}

\includegraphics[width=0.7\textwidth]{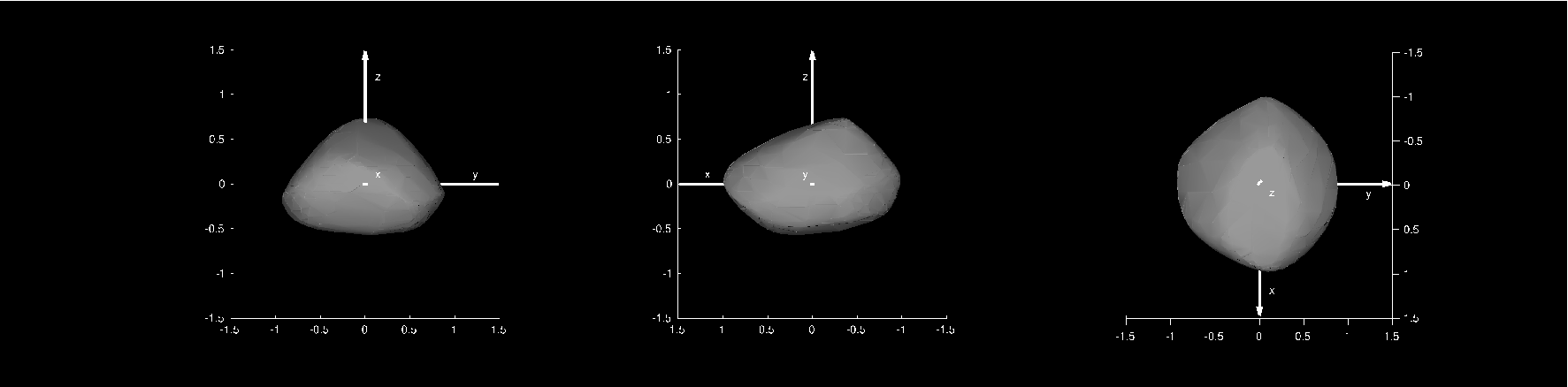}
\caption{Two versions of shape model 2 for (566) Stereoskopia: without shape regularisation (top) and with regularisation 
applied (bottom), using the same starting parameters otherwise. The three views show, from left to right: 
two equatorial views separated by  90\deg  in phase, and the pole-on view. We note that the pole-on silhouettes 
generally agree between two versions of the shape model, as it is the equatorial cross-section that most effectively influences the light curves.}
\label{566reg}
\end{figure*}

 When working with the Gaussian image during the optimisation, the shape
 is not known and we cannot regularise it to obtain the maximum moment of
 inertia around the rotation axis. However, even without computing the
 inertia tensor precisely, the shapes that are not rotating along the
 shortest axis are elongated along the $z$ axis. This elongation means
 that the cross-section viewed from the pole direction is smaller than
 that from the equatorial view ($xy$ plane). Although the 3D shape is not
 available during optimisation, the areas of surface facets and their
 normals are known, and so their projections can be computed. The
 regularisation term $R$ that we introduce is
    \begin{equation}
       R = \frac{\sum\limits_{i=1}^N \sigma_i \, \left( |\hat{\vec{n}}_i \cdot \hat{\vec{x}}| + |\hat{\vec{n}}_i \cdot \hat{\vec{y}}| \right)}{2 \sum\limits_{i=1}^N \sigma_i \, |\hat{\vec{n}}_i \cdot \hat{\vec{z}}|} \,,
    \end{equation}
 where $\hat{\vec{n}}_i$ are unit normals to surface elements of areas
 $\sigma_i$, and $\hat{\vec{x}}, \hat{\vec{y}}, \hat{\vec{z}}$ are unit
 vectors along the coordinate axes; the summation is over all $N$
 surface elements. The more extended the shape along the $z$ axis, the
 higher $R$, and so the penalty function that is added to the $\chi^2$
 is $R^2$ times some weight $w$. 
By increasing $w$, the shape model is forced to have projected areas along the $x$ and $y$ axes that are 
smaller than along the $z$ axis, which makes it more oblate and ensures physically correct rotation.
This way, we were able to produce physically realistic shapes for
(439)~Ohio and (566)~Stereoskopia, for which the standard light-curve
inversion without regularisation led to unphysical shapes. With
regularisation, the fit to the data remained almost the same 
(RMS changed from 0.0073 to 0.0074 for the example shown in Fig. \ref{566reg}), 
meaning that the dimension along the rotation axis was not constrained by the
data, and the regularisation helped to produce realistic shapes without affecting the fit.  
A `negative regularisation' mentioned in the following section means using the
regularisation term $ 1/R^2 $, which produces more stretched shapes when
needed.

\begin{figure*}
\centering
\includegraphics[width=0.7\textwidth]{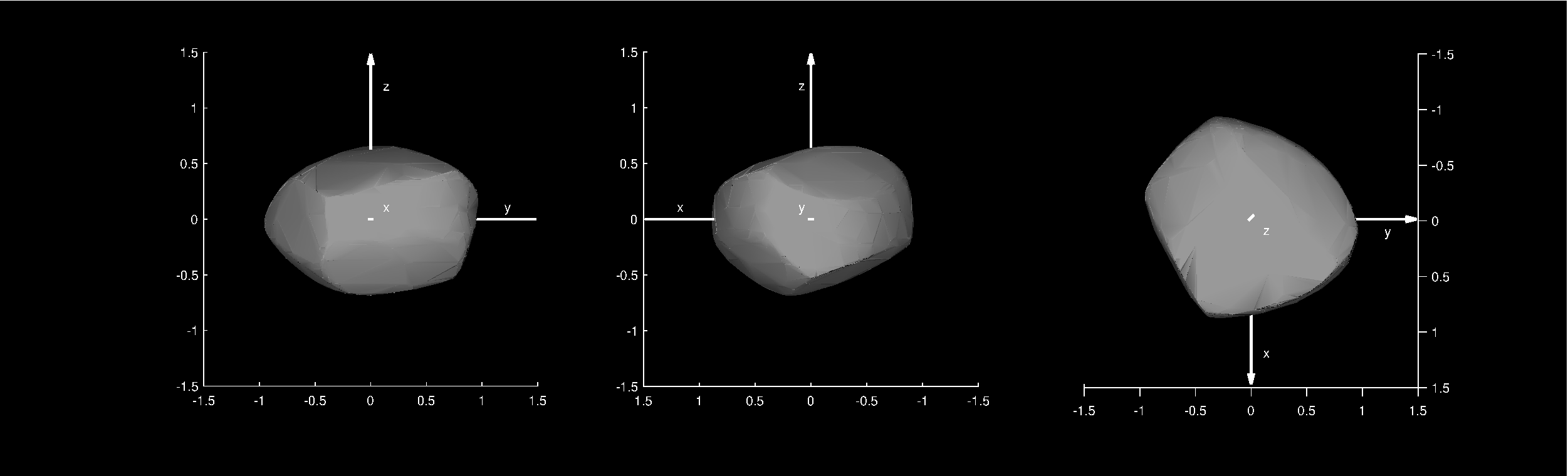}

\includegraphics[width=0.7\textwidth]{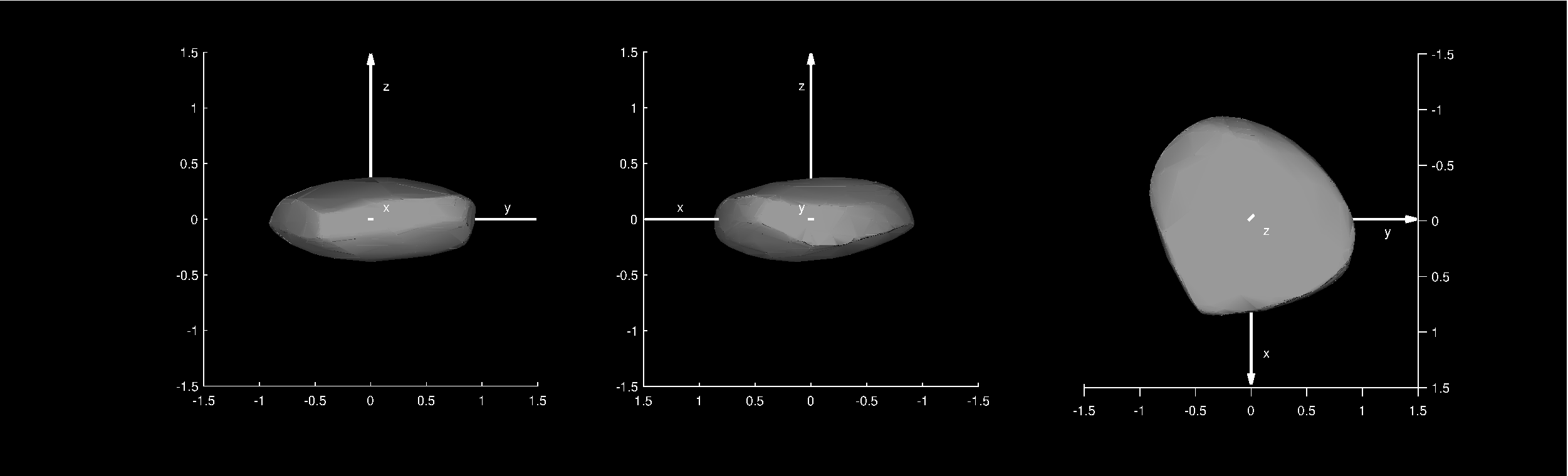}
\caption{Two versions of shape model 2 for (412) Elisabetha. The high-regularisation model 
(bottom) still fits the light curves on formally acceptable level, but is unrealistically flat.
See Section 4 for discussion.}
\label{Elisabetha_reg}
\end{figure*}

 \section{Occultation fitting}

 As mentioned in Sect.~\ref{sec:intro}, shape models derived from
 photometry do not carry information about the size; they are scale-free.
 We scaled them using the same approach as \cite{Durech2011} and recently
 \cite{M2021}. The positions of the observers detecting an occultation were
 projected on the fundamental plane, which is the plane crossing the
 geocenter and is perpendicular to the direction from the asteroid to
 the star. We then computed the orientation of the shape model for the
 mean time of occultation and projected it also to the fundamental plane.
 Because the time span of occultation timings is usually much shorter
 than the rotation period of the  asteroid, the rotation is neglected, and the
 silhouette is assumed to be constant. The size of the asteroid and the
 relative position of its projection on the fundamental plane with
 respect to chords are three parameters optimised to converge to the best
 agreement between the projection of the asteroid shape and occultation chords. The
 goodness of the fit is described by a standard $\chi^2$ measure with timing
 uncertainties taken into account.
 
 The uncertainty of the derived size is affected by many factors: the
 number of occultation events, the number of chords in individual
 occultations and their distribution across the silhouette, the timing errors
 of individual chords, and the uncertainty on the  projection of the asteroid
 (and therefore the uncertainty on its shape and spin state). When there
 are several occultations by one asteroid with many chords covering the
 whole projected disc, the RMS residual in kilometers
 is one estimation of the size uncertainty (see the last column in Table \ref{results}).

 However, the uncertainty on the sizes of the asteroids determined in this way does not take 
 into account the uncertainty on the models themselves. The main problem is that relative light curves 
 supporting the models are largely insensitive to a vertical dimension of the shape model, as the 
 ratio between the biggest and smallest projected area in a given aspect remains similar 
 regardless of this dimension. To tackle this problem, we used the ability of inertia regularisation 
 to influence the vertical stretch of the derived models. We created ten versions of each model, 
 with both positive and negative regularisations, creating flatter and rounder shapes 
 that would still fit the light curves at a similar level to the nominal solution, 
 and at the same time remaining physical (rotating around the shortest axis) and realistic 
 (e.g. not being extremely thin).
 Models fitting the light curves within 10\% RMS of the best-fit solution are accepted following \cite{Kaas2001}. 
 With tens of light curves and thousands of photometric points, the 10\% increase in RMS is really 
 an extreme limit and is the upper boundary for `acceptable' models, as shown by the example in Fig. \ref{Elisabetha_reg}. 
 The upper model is the nominal one, without inertia regularisation, while the bottom one was created with a relatively 
 high level of regularisation. The other model fitted the light curves within 7\% of the best solution (RMS of 0.0117 vs. 0.0110 
 for the best fit) and is already unrealistically flattened. Another approach relates the $\chi^2$ limit  
 to the number of data points \citep[see e.g.][Appendix A]{Hanus2021, Durech2022}.

 Those variations of each model have been fitted to occultations as well, resulting in a pool 
 of plausible sizes. This also revealed the shape versions fitting the occultations with 
 substantially higher RMS, or not fitting some chords at all. In some cases, this strengthened 
 the case for a preference for one of two pole solutions, where all the versions of shape 
 for one pole resulted in a similar value for the size and a consistently small RMS of the fit, 
 while for the other pole the sizes varied substantially, and the RMS was high. Such cases 
 are described in more detail in the following section.
 
 The key point here was to check whether the range of sizes derived in this way would be higher or smaller 
 than the nominal RMS of the occultation fit, which is mainly governed by the occultation timing 
 uncertainties. When it was higher, it replaced the nominal RMS value as a measure of diameter uncertainty
 in Table \ref{results}. 
 However, when the range of sizes was smaller than the RMS of the fit, this meant that the uncertainty here 
 was dominated by the occultation timing errors or the convex approximation of the shape.

\section{Results}

\begin{figure*}
\includegraphics[width=0.33\textwidth]{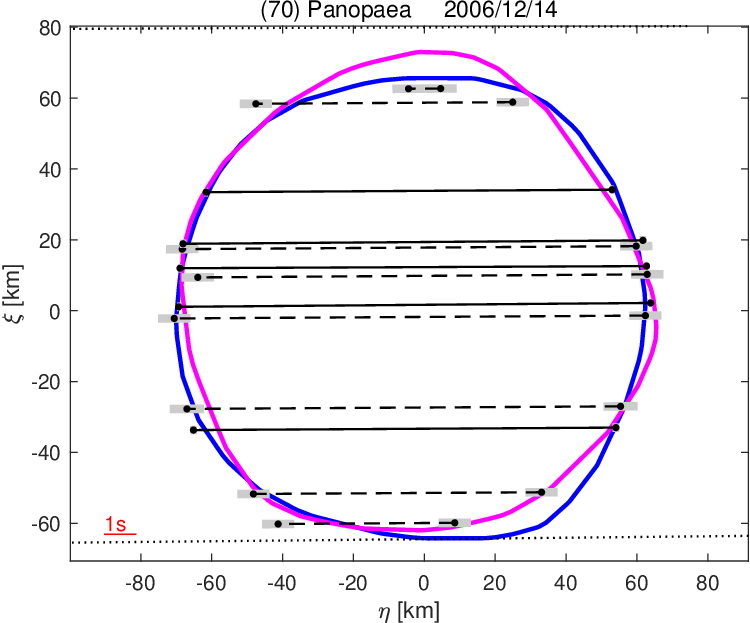}
\includegraphics[width=0.33\textwidth]{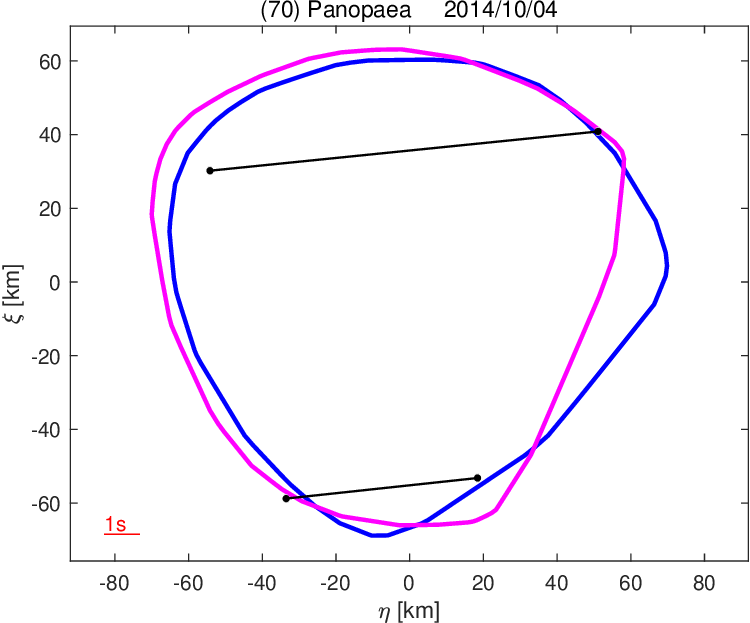}
\includegraphics[width=0.33\textwidth]{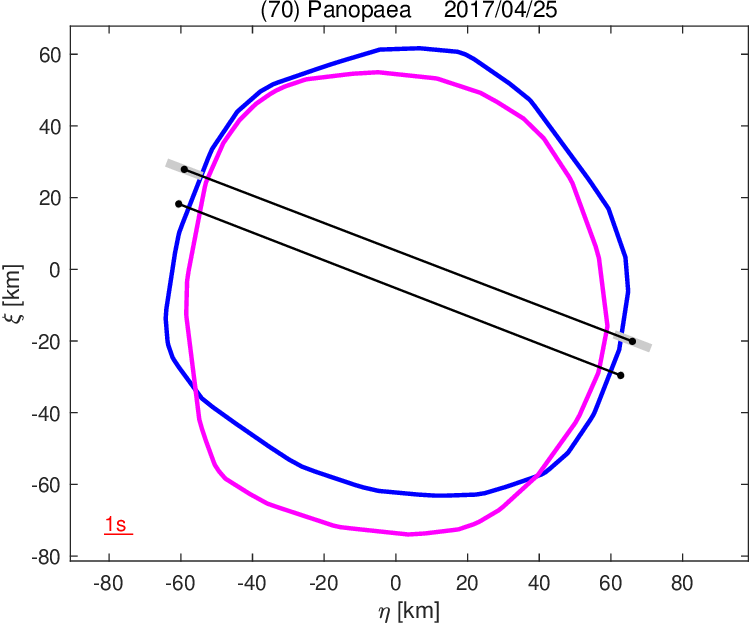}
\caption{(70) Panopaea model fit to occultation chords shown in the fundamental plane. North is up and west is right. 
Dashed lines mark visual observations, as opposed to video or photoelectric 
ones marked with solid lines. Grey segments correspond to timing uncertainties. 
Blue contour: Instantaneous silhouette of shape model 1 (see Table \ref{results}). Magenta: Same but for model 2. 
In this case, occultations show no preference for any of the models.}
\label{70occ}
\end{figure*}

\begin{figure*}
\includegraphics[width=0.33\textwidth]{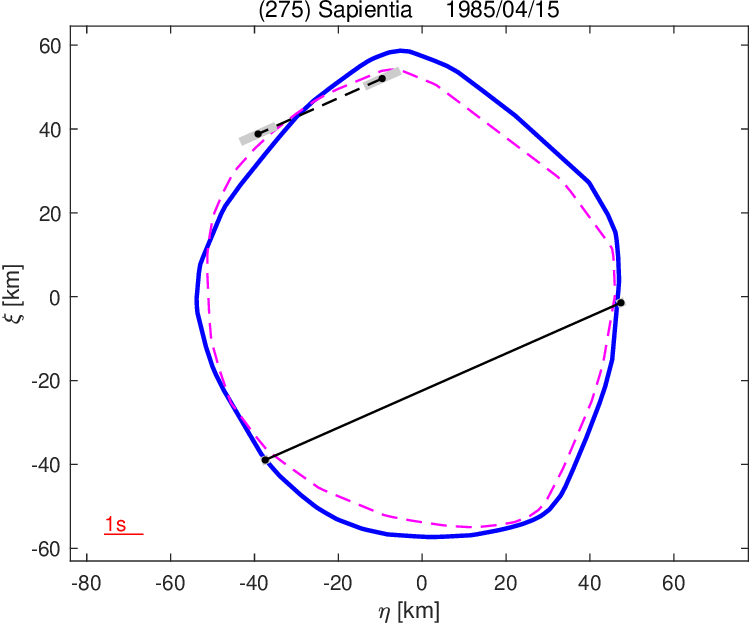}
\includegraphics[width=0.33\textwidth]{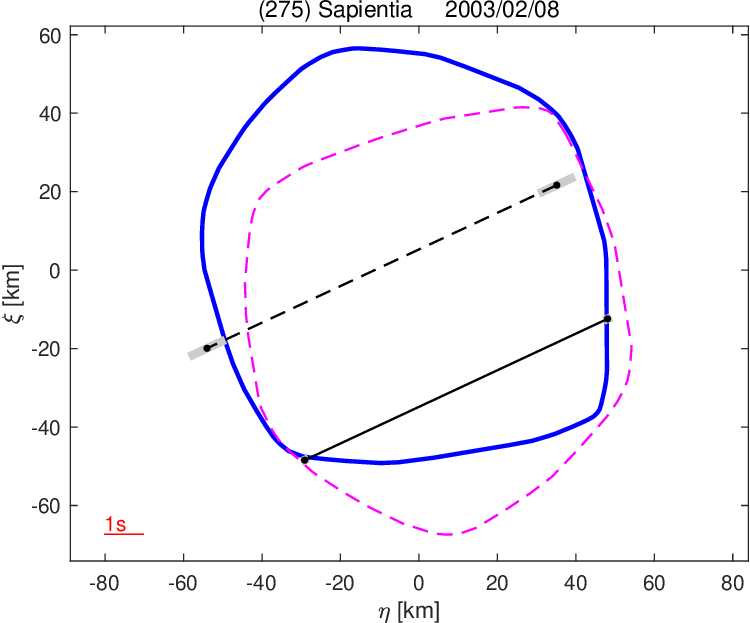}
\includegraphics[width=0.33\textwidth]{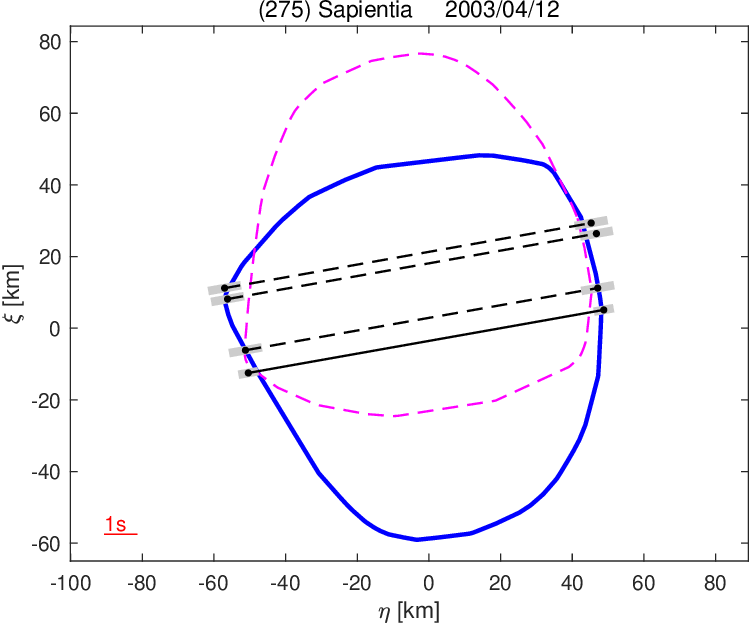}
\includegraphics[width=0.33\textwidth]{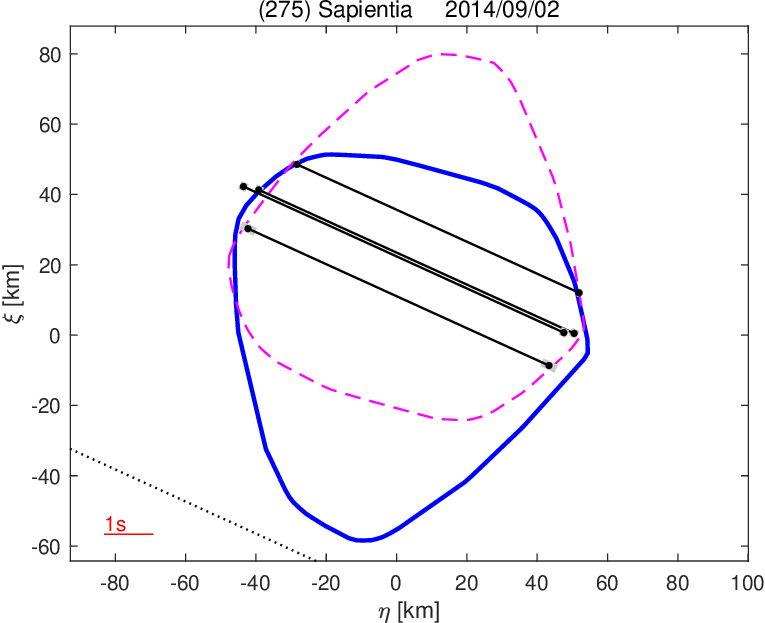}
\includegraphics[width=0.33\textwidth]{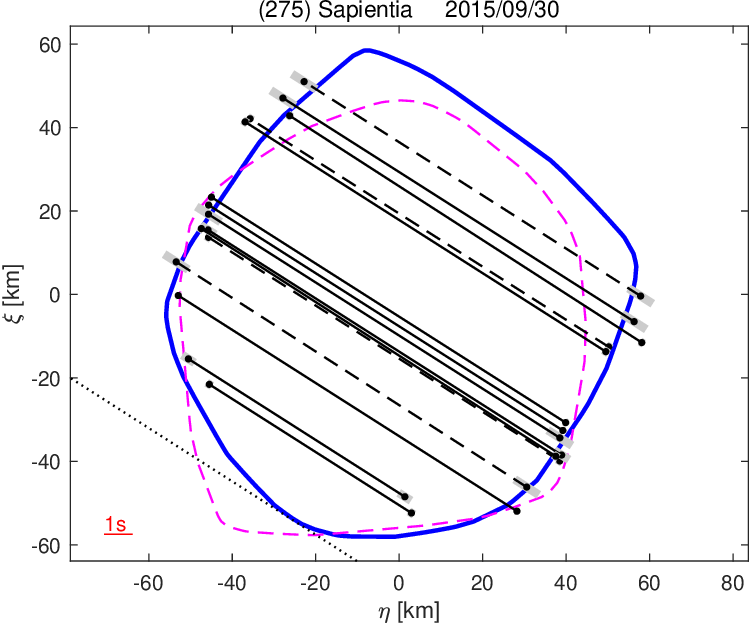}
\includegraphics[width=0.33\textwidth]{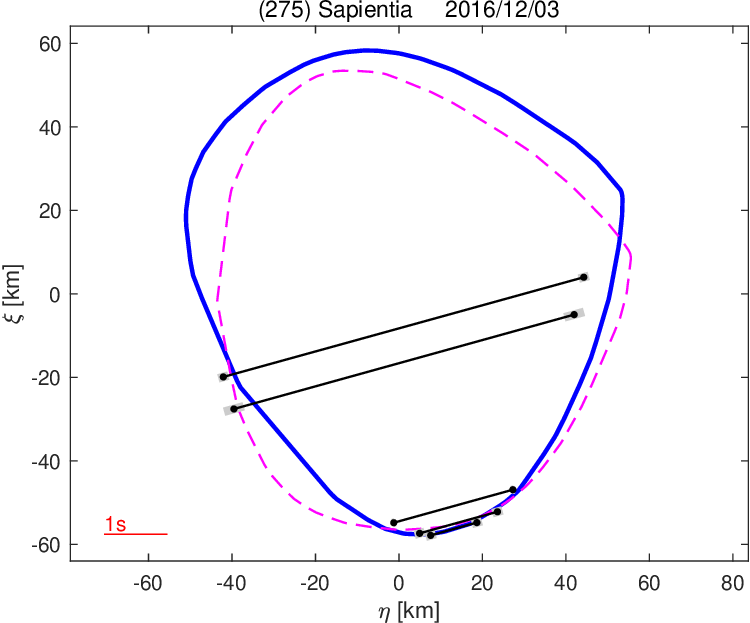}
\includegraphics[width=0.33\textwidth]{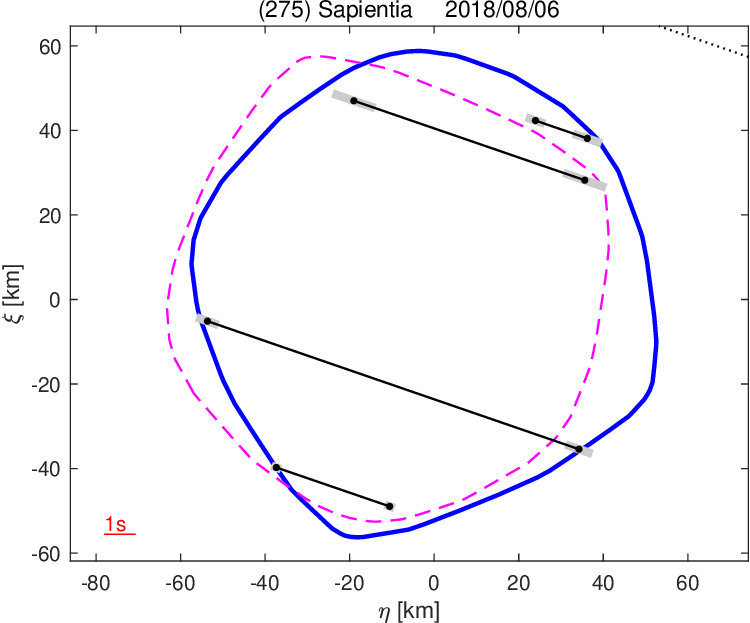}
\includegraphics[width=0.33\textwidth]{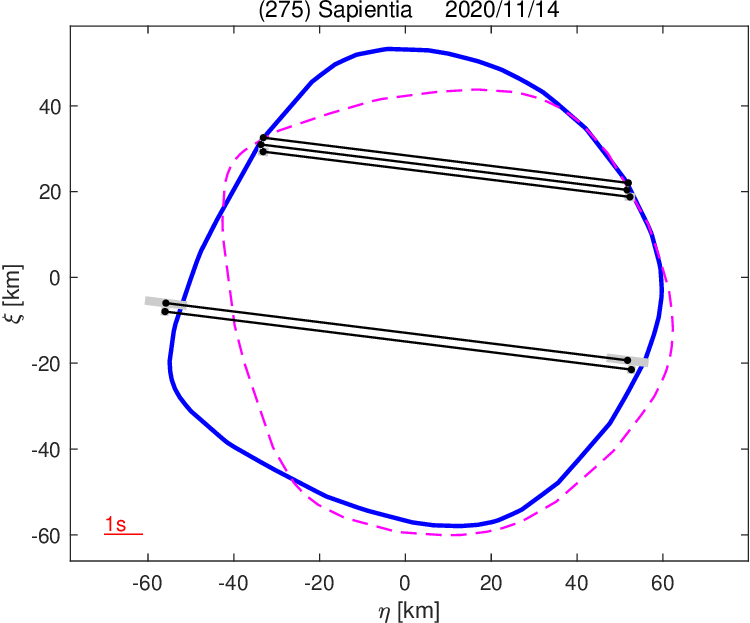}
\includegraphics[width=0.33\textwidth]{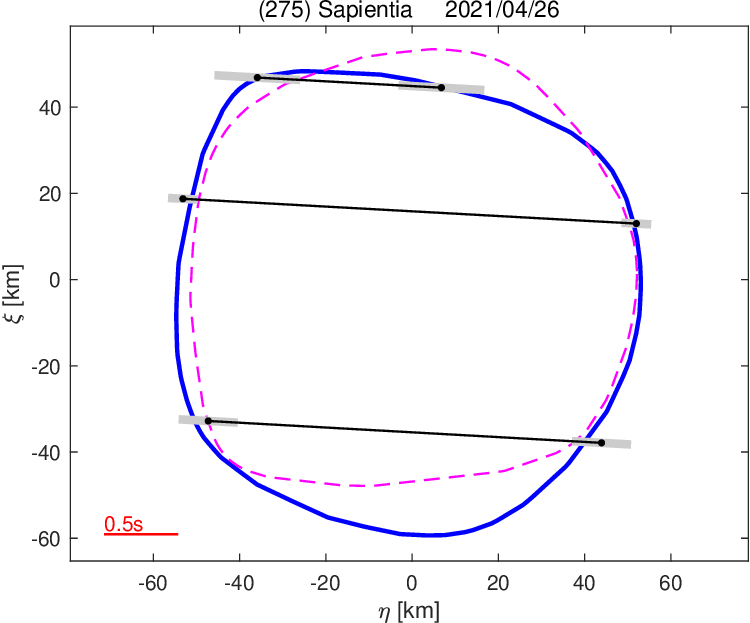}
\caption{(275) Sapientia model fit to occultation chords. Here, model 2 is preferred (blue solid contour).}
\label{275occ}
\end{figure*}

\begin{figure*}
\begin{center}
\includegraphics[width=0.25\textwidth]{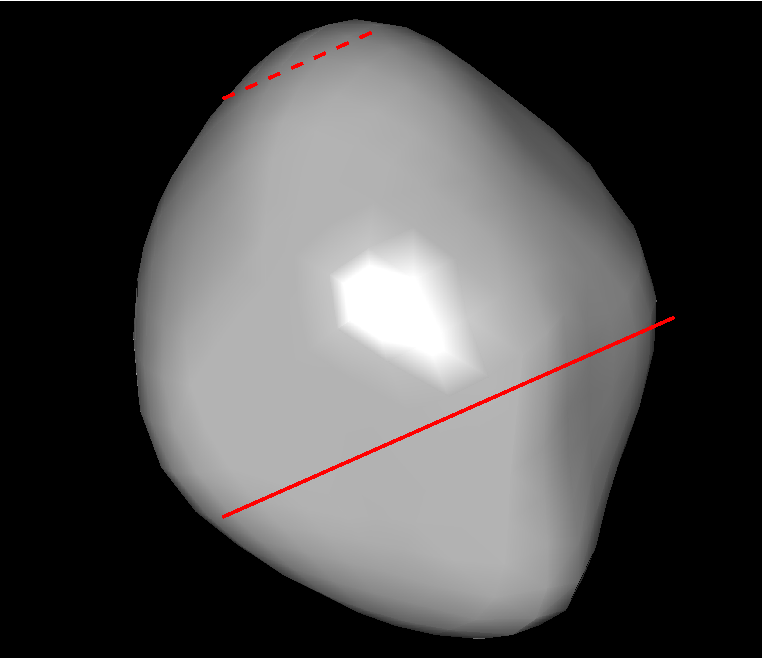}
\includegraphics[width=0.25\textwidth]{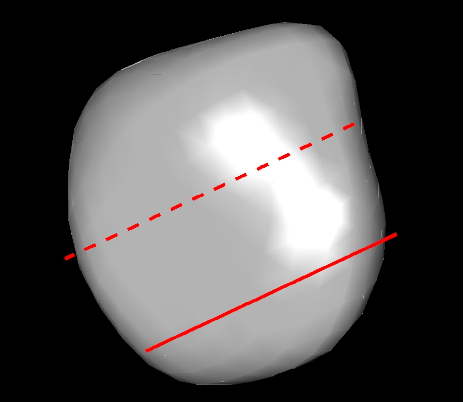}
\includegraphics[width=0.25\textwidth]{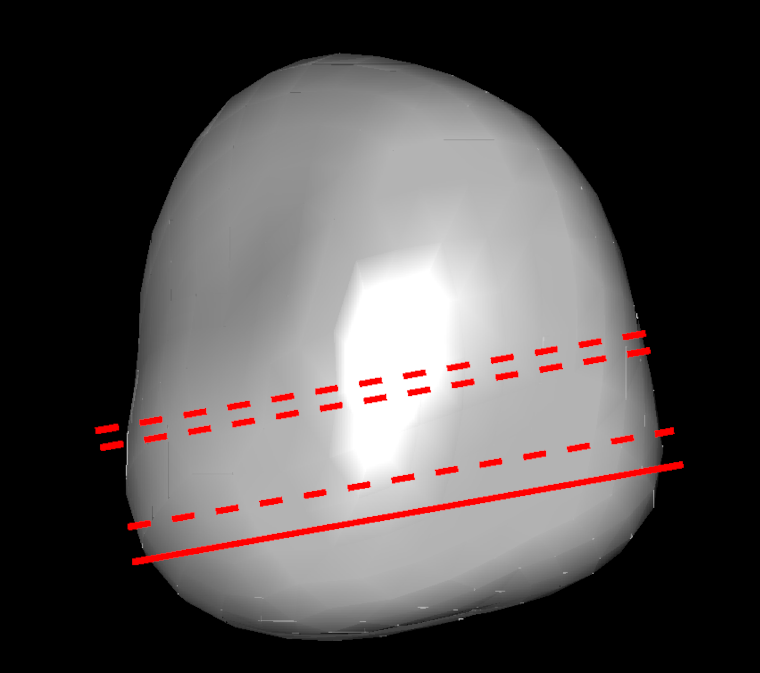}
\includegraphics[width=0.25\textwidth]{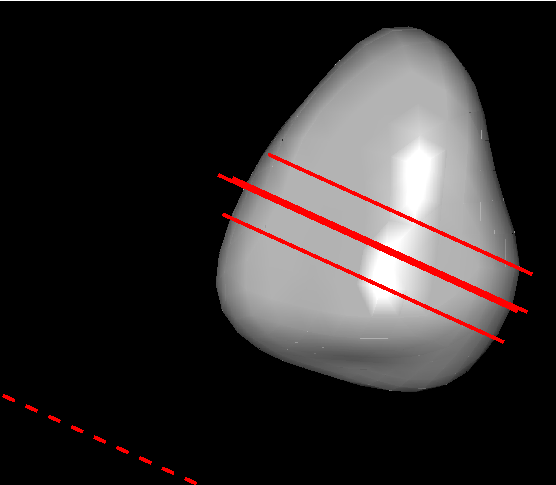}
\includegraphics[width=0.25\textwidth]{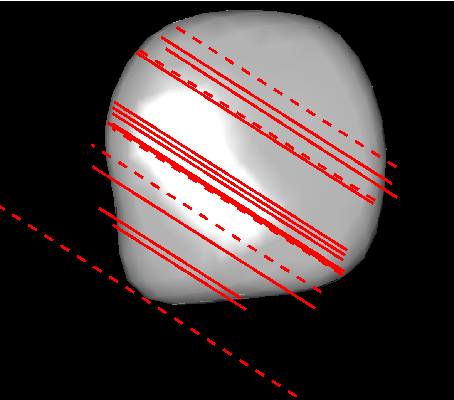}
\includegraphics[width=0.25\textwidth]{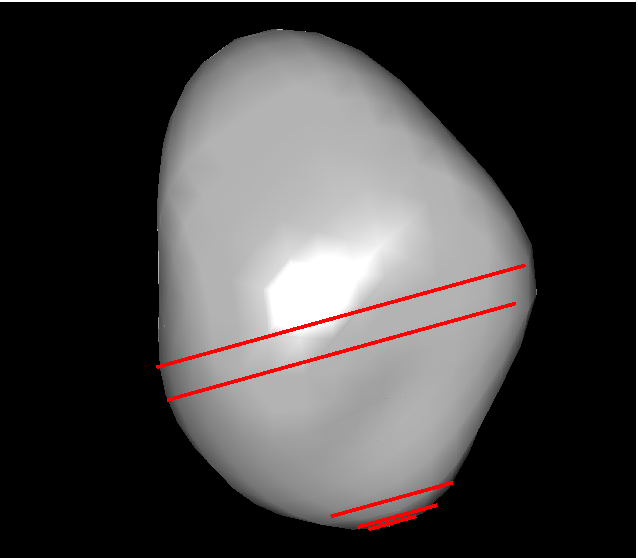}
\includegraphics[width=0.25\textwidth]{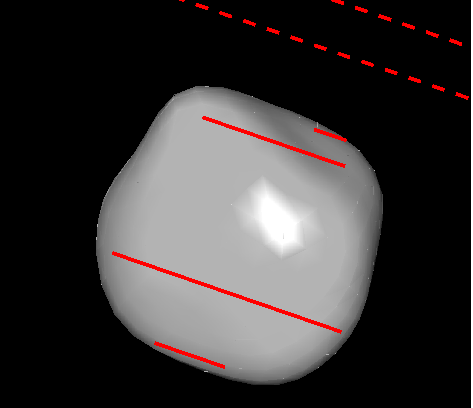}
\includegraphics[width=0.25\textwidth]{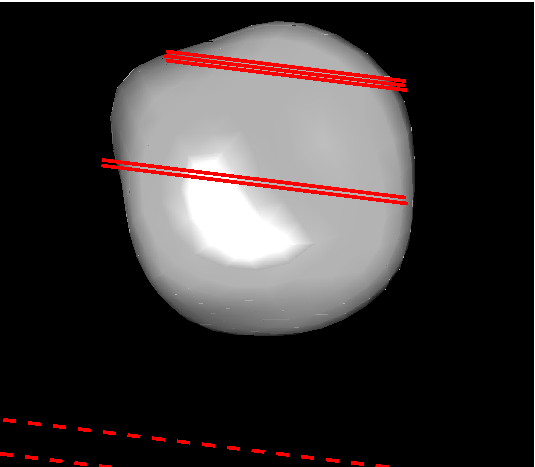}
\includegraphics[width=0.25\textwidth]{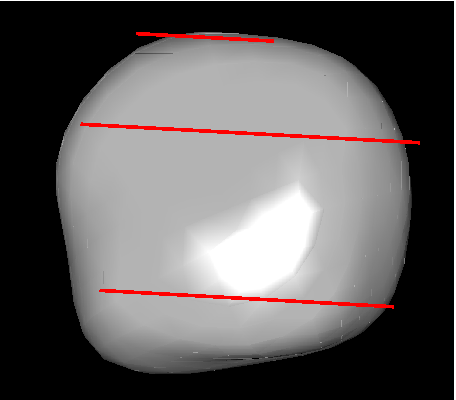}
\end{center}
\caption{(275) Sapientia, ADAM solution pole 1, shown with occultation chords used to construct the model. Occultation events are the same as in Fig.~\ref{275occ}.}
\label{275ADAM1}
\end{figure*}

 We used the rich datasets of dense light curves from our campaign (see Figures
 \ref{70composit2017}-\ref{806composit2021} in the Appendix) for spin and shape modelling with the classical light-curve
 inversion method ---or its slight variation with shape regularisation--- in order to
 make non-physical models rotate in a physical manner around the axis of
 largest inertia (see Section 3). Table \ref{results} summarises the spin
 parameters and diameters determined in this work. The fit of the shape
 models to occultation chords can be seen in Figures \ref{70occ} -
 \ref{806occ}. A good fit has manifold benefits: primarily it serves to
 scale the models in kilometers, but it also verifies the shape
 features, and often clearly points to the solution that must be close to
 the real one, making its mirror counterpart less probable. 
 We should note that the asteroid sizes determined here are
 diameters of volume-equivalent spheres, while sizes determined from
 infrared measurements are for surface-equivalent spheres. In any case, the
 difference between the two diameters for the same shape model is at the
 level of 5\% or less, which is smaller than the uncertainty on the
 models themselves, especially with NEATM spherical shape approximation.
 It is therefore justified to directly compare sizes determined with the two
 methods, without any scaling factors.

 In parallel, in a few justified cases (275 Sapientia, 464 Megaira, 530 Turandot) where the occultation results suggest 
 the presence of non-convex shape features, we additionally used the all-data asteroid modelling \citep[ADAM,][]{Viikinkoski2015}  method 
 for shape reconstruction. 
 This technique uses both light curves and occultation timings as input data in a single optimisation process. The results 
 are presented in Figures \ref{275ADAM1}, \ref{275ADAM2}, \ref{464ADAM}, and \ref{530ADAM} and in Table \ref{results}. 
 They can be compared with the respective shape models constructed with 
 traditional convex inversion without occultations; and fitted to occultations later (Figures \ref{275occ}, \ref{464occ}, and \ref{530occ}). 
 The determined sizes and spin parameters are consistent between the two methods, while ADAM shapes show a slightly better fit to the occultations, 
 as expected. The size uncertainties for ADAM shape models presented in Table \ref{results} come from the scatter of solutions 
 and are clearly underestimated.

 For each model, Table \ref{results} presents the spin axis ecliptic
 coordinates, with the sidereal period of rotation, together with the
 uncertainties on each parameter. The table also contains details of the
 light-curve datasets, and the volume-equivalent diameters with 
 uncertainties based on the deviation of the shape model silhouettes
 from the occultation chord ends, or on the scatter of sizes for 
 various versions of the same shape, whichever source dominated the error budget. 
 The preferred pole solutions after occultation fitting 
 are marked in boldface. In Figures \ref{70occ} - \ref{806occ}, lesser preferred 
 solutions are marked with dashed magenta contours, and the solid
 blue contours mark the preferred solutions. In the case of only slightly
 preferred solutions, both contours are solid lines, but the preferred
 solution is still shown in blue. In some cases, the preference seems to be based on single chords, 
 but it is the duration rather than the chord absolute position that is the deciding factor in such cases
 The duration of the occultation events is usually recorded correctly, while the gross timing error
 might shift the whole chord, but does not change its length. There are of course exceptions to this rule; 
 see the discussion on (275) Sapientia in Section 5.
 In the cases with no preference for the spin
 solution,  both coloured contours are again solid; see the text
 and figure captions for details.

 Additionally,   
Table \ref{diam_lit} contains the 
range of previously determined sizes, with the source references. Please
note the diversity of the literature diameters, differing by 50\%
for some targets between different works.  Below we describe all
targets in more detail.

\begin{table*}
\caption{Spin parameters and sizes of asteroid models obtained in this work. 
 The columns contain asteroid name, J2000 ecliptic coordinates $\lambda_p$, $\beta_p$ of the spin axis solution, and the sidereal rotation period $P$, 
 with the mirror pole solution in the second row. 
 The following columns provide the main characteristics of the light-curve dataset (same for both pole solutions): observing span in calendar years, 
 number of apparitions ($N_{app}$) and number of light curves ($N_{lc}$.)
 The last two columns give volume-equivalent diameter $D$ and its RMS residual from the stellar occultation fitting. 
 Boldface highlights the solution preferred by means of occultation fits. See Section 4 for the discussion on diameter uncertainties.
}
\label{results}
\begin{tabular}{lllllcccr}
\hline
    Asteroid   & \multicolumn{2}{c}{Pole}              &   P           & Observing span & $N_{app}$ & $N_{lc}$ &   D    & D RMS  \\
               & $\lambda_p [\deg]$ & $\beta_p [\deg]$ & [hours]       &      (yr)      &           &          &  [km]  & [km]   \\
\hline                                                                                          

 (70) Panopaea & $ 42 \pm 5 $ & $+27 \pm 3 $ & $15.80440 \pm 0.00002$  &  1980 -- 2019  &  7        &  122     &   128  $\pm$  7  & 7 \\

               & $240 \pm 6 $ & $+26 \pm 4 $ & $15.80439 \pm 0.00001$  &  &  &  & 128$^{+7}_{-11}$  & 7  \\

\hline                                                                                  

(275) Sapientia&  $85 \pm 11$ & $-10 \pm 18$ & $14.93045 \pm 0.00005$  & 1998 -- 2018   &  7        &   38     &  98$^{+6}_{-11}$ & 6 \\

               & ${\bf 264 \pm 4}$ & ${\bf  -1 \pm 20}$ & ${\bf 14.93045 \pm 0.00005}$ &  &  &  &{\bf 103$^{+6}_{-7}$} & 6 \\

(275) Sapientia (ADAM)&  $82 \pm 8$ & $-11 \pm 17$ & $14.93044 \pm 0.00005$  & 1998 - 2018    &  7        &   38     & 100 $\pm$ 1 & - \\

                      & ${\bf 260 \pm 7 }$ & ${\bf -2 \pm 20}$ & ${\bf 14.93044 \pm 0.00005}$  &  &  &  &{\bf 100 $\pm$ 1} &  - \\

\hline                                                                                  
(286) Iclea    & ${\bf 31 \pm 5}$ & ${\bf +13 \pm 4}$ & ${\bf 15.36120 \pm 0.00004}$  & 2002 -- 2019 &  9  & 52 & {\bf 86$^{+13}_{-7}$} & {\bf 2} \\
               & $196 \pm 5 $ & $+44 \pm 7$  & $15.36114 \pm 0.0007$   &  &  &  & 69$^{+3}_{-12}$ & 3\\
\hline                                                                                  
(326) Tamara   &  ${\bf 79 \pm 1} $ & ${\bf -3 \pm  3}$ & ${\bf 14.46130 \pm 0.0005}$   & 1981 -- 2019   &   9  & 69 &  {\bf 77$^{+5}_{-10}$} & {\bf 5} \\
               & $242 \pm 1 $ & $-36 \pm  3$ & $14.46136 \pm 0.0004$   &  &  &  & 81 $\pm$ 9 & 9 \\ 
\hline                                                                                  
(412) Elisabetha& $ 3 \pm 20$ & $+17 \pm 7 $ & $19.65610 \pm 0.00003$  & 1990 -- 2021   &   7       & 77        &  119 $\pm$ 9 & 9 \\
               & ${\bf 191 \pm 23}$& ${\bf +52 \pm 8}$ & ${\bf 19.65618 \pm 0.00004}$  &  &  &  &  {\bf97$^{+7}_{-14}$} & {\bf 4} \\
\hline                                               
(426) Hippo    & $ 62 \pm 55$& $ -49 \pm 20$ & $ 67.5038 \pm 0.0005$  & 1993 -- 2021 & 7 & 103  &  129$^{+19}_{-8}$ & 6 \\
               & ${\bf 223 \pm 80}$& $ {\bf -89 \pm 17}$ & $ {\bf 67.5041 \pm 0.0005} $  &  &  &  &  {\bf 122 $\pm$ 4} &  {\bf 4}\\
\hline                                                                                  
(439) Ohio     & ${\bf 308 \pm 50}$ & ${\bf -61 \pm 7}$ & ${\bf 37.46726 \pm 0.00005}$  & 1984 -- 2022 &  10  & 66 & {\bf 74$^{+3}_{-8}$} & {\bf 2} \\
\hline                                                                                          
(464) Megaira  & $ 47 \pm 2 $ & $ +16 \pm 5 $ & $ 12.878572 \pm 0.000002$ &  1979 -- 2020 &  9  &  85  & 76$^{+3}_{-6}$ &  3 \\
               & $236 \pm 5 $ & $+38 \pm 10$ & $12.878572 \pm 0.000002$ &  &  &  &  76$^{+2}_{-11}$ & 2\\
(464) Megaira (ADAM)&$ 45 \pm 3 $ & $+13 \pm 5 $ & $12.878573 \pm 0.000001$ &  1979 -- 2020 &  9  &  85  & 75 $\pm$ 1 & -\\
               & $234 \pm 4 $ & $+34 \pm 14$ & $12.878573 \pm 0.000001$ &  &  &  &  79 $\pm$ 1 & - \\
\hline                                                                                  
(476) Hedwig   & ${\bf 49 \pm 4}$ & ${\bf +60 \pm 7}$ & ${\bf 27.2403 \pm 0.0006}$  & 1982 -- 2018 &  6  & 58 & {\bf 116$^{+6}_{-16}$ } & {\bf 6} \\
               & $218 \pm 4 $ & $+36 \pm 4$  & $27.2404 \pm 0.0005$   &  &  &  & 122 $\pm$ 10 & 10\\
\hline                                                                                  
(524) Fidelio  & $52 \pm 50$ & $+76 \pm 6$ & $14.171031 \pm 0.000005$  & 2005 -- 2019 &  6  & 31 &  66 $\pm$ 5 & 5 \\
               & ${\bf 186 \pm 25}$ & ${\bf +54 \pm 10}$ & ${\bf 14.171042 \pm 0.000005}$   &   &   &  &  {\bf 67 $\pm$ 3} & {\bf 3} \\
\hline                               
(530) Turandot & $42 \pm  3$ & $+28 \pm 5$ & $19.95240 \pm 0.00009$    & 1986 -- 2019 &  8  & 62 & 89 $\pm$ 11 & 11\\
               & ${\bf 226 \pm 9}$ & ${\bf +54 \pm 10}$ & ${\bf 19.95231 \pm 0.000009}$   &   &   &  &  {\bf 89 $\pm$ 6} & {\bf 6}\\
(530) Turandot (ADAM)& $39 \pm  6$ & $+31 \pm 11$ & $19.95239 \pm 0.00009$   & 1986 -- 2019 &  8  & 62 & 89 $\pm$ 2 & - \\
               & ${\bf 223 \pm 9}$ & ${\bf +60 \pm  9}$ & ${\bf 19.95232 \pm 0.000009}$   &   &   &  &  {\bf 89 $\pm$ 2} & - \\
\hline                                                                                          
(551) Ortrud   & $149 \pm 35 $& $-63 \pm  9 $& $17.41924 \pm 0.00003$ & 2006--2021 &  7  &  88  &  85 $\pm$ 11 & 11\\
               & $305 \pm 30 $& $-66 \pm  8 $& $17.41921 \pm 0.00002$ &  &  &  &  86 $\pm$ 10 & 10\\ 
\hline
(566) Stereoskopia& ${\bf 164 \pm 15}$ & ${\bf -2 \pm 2}$ & ${\bf 12.08466 \pm 0.00006}$  & 1990 -- 2022 &  6  & 35 & {\bf 148 $\pm$ 8} & {\bf 8}\\
               & $338 \pm 9 $ & $-13 \pm 2$  & $12.08463 \pm 0.00003$   &  &  &  & 148 $\pm$ 11 & 11\\
\hline                                                                                                 
(657) Gunlod   & $127 \pm 25$ & $ +61 \pm 6$ & $ 15.92872 \pm 0.00003$  & 1984 -- 2022 &  6  & 47   &  37 $\pm$ 3  & 3 \\
               & ${\bf 252 \pm 50}$ & ${\bf +51 \pm 20}$ & ${\bf 15.92870 \pm 0.00004}$  & 1984 -- 2022 &  6  & 47  & {\bf 39$^{+2}_{-1}$} & {\bf 1} \\ 
\hline                                                                                  
(738) Alagasta &  $67 \pm 7 $ & $-47 \pm  5$ & $17.8888 \pm 0.0005$    & 2015 -- 2020   &   5       & 41       &  56 $\pm$ 6 & 6\\
               & $246 \pm 10$ & $-41 \pm  8$ & $17.8889 \pm 0.0006$    &  &  &  &  54  $\pm$ 4 & 4\\
\hline                                                                                  
(806) Gyldenia &  $34 \pm 12$ & $+39 \pm  2$ & $16.85695 \pm 0.00003$  & 2010 -- 2021   &   8       & 63       &  57 $\pm$ 8 & 8\\
               & $235 \pm 15$ & $+42 \pm  5$ & $16.85699 \pm 0.00007$  &  &  &  &  55  $\pm$ 4 & 4\\
\hline
\end{tabular}
%\end{small}
\end{table*}
%$

\begin{table*}[h!]
\begin{center}
\noindent
\caption{{\small Previously published diameters for the asteroids studied here.
Minimum and maximum diameters are given with their uncertainties and references.
For the full list, see the MP3C database.
 }}
\begin{tabular}{lcccc}
\hline
 Asteroid         & D$_{\text{min}}$    & D$_{\text{min}}$   & D$_{\text{max}}$    & D$_{\text{max}}$ \\
                  &      [km]           &   reference        &      [km]           &   reference      \\
\hline
 (70) Panopaea    & 105.17 $\pm$  2.797 & \cite{Ryan2010}    & 162.63 $\pm$  1.280 & \cite{Masiero2012} \\
 (275) Sapientia  &  89.05 $\pm$ 25.310 & \cite{Masiero2020} & 124.59 $\pm$ 35.830 & \cite{Masiero2021} \\
 (286) Iclea      &  81.31 $\pm$  2.762 & \cite{Ryan2010}    & 125.20 $\pm$ 40.850 & \cite{Masiero2021}  \\
 (326) Tamara     &  63.30 $\pm$ 14.530 & \cite{Masiero2021} & 116.98 $\pm$ 36.530 & \cite{Masiero2021} \\
 (412) Elisabetha &  76.38 $\pm$  2.114 & \cite{Ryan2010}    & 111.12 $\pm$ 22.220 & \cite{AliLagoa2018} \\
 (426) Hippo      & 107.33 $\pm$  4.405 & \cite{Ryan2010}    & 137.56 $\pm$  1.080 & \cite{Masiero2012} \\
 (439) Ohio       &  69.53 $\pm$  2.060 & \cite{Ryan2010}    &  86.86 $\pm$  0.750 & \cite{Masiero2012} \\
 (464) Megaira    &  56.93 $\pm$ 15.508 & \cite{Masiero2017} &  91.81 $\pm$ 31.750 & \cite{Masiero2021} \\
 (476) Hedwig     & 106.15 $\pm$ 27.050 & \cite{Nugent2016}  & 138.49 $\pm$  0.970 & \cite{Masiero2012} \\
 (524) Fidelio    &  61.71 $\pm$ 26.030 & \cite{Nugent2016}  &  83.26 $\pm$  4.093 & \cite{Ryan2010} \\
 (530) Turandot   &  75.88 $\pm$ 19.520 & \cite{Masiero2021} &  89.50 $\pm$ 17.900 & \cite{AliLagoa2018} \\
 (551) Ortrud     &  65.91 $\pm$ 14.860 & \cite{Nugent2016}  &  86.04 $\pm$  4.731 & \cite{Ryan2010} \\
 (566) Stereoskopia & 134.00 $\pm$ 6.627 & \cite{Masiero2011} & 190.08 $\pm$ 7.909 & \cite{Ryan2010} \\
 (657) Gunlod     &  31.44 $\pm$  8.570 & \cite{Masiero2020} &  46.59 $\pm$  3.149 & \cite{Ryan2010} \\
 (738) Alagasta   &  55.36 $\pm$  1.867 & \cite{Ryan2010}    &  77.94 $\pm$ 24.740 & \cite{Nugent2016} \\
 (806) Gyldenia   &  56.43 $\pm$  3.098 & \cite{Ryan2010}    &  83.10 $\pm$  0.740 & \cite{Masiero2012} \\
\hline
\end{tabular}
\label{diam_lit}
\end{center}
\end{table*}

\subsection{(70) Panopaea}

 Our result for the spin and shape of Panopaea (see Table \ref{results}) closely agrees with the one recently published by \citet{Hanus2021}.
 Three multi-chord occultations were available for Panopaea, one of them with a particularly rich set of 13 positive chords 
 (see Figure \ref{70occ}). However, many of these were visual observations, 
 and the chords had unrealistically small error bars. All errors have been changed here to 0.05 seconds for photoelectric 
 and video observations, and to 0.5 seconds for visual observations. Both shape solutions are consistent with the 
 occultations, resulting in a size determination of 128 km with very small uncertainty of a few percent (Table \ref{results}). 
 Previous literature diameter determinations for Panopaea varied widely from 105 km to 163 km (Table \ref{diam_lit}), 
 and were mostly based on the NEATM model \citep{Harris1998}, which approximates the body shape with a sphere.

\subsection{(275) Sapientia}
 
 There were seven pre-existing occultation observations for this target  (one with 16 chords), and our  Neglected
Asteroids campaign adds another two events. 
From the full set of nine events, four chords had to be removed because of their clear 
 inconsistency with the remaining ones: there must have been large errors on either the duration or the absolute timing.
 The fitting to occultations clearly points at pole 2 as the preferred solution  (Table \ref{results}), which is especially evident 
 from the fit to occultations from years 2015, 2020, and 2021 (see Fig. \ref{275occ}). The size is determined 
 with high accuracy (7\%).

 With such a rich set of occultations, and some of them suggesting non-convex shape features, the ADAM method was also applied. 
 The resulting shape for pole 2 shows a slightly better fit to all occultations 
 simultaneously (see Figs. \ref{275ADAM1} and \ref{275ADAM2}), but the fit is still not perfect. In particular, the four-chord event from 
 the year 2018 cannot be fittted by any of the models, even though here we allow the occultations to influence the shape. 
 The northernmost chord probably has an underestimated duration (this event is actually annotated `short low drop event in noisy recording')
 or it points to some small shape feature below the resolution of the ADAM shape reconstruction procedure. 
 The size of Sapientia found with the ADAM method (100 km) is consistent with that described above, 
 and the scatter of possible sizes shrank to a mere 1\%. Previous size determinations for Sapientia were somewhat less discrepant than for Panopaea, 
 ranging from 89 to 124 km (Table \ref{diam_lit}).

\subsection{(286) Iclea}

 Although the fit to the only available three-chord occultation for Iclea only slightly favours the pole 1 solution (see Fig. \ref{286occ}), 
 the sizes from the fitting for both solutions are clearly discrepant: the size for pole 2 solution is much smaller than the size 
 for the pole 1 solution (69 vs. 86 km), and is also inconsistent with all the previous sizes determined using infrared data (81--125 km, see Table \ref{diam_lit}). 
 Also, pole 1 spin parameters are confirmed by the results of \citet{ATLAS}; this is the only solution for the pole of Iclea. 
 Considering all of the above, we think that our solution for pole 2 
 (Table \ref{results}) can be safely rejected.

\subsection{(326) Tamara}

  There exists an extremely
wide range of literature diameters for Tamara (see Table \ref{diam_lit}): 
 from 63 km to as much as 117 km, an almost two-fold difference. This might be due to the relatively unusual 
 orbit of Tamara given that it is a main-belt asteroid, with an eccentricity of 0.19, and an inclination of 23$\deg$. As a consequence, thermal 
 measurements of Tamara must have been taken at substantially varied heliocentric distances, complicating thermal analysis. 
 Our result, being independent of thermal aspects, resolves these inconsistencies, pinpointing the diameter of this body to 
  77$^{+5}_{-10}$ km. The first pole solution is preferred based on the quality of the occultation fit (Fig \ref{326occ}), 
  with the RMS being almost twice as low as in any version of model 2. 
 Also, the spin solution found here is in good agreement with that from \citet{Hanus2021}.

%$

\subsection{(412) Elisabetha}

 The model 2 solution for Elisabetha is a much better fit to the occultations than any version of model 1 (compare the events from years 2011 and 2016 
 between the two poles in Fig. \ref{412occ}), and therefore model 2 is our preferred solution (Table \ref{results}). Our size determination 
 (97$^{+7}_{-14}$ km) confirms previous determinations (76--112 km), which tended towards larger values (see Table \ref{diam_lit}, and MP3C database). 

%$
\subsection{(426) Hippo}

 For Hippo we obtained two pole solutions from light-curve inversion; however, the second pole (see Table \ref{results}) was at the verge of
 rejection on the basis of the RMS fit to the light curves, which was almost 10\%
 higher than for the first pole. See Section 4 for a discussion of 
 this threshold.
Our spin solutions roughly agree with those published by \cite{Hanus2021} and \cite{Durech2019}, 
 in terms of pole latitude and period 
 but disagree with the pole longitude for model 2 presented by these authors.

 We decided to fit both shape solutions to the rich set of occultations;
 see Fig. \ref{426occ}. Surprisingly, it is the shape connected to pole
 solution 2 (blue contour in Fig. \ref{426occ}) that fits most of the occultations better than solution 2, 
 especially the richest one from 27 December 2021. 
 We therefore consider the pole 1 solution to be lesser probable (dashed magenta contour), even
 though it provided a formally better fit to light curves. The size determined
 here (122 $\pm$ 4 km) is close to the middle of the size range found in the literature (Table \ref{diam_lit}).

\subsection{(439) Ohio}
 
 Our research into the asteroid Ohio is a success in multiple ways. First, we have been able to correct a previously accepted rotation period, 
 identifying a period of around 37.46 hours as the correct one \citep{M2015}, instead of 19.2 hours which persisted for decades 
 in the asteroid Light-Curve Database \citep{Warner2009}. 
 Later, after gathering sufficient light-curve data over multiple apparitions, we managed to obtain its physically rotating model, 
 which was possible thanks to regularisation added to light-curve inversion routines. There was only a single pole solution (Table \ref{results}) on the level of light-curve inversion, as the mirror pole solution gave a much poorer fit to light curves. This is facilitated 
 by a relatively high orbital inclination of 19$\deg$ for Ohio. Our spin solution is $\sim$30$\deg$ and $\sim$50$\deg$ away in spin
axis latitude and longitude, respectively, from the pole 2 solution 
 provided by \citet{Durech2018}. 

 Later, in March 2022, we coordinated observers around two stellar occultation events (resulting in 6 and 9 positive chords, 
 including one grazing event observed by the first author), where there were no previously 
 recorded multi-chord events (i.e. containing more than three positive chords). In the end, our shape model was fitted to these occultations, presenting a very good-quality fit 
 (see Figure \ref{439occ}). Thanks to the quality of both the model and the occultation timings, the shape was precisely scaled 
 with small residual RMS. The only small discrepancies between the model and the chords from the last occultation 
 probably reveal the shape resolution limits of a classical light-curve inversion.

\subsection{(464) Megaira}
 
 The fit to occultations for Megaira is puzzling: Model 2 gives a somewhat better fit than model 1 (Fig. \ref{464occ}), but it always 
 leaves the southernmost chord outside the profile. Some variations of the shape for model 1 can fit that chord, 
 but they then lead to a clearly poorer fit to the remaining chords. The situation is similar when modelling this target with the ADAM algorithm 
 (Fig. \ref{464ADAM}). In summary, model 2 is slightly preferred, but cannot fit all the chords. 
 Previously published sizes range from 57 to 92 km. Our determination results in sizes of 75--79 km. 

\subsection{(476) Hedwig}

 Both pole solutions for Hedwig agree with those found by \citet{Hanus2021}, but
the pole 1 solution is clearly the preferred one by the fit to almost all of the six occultations, especially to the event from the year 2000 
 (see Fig. \ref{476occ}). The diameter we determine (116$^{+6}_{-16}$ km) is around the mean of the range of previous determinations 
 (106--138 km, Table \ref{diam_lit}.)

\subsection{(524) Fidelio}

 Pole 2 (see Table \ref{results}) is preferred by occultations here (Fig. \ref{524occ}). 
 This solution is also confirmed by the value for $\beta_p$ found by \citet{ATLAS} and the full spin solution by \citet{Hanus2021}. 
 The size determined here (67 $\pm$3 km) is closer to the lower end of the range of literature sizes (62--83 km). 

\subsection{(530) Turandot}

 The occultation set for Turandot contained a few discrepant chords, and so we had to remove four chords from the first event 
 in the year 2006, and completely leave out the second three-chord event from that year due to strong mutual inconsistencies among 
 the chords. The remaining set allowed us to clearly identify one of two pole solutions as the preferred one (see Fig. \ref{530occ}), especially in the 
 ADAM version of the model. With the ADAM model for pole 2, the southernmost chord of the 2006 event is finally fitted, revealing a non-convex 
 feature on the shape (Fig. \ref{530ADAM}). Both methods gave the same size for the two models, of namely 89 km, while previous determinations were in 
 the narrow range from 76 to 89 km. Our spin latitude found with the ADAM method is almost 30\deg \ away from the value found by \citet{Hanus2021} using convex inversion.

\subsection{(551) Ortrud}
 
 The occultation event for Ortrud from July 2021 contains problematic, mutually inconsistent chords (Fig. \ref{551occ}). Because it is hard 
 to identify the erroneous chords, we keep them all for the scaling. This results in larger RMS residuals for the diameter (12\%). 
 Previous determinations ranged from 66 to 86 km (see Table \ref{diam_lit}), while ours are around 85 km, confirming the upper values. 

\subsection{(566) Stereoskopia}

 One of the best covered occultations for Stereoskopia was observed within our campaign (see the event from 2021 in Fig. \ref{566occ}). 
 As the shape solutions tended to be non-physical and unstable here, we used the regularisation, and tried a few versions 
 of the shape model for each of the two pole solutions. We adopted the one that gave the best fit to occultations. 
 All the shape fits for pole 1 were clearly superior to any shape fits for pole 2. 
 Still, no shape was able to fit all the chords from the event in 2021. This might be due to insufficient information 
 on the shape contained in the light curves of this low-pole asteroid (see very low $|\beta_p|$ values in Table \ref{results}). 
 Some light curves were obtained in pole-on geometries, and therefore showed hardly any brightness variations.

 There were large discrepancies in literature diameter determinations for Stereoskopia, which range from 134 to 190 km. 
 The value found here (148 km) is consistent with lower of these values, and in spite of the imperfect fit, has an RMS of only 5\%--7\%.

\subsection{(657) Gunlod}

 The values for both pole coordinates for Gunlod are roughly
 consistent with the ones determined previously by \citet{Durech2019},
 although the poles found here have notably higher ($\sim$20\deg) values of
 $\beta_p$, which are closer to the values found by \citet{Hanus2021}. The fit
 to the only multi-chord occultation by Gunlod prefers pole 2
 solution (see Fig. \ref{657occ}), however that preference is based on
 only one chord, and a smaller RMS value for the diameter (Table \ref{results}).   
 Literature size determinations for Gunlod range from 31 to 46 km, while this work 
 places its diameter in the middle of this range (37--39 km).

\subsection{(738) Alagasta}

 The fit to occultations shows that the solution for pole 2 of Alagasta is slightly better, but still imperfect (Fig. \ref{738occ}). 
 Sizes from the literature are from 55 to 78 km, while the size determined here is 55 $\pm$ 5 km, confirming 
 the values at the lower end of this range.

\subsection{(806) Gyldenia}

 The spin solution for asteroid Gyldenia is within a few degrees of that found by \citet{Durech2019}.  
 The pole 2 solution seems to better fit the only available occultation (Fig. \ref{806occ}), as also evidenced by the smaller 
 RMS residual for the volume-equivalent size determined in this way (Table \ref{results}). Still, this solution also fails to fit the longest occultation chord. 
 On the other hand, pole 1 has difficulty in fitting the short chord at the edge of the shape; although the fitting of such small chords is problematic  in general. 
 Previous diameter determinations range from 56 to 83 km (Table \ref{diam_lit}), and the diameter presented here is 55--57 km, which is consistent with the lowest 
 of the published values.

\section{Conclusions and future prospects}

 Our project and comprehensive approach has provided almost 50 high-quality scaled models 
 \citep{M2018, M2019, M2021} for  poorly studied but large and ubiquitous asteroids, including 16 presented here. 
 Two observing campaigns ---for light curves and occultations--- were carried out by a wide network of collaborators and involved many amateur astronomers, bringing new people to the field. We are going to continue 
 the occultation campaign, and to a lesser extent, also the one for time-series photometry, 
 as the latter is gradually being replaced by wide ground-based and space-based sky surveys.
 As so many asteroids have poorly determined sizes, stellar occultations are the best technique to reliably determine the sizes  
 of a large number of targets from all groups, especially in the era of the Gaia mission.
 
 Our precise determinations of spins, shapes, 
 and sizes for many, previously largely omitted targets enrich and complement the available sample of modelled
 asteroids\footnote{https://astro.troja.mff.cuni.cz/projects/damit/} \citep{DAMIT}. 
 Our set includes asteroids belonging to a few asteroid families that are 
 believed to form in catastrophic disruptions of larger parent bodies. 
 Current theories of collisional evolution of the Solar System predict a steady number of existing asteroid families
 throughout history, but observational evidence contradicts this prediction \citep{Broz2013}. The key to solving this
 inconsistency is to determine family ages by deciphering the drift rate of family members from the centre. This can be done by
 studying the V-shaped dependence of the inverse size 1/D on the proper semimajor axis $a_p$ \citep{Vokrouhlicky2006}. 
 Among the many potential applications, our improved size determinations and spin properties can  facilitate better 
 predictions of the thermally driven drift of the asteroids within their families.

\begin{acknowledgements}

We thank an anonymous referee, whose comments led to a substantial improvement of this paper.

This work was supported by the National Science Centre, Poland, through grant no. 2020/39/O/ST9/00713

This project has been supported by the Lend{\"u}let grant LP2012-31 of the Hungarian Academy of Sciences and by the KKP-137523 
{\'E}lvonal grant of the Hungarian Research, Development and Innovation Office (NKFIH).

The work of J\v{D} and JH was supported by the grant 20-08218S of the Czech Science Foundation,  
and by the Erasmus programme of the European Union under grant number 2020-1-CZ01-KA203-078200.

A. P{\'a}l and R. Szak{\'a}ts has received funding from the K-138962
grant of the National Research, Development and Innovation Office (NKFIH,
Hungary).

Erika Pak\v{s}tien{\.e} and R\={u}ta Urbonavi\v{c}i\={u}t{\.e} acknowledge the Europlanet 2024 RI project funded by the European Union's Horizon 2020 
Research and Innovation Programme (Grant agreement No. 871149). 

Data from Observat{\'o}rio Astron{\^o}mico do Sert{\~a}o de Itaparica (OASI, Itacuruba) have been obtained with the 1-m telescope, 
a facility operated by the IMPACTON project of the Observat{\'o}rio Nacional, Brazil. F.M., E.R., M.E., W.M., W.P. and J.M. would like to thank 
CNPq, FAPERJ (E-26/201.877/2020 and E-26/204.602/2021) and CAPES for their support through diverse fellowships (Brazilian funding agencies). 
Support by CNPq (310964/2020-2) and FAPERJ (E-26/202.841/2017 and E-26/201.001/2021) is acknowledged by D.L.

We acknowledge financial support from the State Agency for Research of
the Spanish MCIU through the ``Center of Excellence Severo Ochoa'' award to
the Instituto de Astrof{\'i}sica de Andaluc{\'i}a (SEV-2017-0709). Funding from
Spanish projects PID2020-112789GB-I00 from AEI and Proyecto de Excelencia
de la Junta de Andaluc{\'i}a PY20-01309 is acknowledged.

This work uses data obtained from the Asteroid Light-Curve Data Exchange Format (ALCDEF) database, which is supported by funding from NASA grant 80NSSC18K0851. 

This work is based on data provided by the Minor Planet Physical Properties Catalogue (MP3C) of the Observatoire de la Côte d'Azur.

Based on observations made with the Mercator Telescope, operated on the
island of La Palma by the Flemmish Community, at the Spanish Observatorio
del Roque de los Muchachos of the Instituto de Astrof{\'i}sica de Canarias.
Based on observations obtained with the MAIA camera, which was built by
the Institute of Astronomy of KU Leuven, Belgium, thanks to funding from
the European Research Council under the European Community's Seventh
Framework Programme (FP7/2007-2013)/ERC grant agreement no 227224
(PROSPERITY, PI: Conny Aerts) and from the Research Foundation - Flanders
(FWO) grant agreement G.0410.09. The CCDs of MAIA were developed by e2v
in the framework of the ESA Eddington space mission project; they were
offered by ESA on permanent loan to KU Leuven \citep{Raskin2013}. This
article is based on observations obtained at the Observat{\'o}rio
Astron{\^o}mico do Sert{\~a}o de Itaparica (OASI, Itacuruba) of the
Observat{\'o}rio Nacional, Brazil.

The Joan Or{\'o} Telescope (TJO) of the Montsec Astronomical Observatory (OAdM)
       is owned by the Catalan Government and operated by the Institute for Space Studies of Catalonia (IEEC).
       This article is based on observations made in the Observatorios de Canarias del IAC with the
       0.82m IAC80 telescope operated on the island of Tenerife by the Instituto de Astrof{\'i}sica de Canarias (IAC)
       in the Observatorio del Teide.
This article is based on observations made with the SARA telescopes (Southeastern Association for Research in Astronomy),
whose nodes are located at the Observatorios de Canarias del IAC on the island of La Palma in the Observatorio
del Roque de los Muchachos; Kitt Peak, AZ under the auspices of the National Optical Astronomy Observatory (NOAO);
and Cerro Tololo Inter-American Observatory (CTIO) in La Serena, Chile.

This project uses data from the SuperWASP archive. The WASP project is currently funded and operated by Warwick University
and Keele University, and was originally set up by Queen's University Belfast, the Universities of Keele, St. Andrews,
and Leicester, the Open University, the Isaac Newton Group, the Instituto de Astrofisica de Canarias,
the South African Astronomical Observatory, and by STFC.
TRAPPIST-South is a project funded by the Belgian Fonds
(National) de la Recherche Scientifique (F.R.S.-FNRS) under
grant PDR T.0120.21. TRAPPIST-North is a project funded by
the University of Li{\`e}ge, in collaboration with the Cadi Ayyad
University of Marrakech (Morocco).

Funding for the Kepler and K2 missions are
provided by the NASA Science Mission Directorate.
The data presented in this paper were
obtained from the Mikulski Archive for Space
Telescopes (MAST). STScI is operated by the
Association of Universities for Research in Astronomy,
Inc., under NASA contract NAS5-
26555. Support for MAST for non-HST data is
provided by the NASA Office of Space Science
via grant NNX09AF08G and by other grants
and contracts.

Data from Pic du Midi Observatory have been obtained with the 0.6-m telescope, a facility operated by
Observato{\'i}re Midi Pyr{\'e}n{\'e}es and Association T60, an amateur association.

This research is partly based on observations obtained with the 60-cm Cassegrain telescope (TC60) and 
90-cm Schmidt-Cassegrain telescope (TSC90) of the Institute of Astronomy of the Nicolaus Copernicus University in Toru{\'n}.

We acknowledge the contributions of the occultation observers
who have provided the observations in the dataset. Most of
those observers are affiliated with one or more of:
European Asteroidal Occultation Network (EAON), International Occultation Timing Association (IOTA),
International Occultation Timing Association  European Section (IOTA/ES), Japanese Occultation Information Network (JOIN), and
Trans Tasman Occultation Alliance (TTOA).

\end{acknowledgements}

 \bibliographystyle{aa}
\bibliography{bibliography}

%\clearpage

%==================================================================

\clearpage

\begin{appendix}
\section{Additional figures}
\begin{figure*}
\begin{center}
\includegraphics[width=0.25\textwidth]{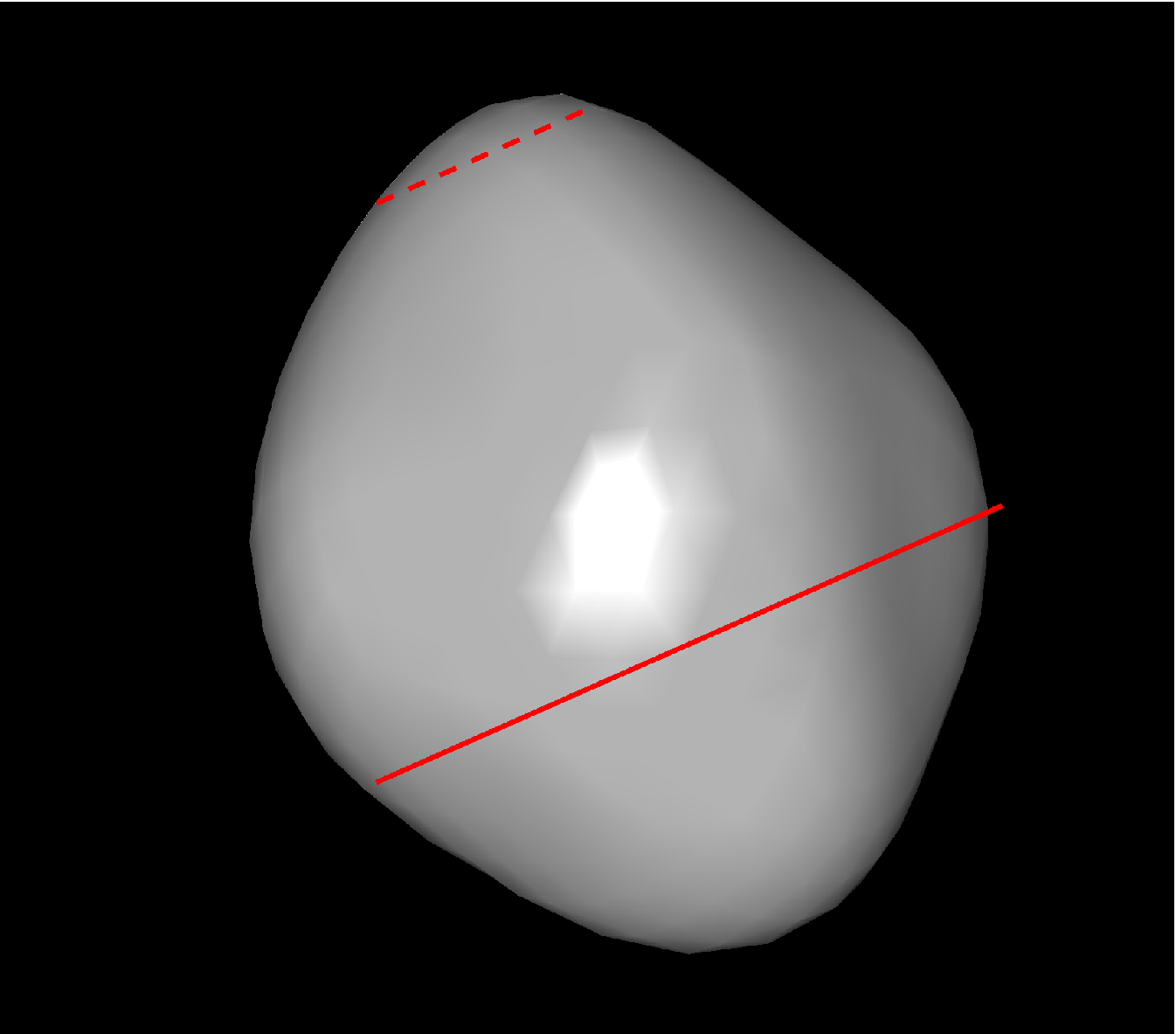}
\includegraphics[width=0.25\textwidth]{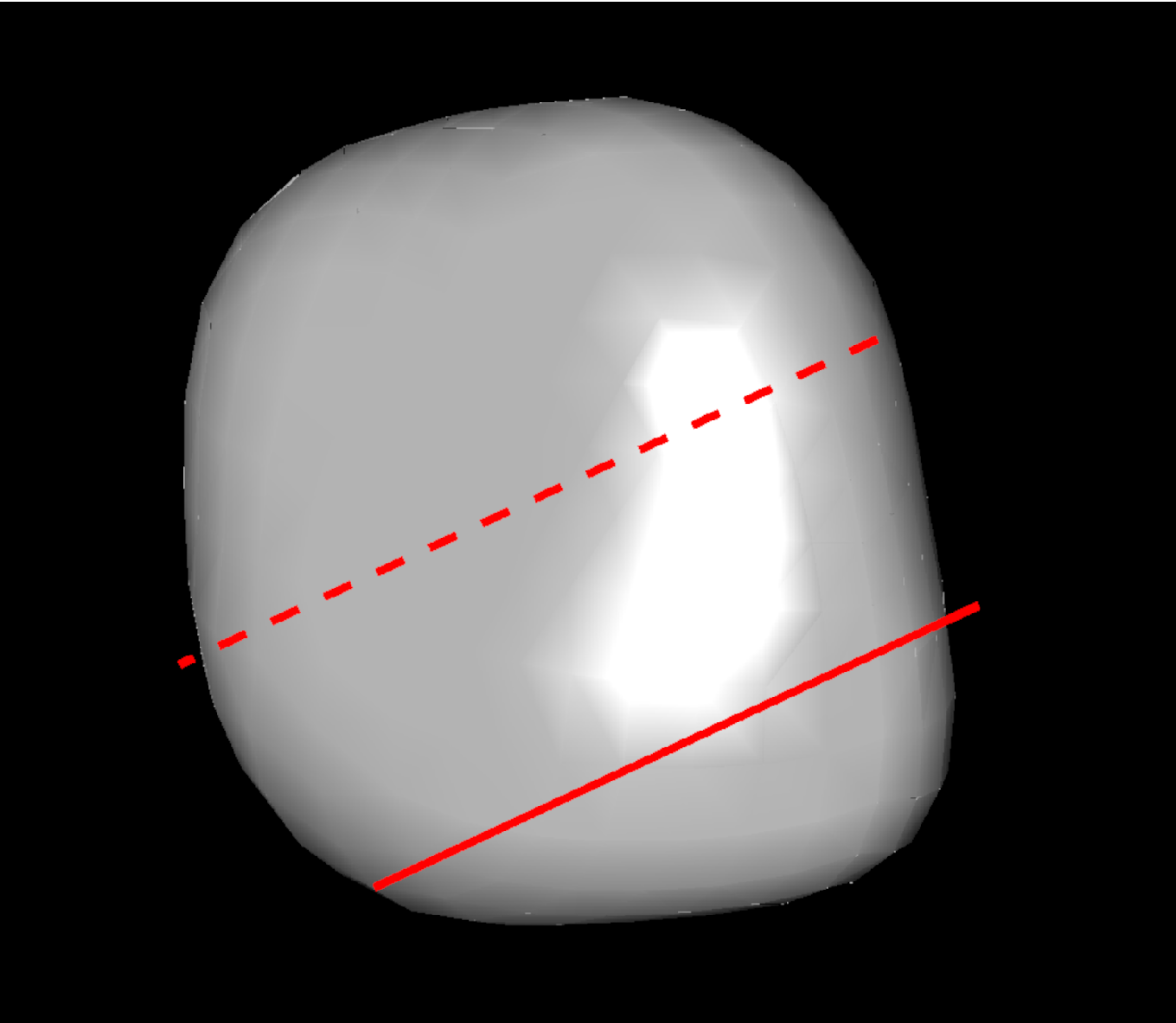}
\includegraphics[width=0.25\textwidth]{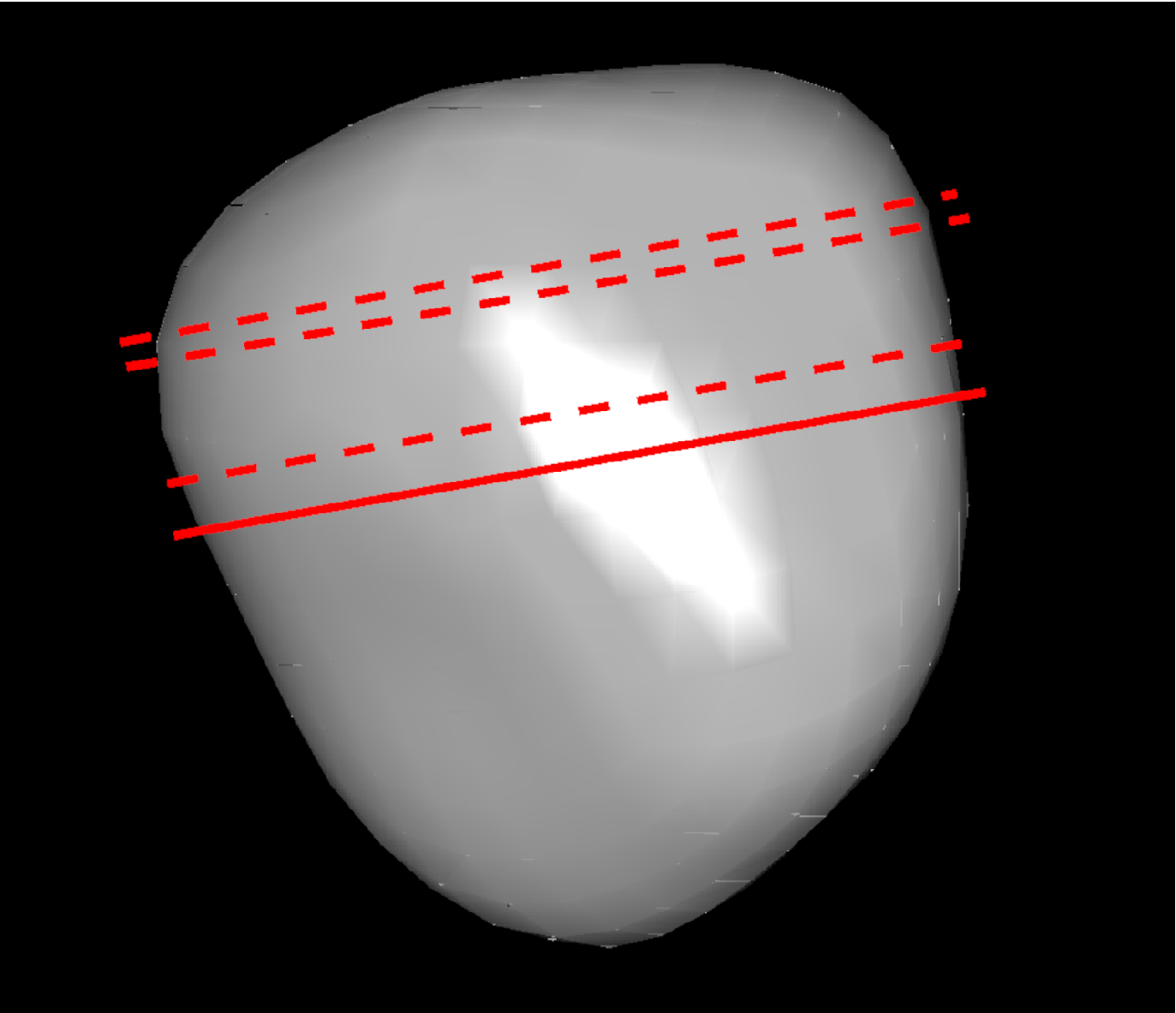}
\includegraphics[width=0.25\textwidth]{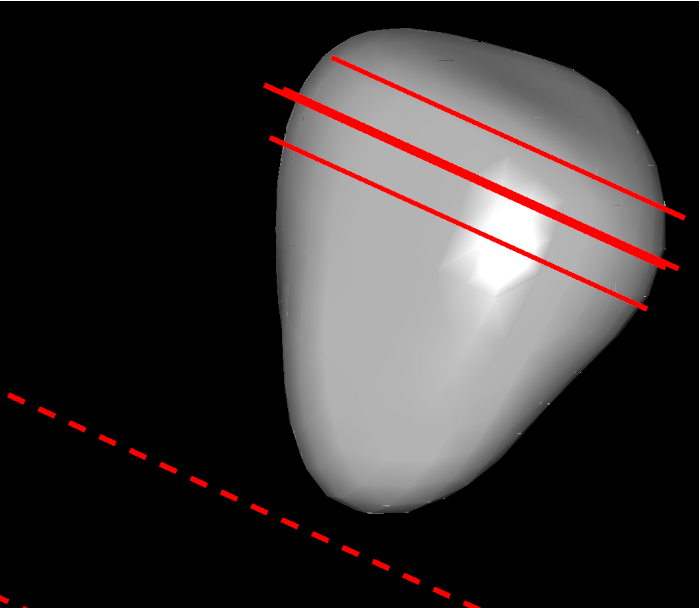}
\includegraphics[width=0.25\textwidth]{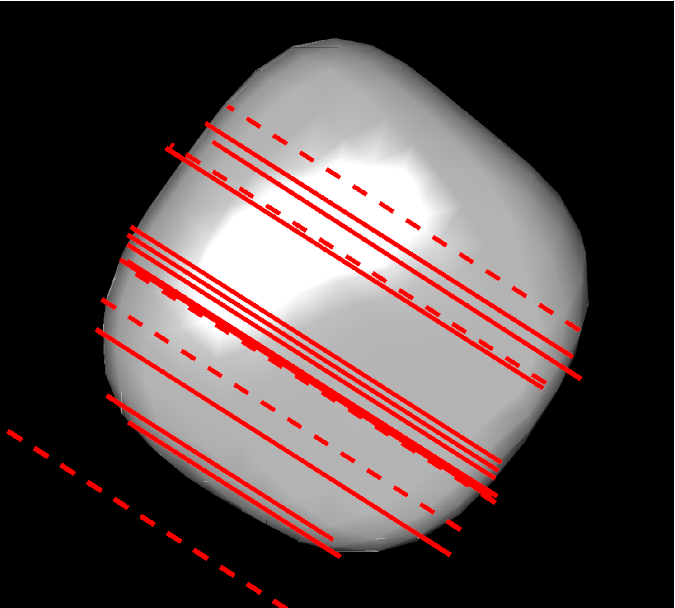}
\includegraphics[width=0.25\textwidth]{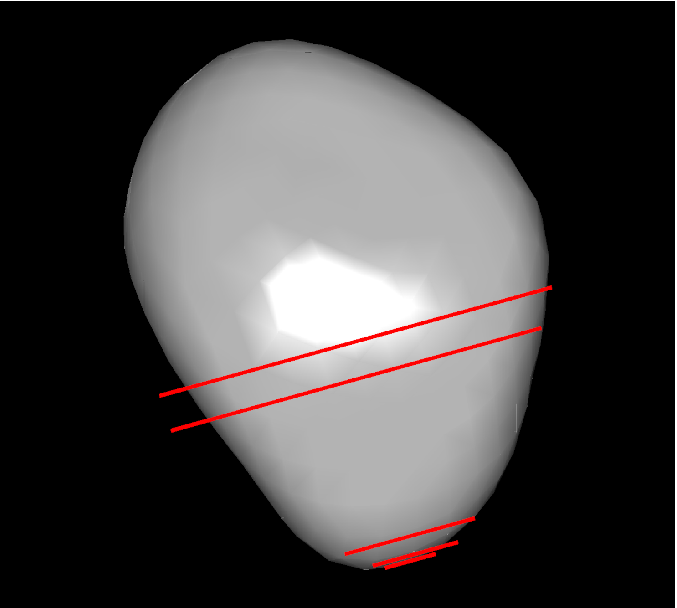}
\includegraphics[width=0.25\textwidth]{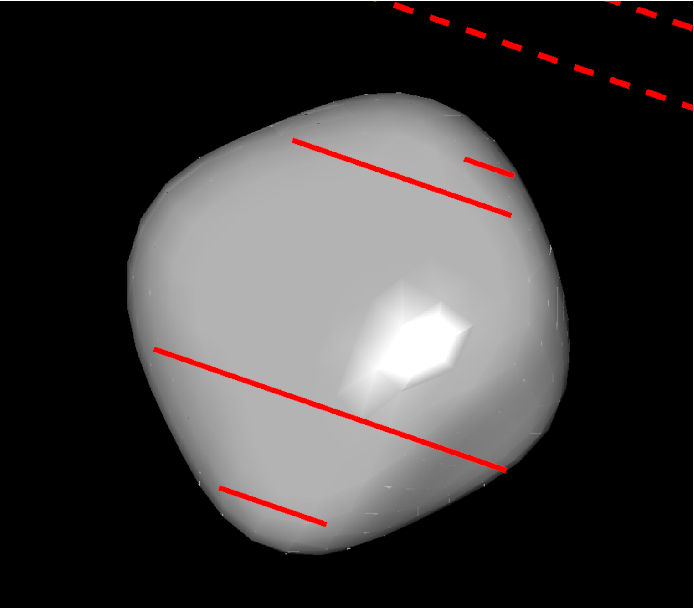}
\includegraphics[width=0.25\textwidth]{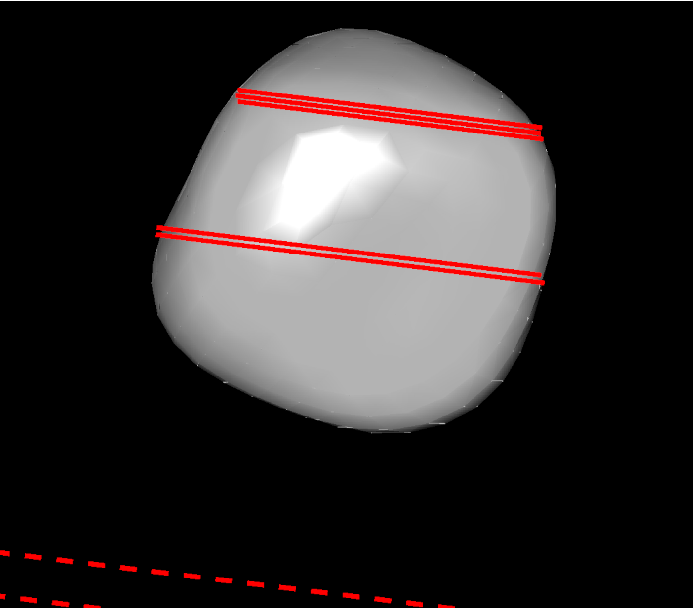}
\includegraphics[width=0.25\textwidth]{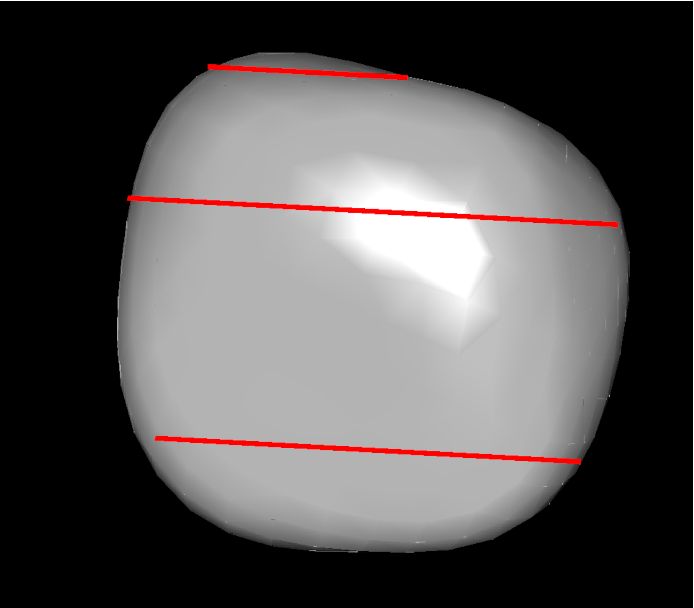}
\end{center}
\caption{(275) Sapientia ADAM solution pole 2, shown with occultation chords used to construct the model. Occultation events are the same as in Fig.~\ref{275occ}.}
\label{275ADAM2}
\end{figure*}

%\clearpage

\begin{figure*}
\begin{center}
\includegraphics[width=0.33\textwidth]{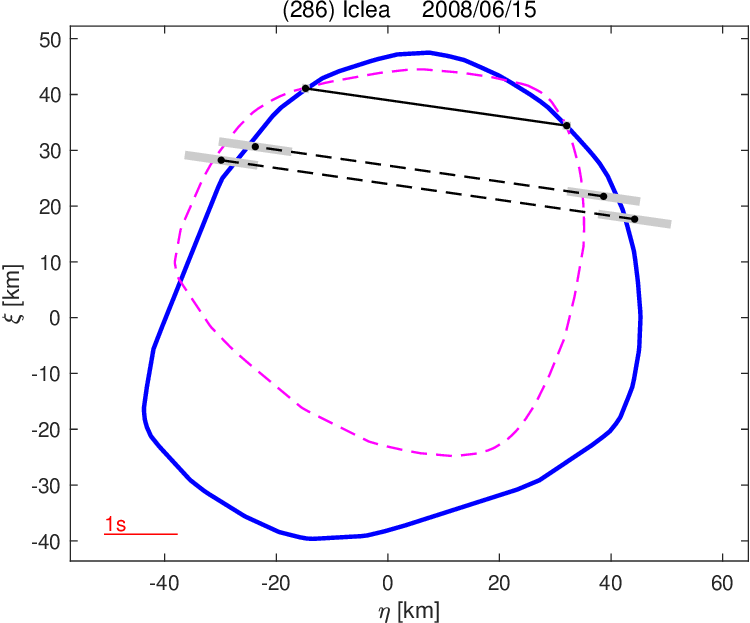}
\end{center}
\caption{(286) Iclea model fit to occultation chords. Pole 2 (dashed contour) is rejected as inconsistent with all previous size determinations.}
\label{286occ}
\end{figure*}

\begin{figure*}
\includegraphics[width=0.33\textwidth]{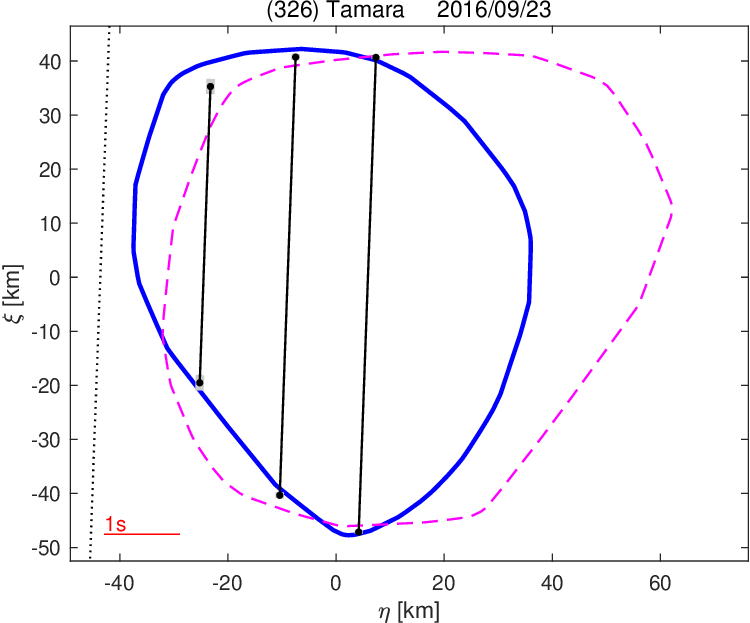}
\includegraphics[width=0.33\textwidth]{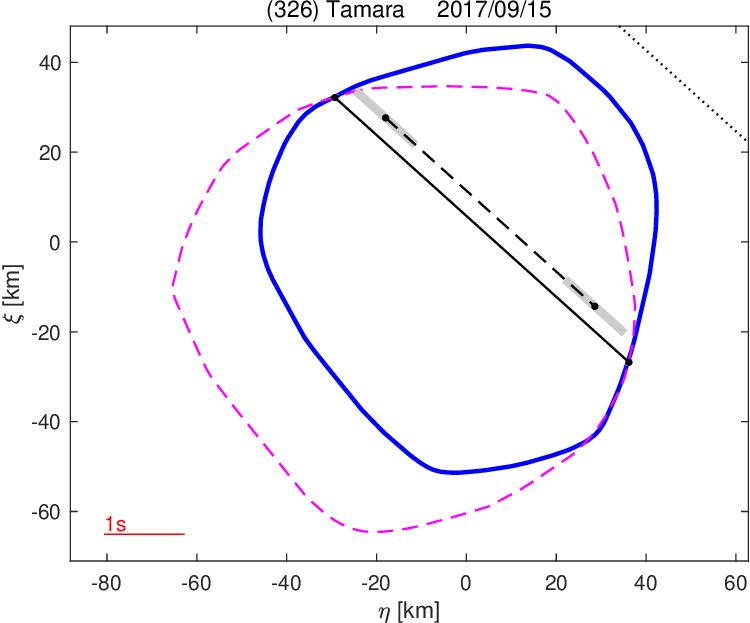}
\includegraphics[width=0.33\textwidth]{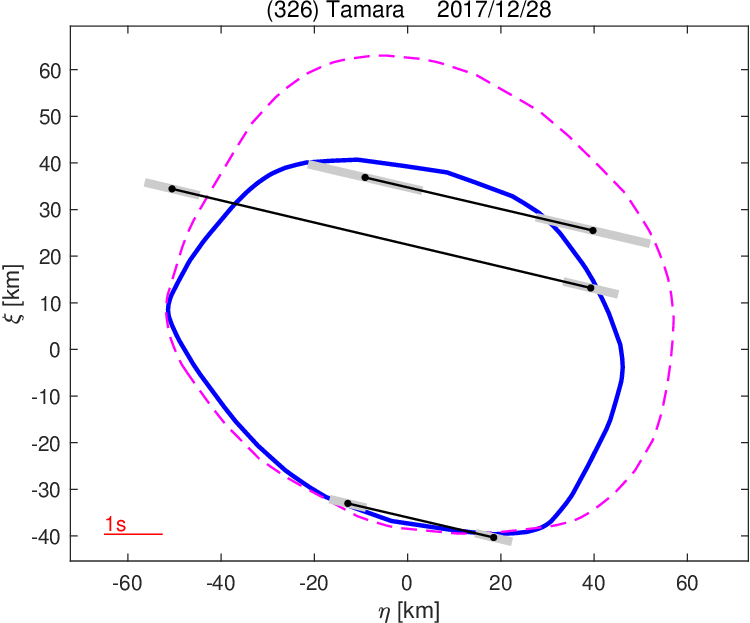}
\includegraphics[width=0.33\textwidth]{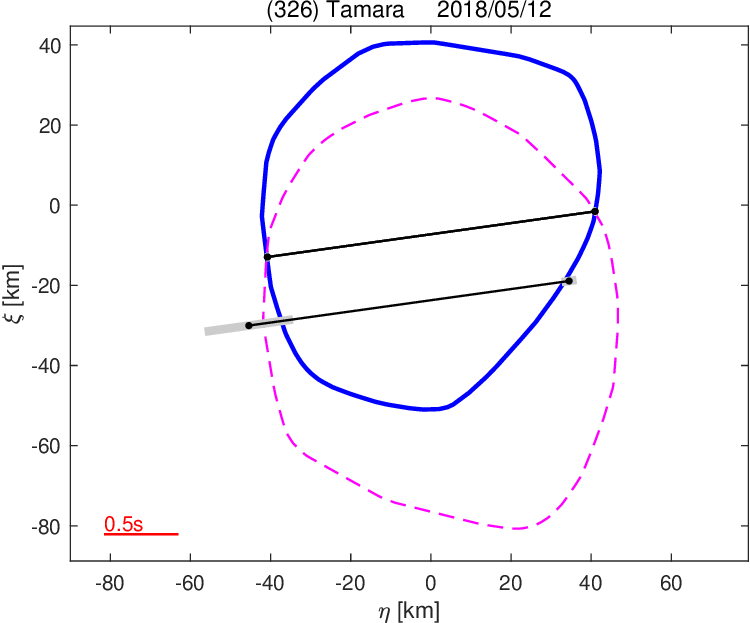}
\includegraphics[width=0.33\textwidth]{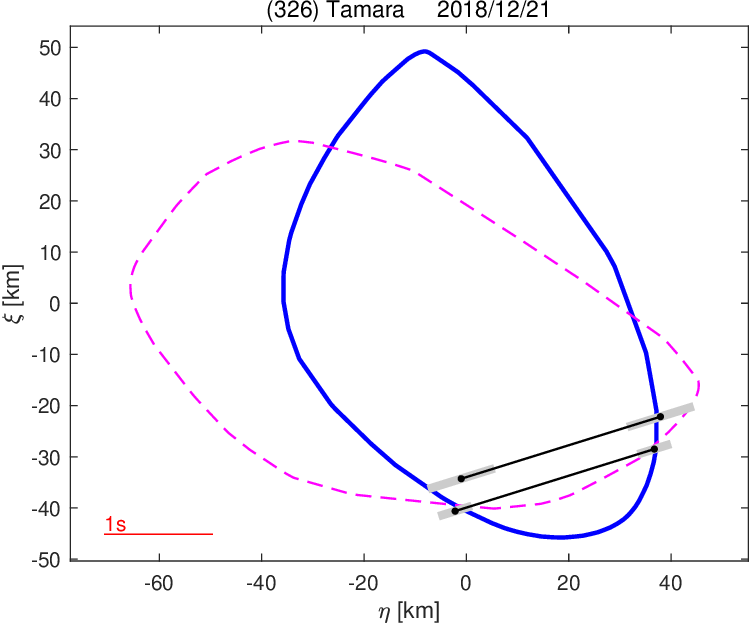}
\caption{(326) Tamara model fit to occultation chords. Pole 1 is preferred (solid contour).}
\label{326occ}
\end{figure*}

\begin{figure*}
\includegraphics[width=0.33\textwidth]{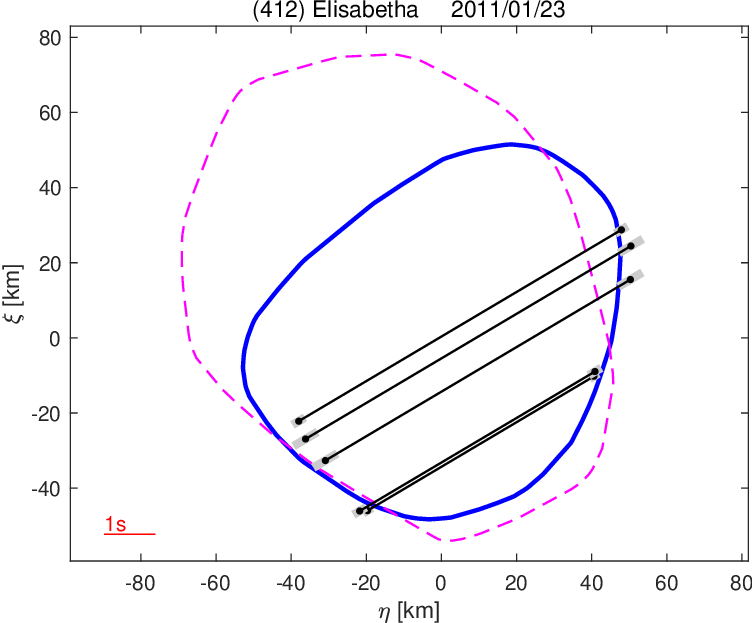}
\includegraphics[width=0.33\textwidth]{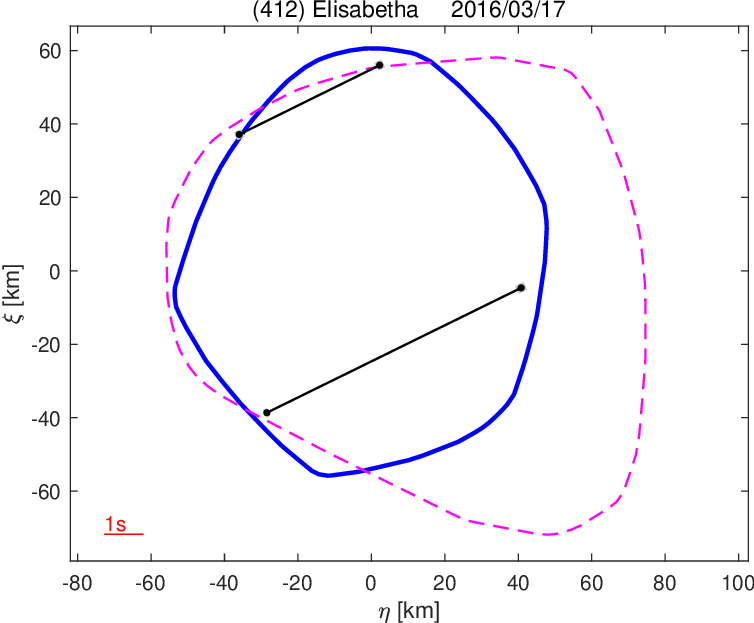}
\includegraphics[width=0.33\textwidth]{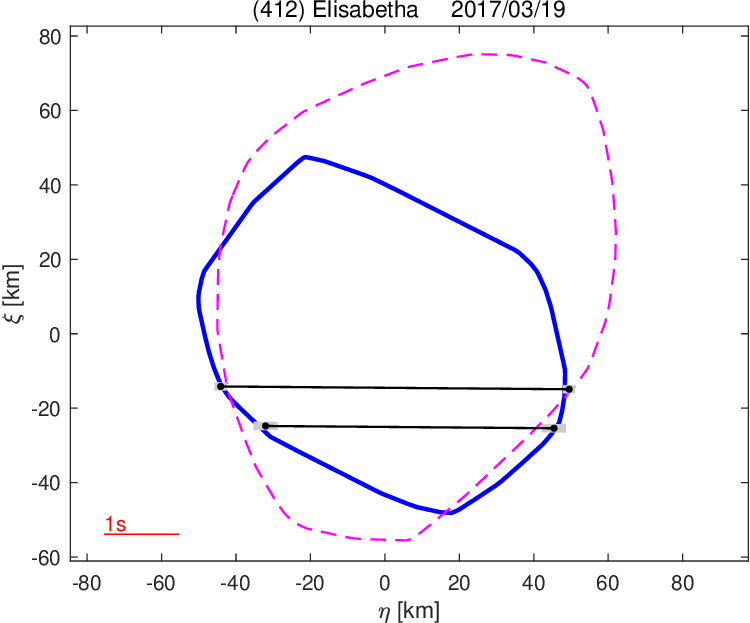}
\includegraphics[width=0.33\textwidth]{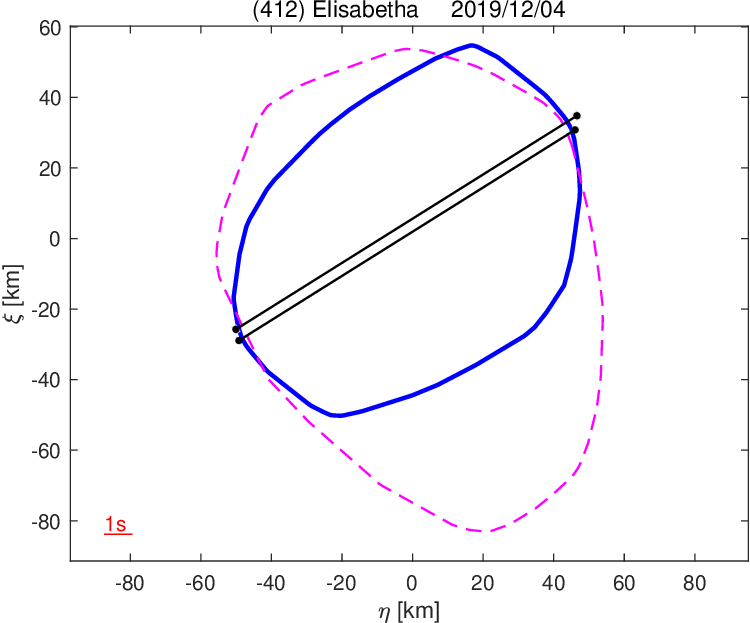}
\caption{(412) Elisabetha model fit to occultation chords. Pole 2 is preferred (solid contour).}
\label{412occ}
\end{figure*}

\begin{figure*}
\includegraphics[width=0.33\textwidth]{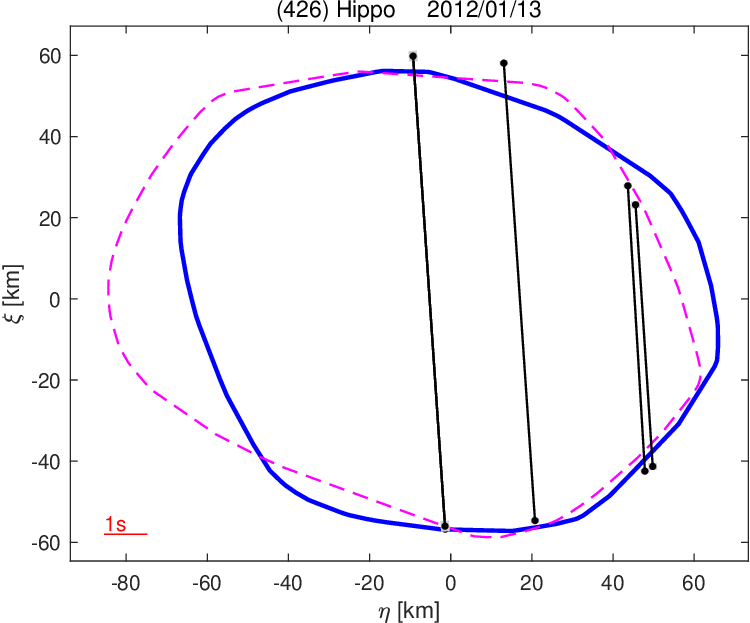}
\includegraphics[width=0.33\textwidth,clip]{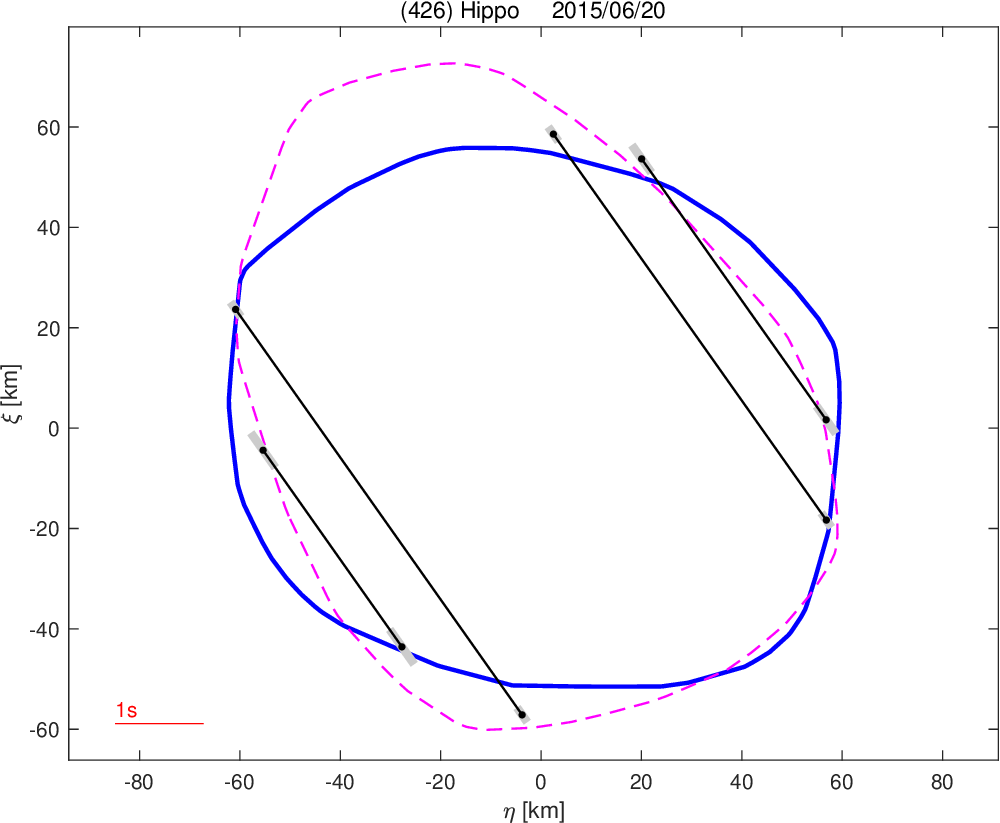}
\includegraphics[width=0.33\textwidth]{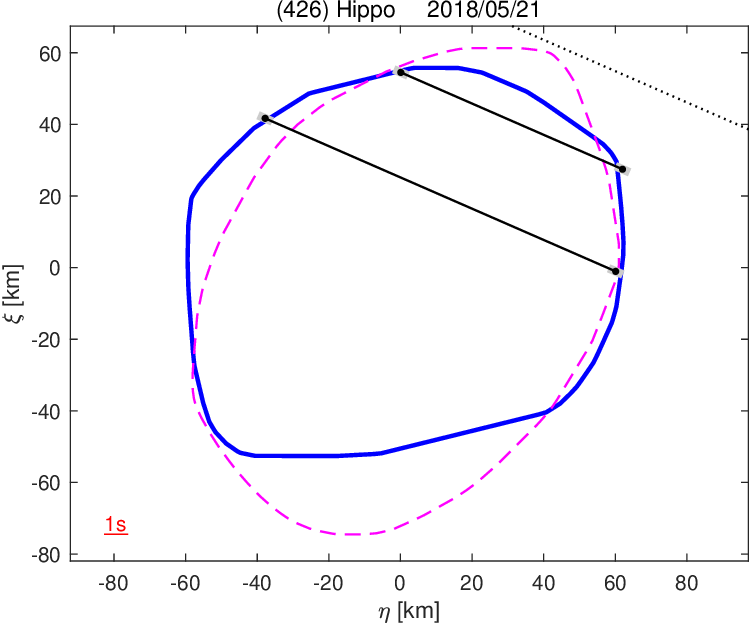}
\includegraphics[width=0.33\textwidth]{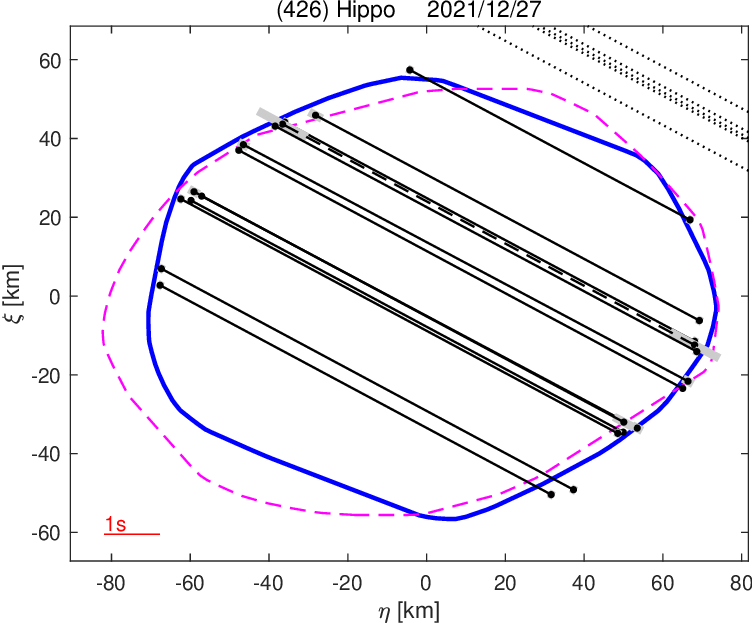}
\includegraphics[width=0.33\textwidth]{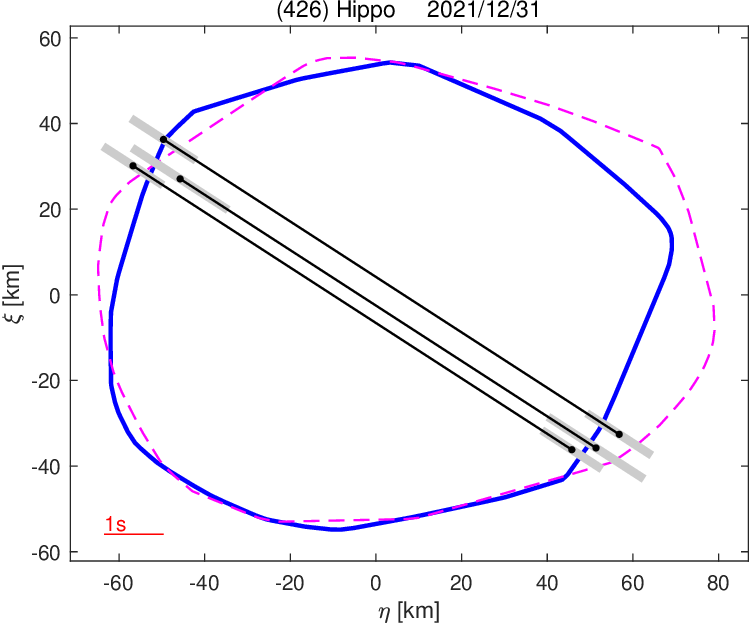}
\includegraphics[width=0.33\textwidth]{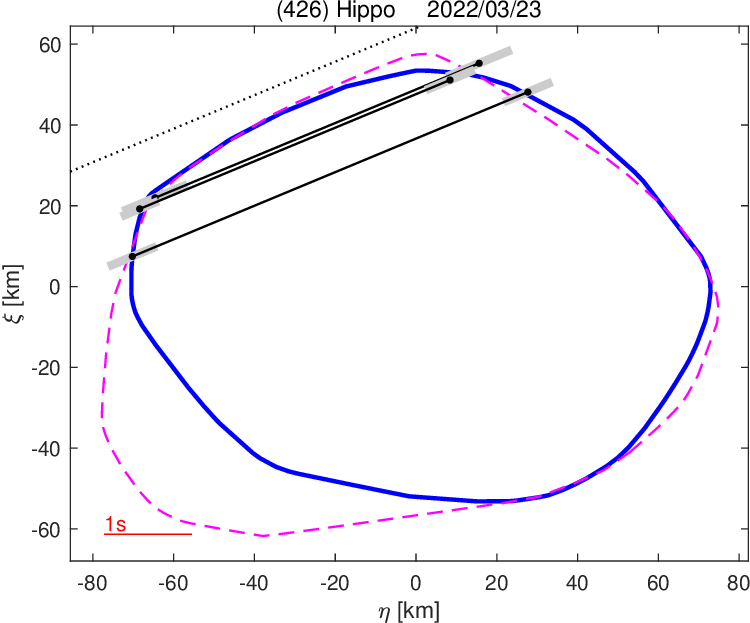}
\caption{(426) Hippo model fit to occultation chords. Pole 2 is preferred (solid contour).}
\label{426occ}
\end{figure*}

\begin{figure*}
\includegraphics[width=0.33\textwidth]{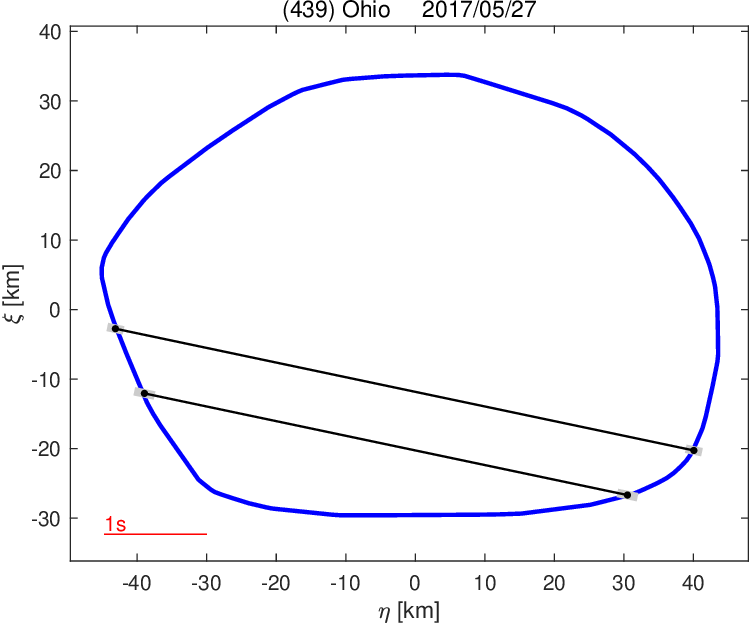}
\includegraphics[width=0.33\textwidth]{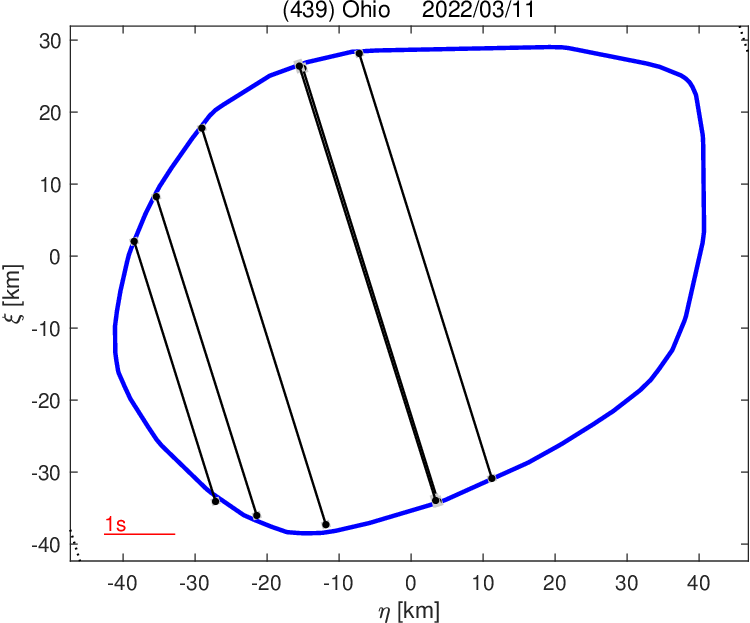}
\includegraphics[width=0.33\textwidth]{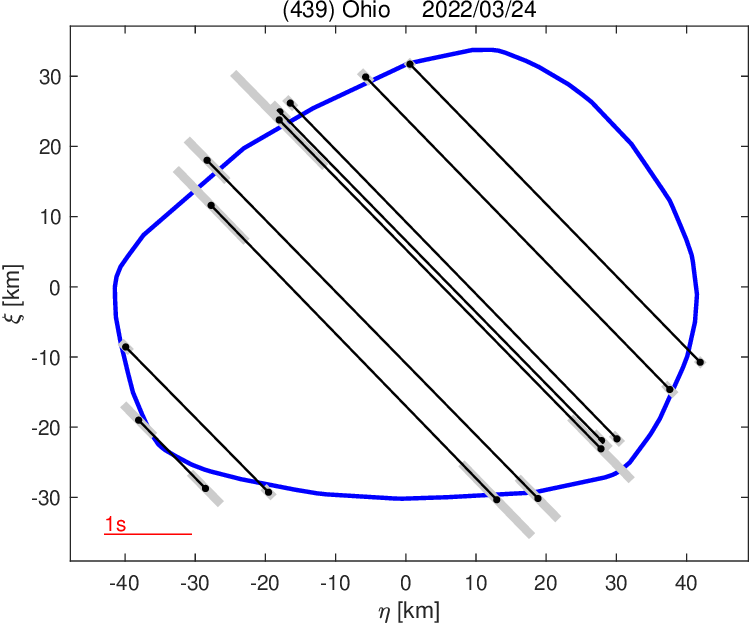}
\caption{(439) Ohio model fit to occultation chords.}
\label{439occ}
\end{figure*}

\begin{figure*}
\begin{center}
\includegraphics[width=0.33\textwidth]{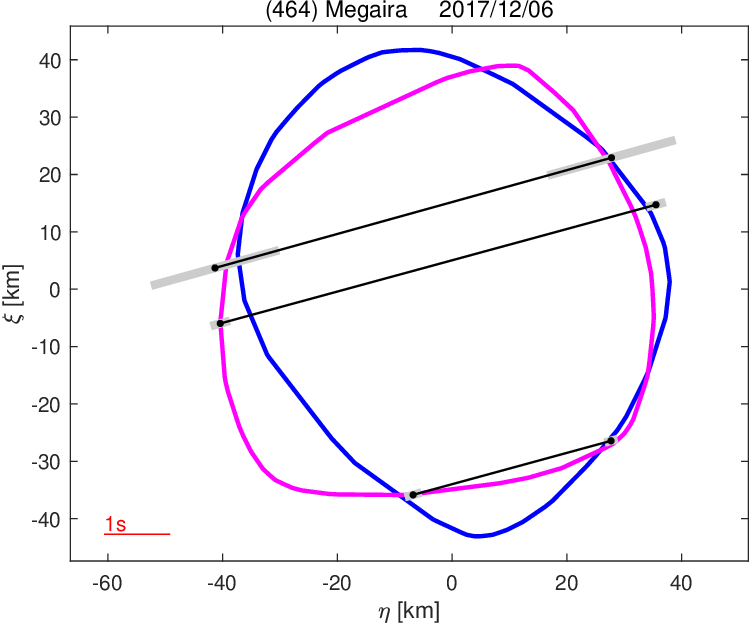}
\includegraphics[width=0.33\textwidth]{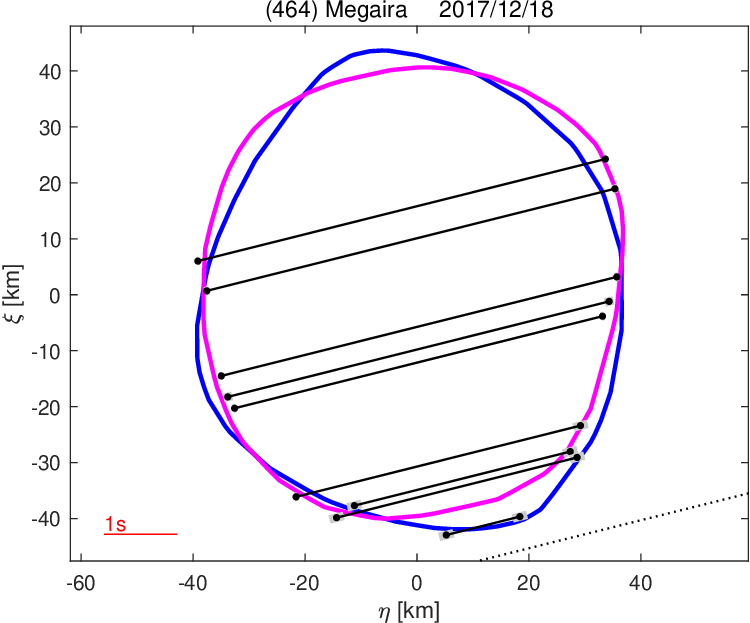}
\end{center}
\caption{(464) Megaira model fit to occultation chords: pole 1 (blue), and pole 2 (magenta).}
\label{464occ}
\end{figure*}

\begin{figure*}
\begin{center}
\includegraphics[width=0.3\textwidth]{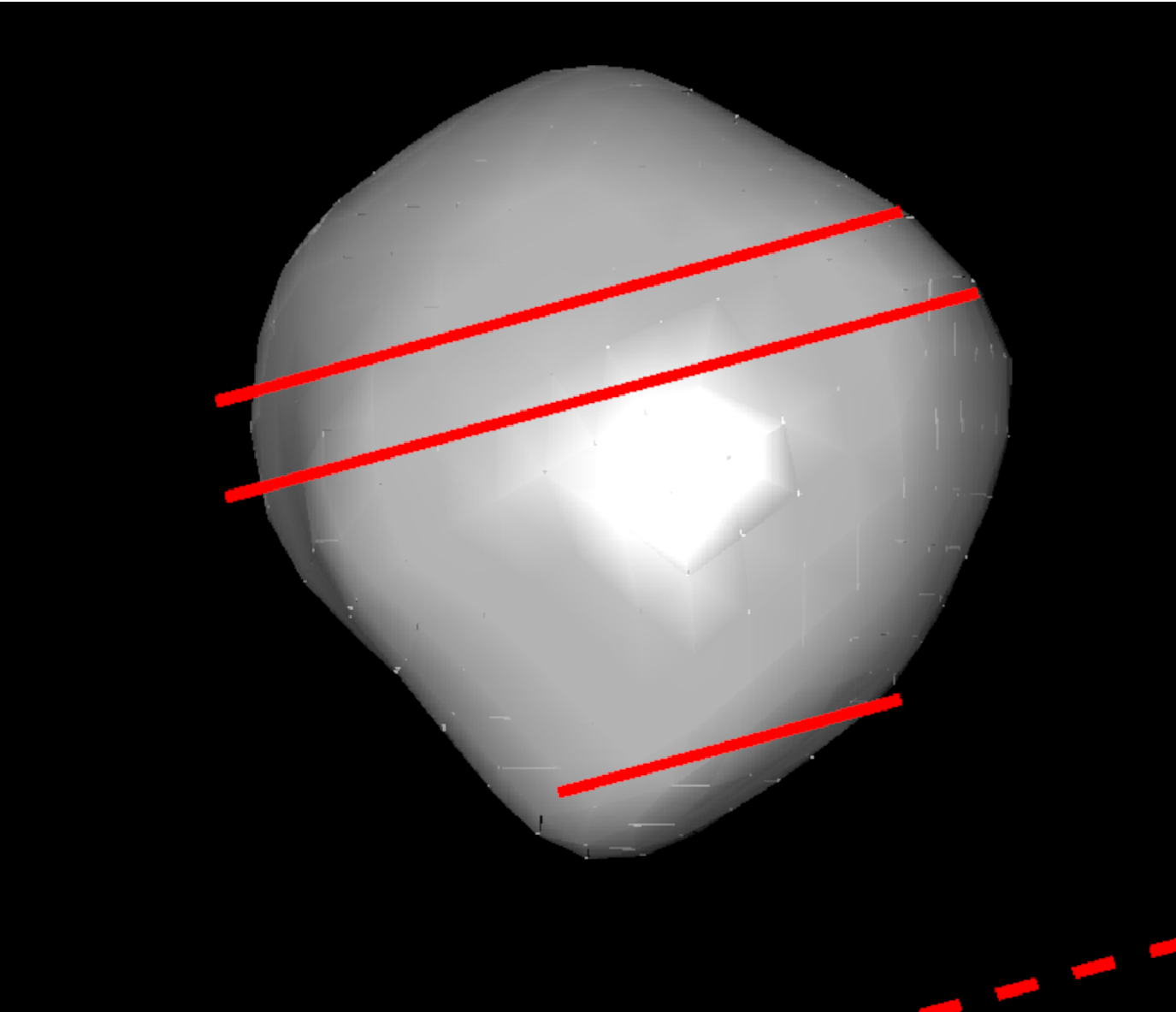}
\includegraphics[width=0.3\textwidth]{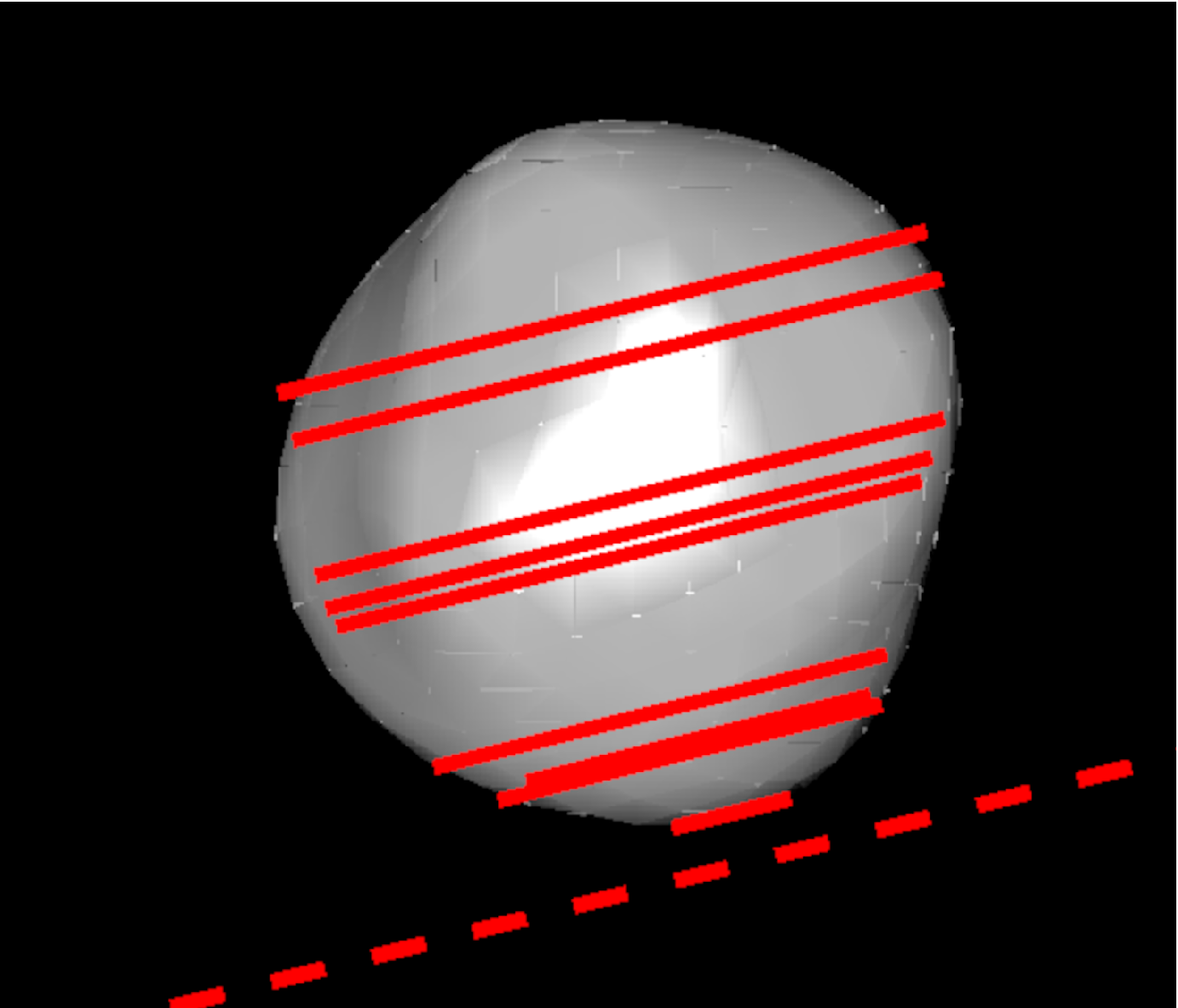}

\includegraphics[width=0.3\textwidth]{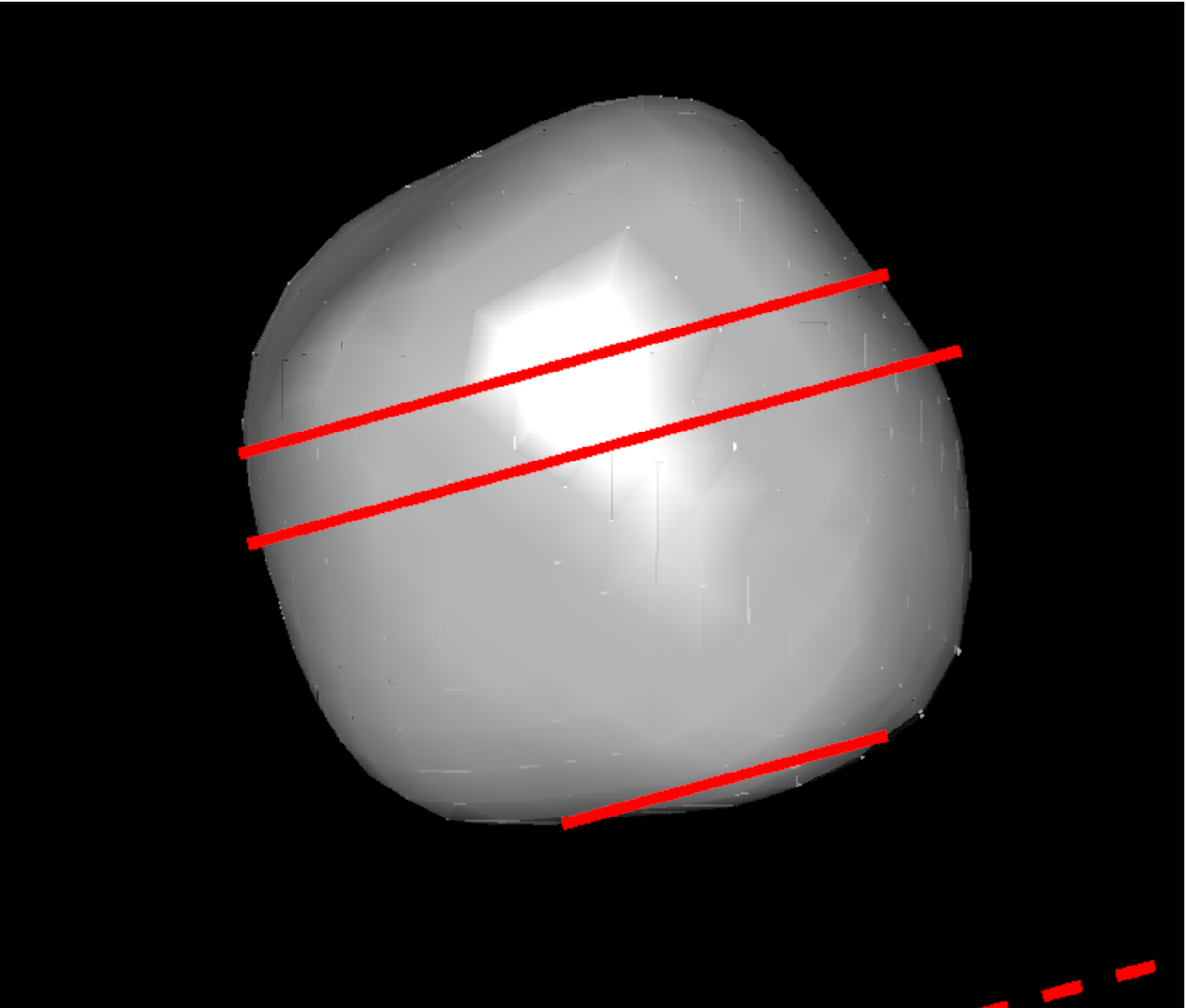}
\includegraphics[width=0.3\textwidth]{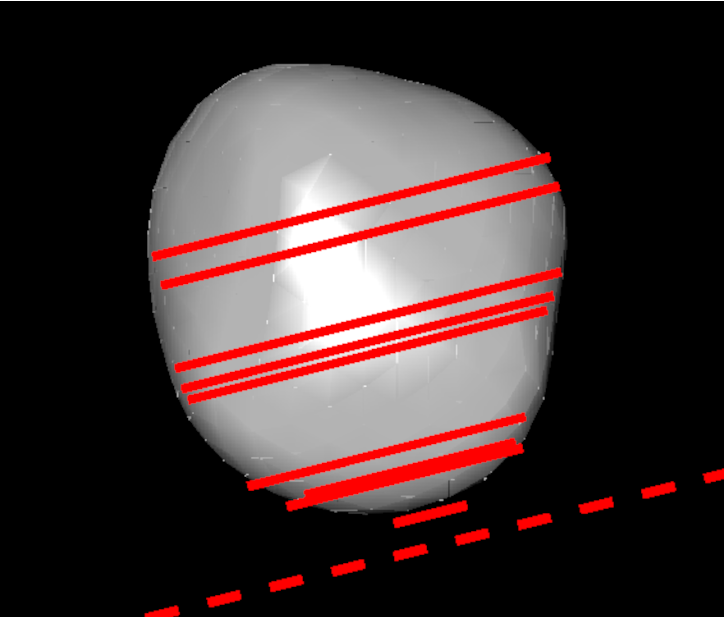}
\end{center}
\caption{(464) Megaira ADAM solution pole 1 (top row), pole 2 (bottom row), shown with occultation chords used to construct the models. 
Occultation events same as in Fig. \ref{464occ}.}
\label{464ADAM}
\end{figure*}

\begin{figure*}
\includegraphics[width=0.33\textwidth]{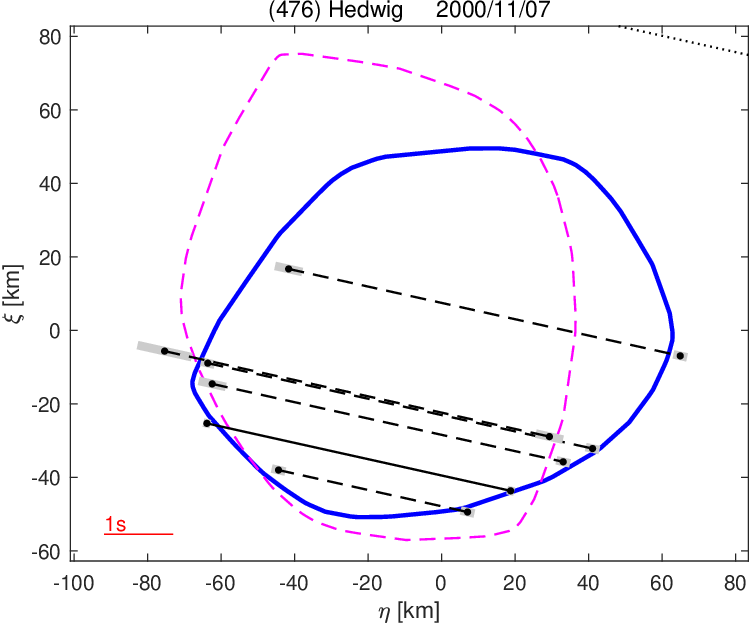}
\includegraphics[width=0.33\textwidth]{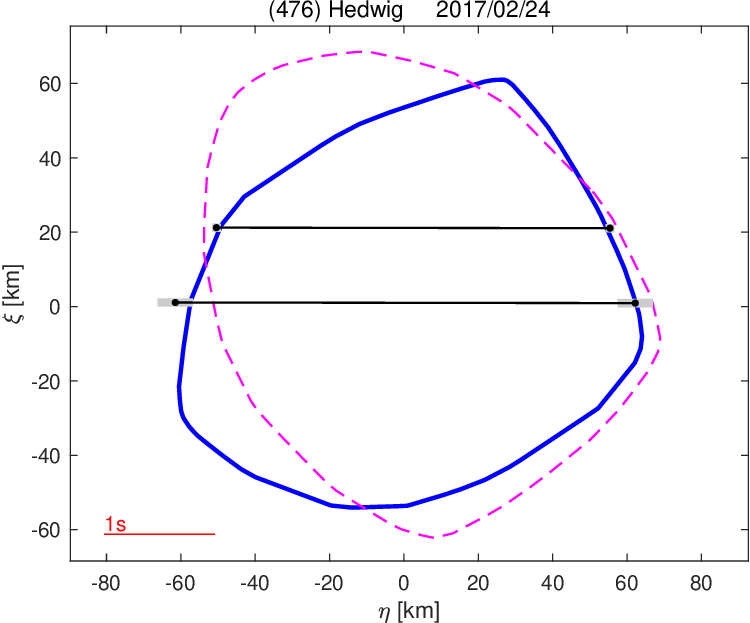}
\includegraphics[width=0.33\textwidth]{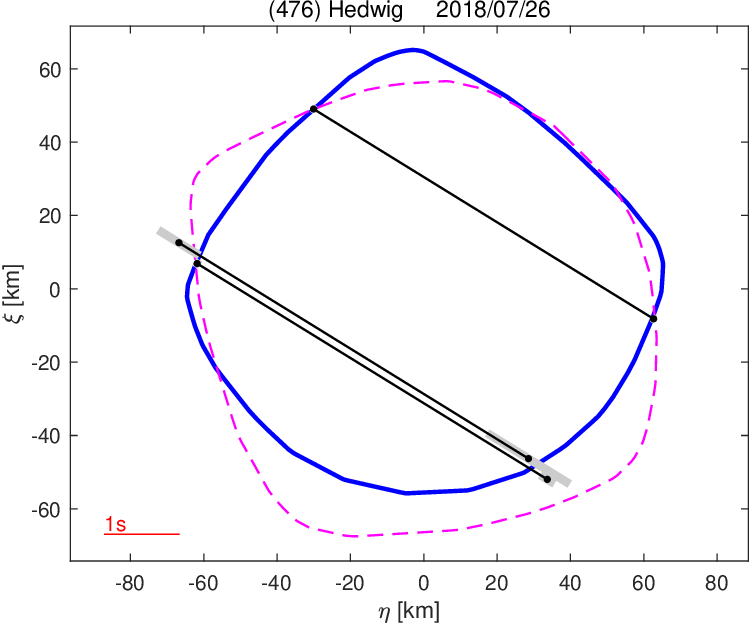}
\includegraphics[width=0.33\textwidth]{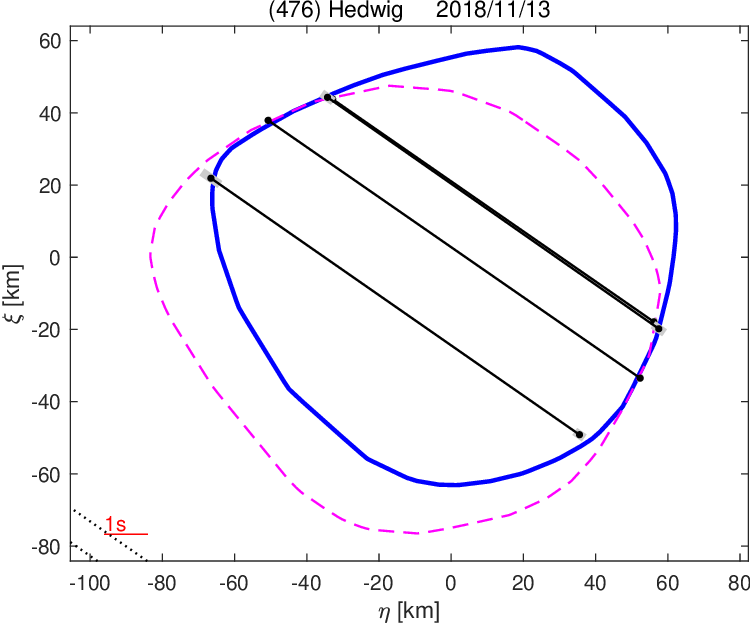}
\includegraphics[width=0.33\textwidth]{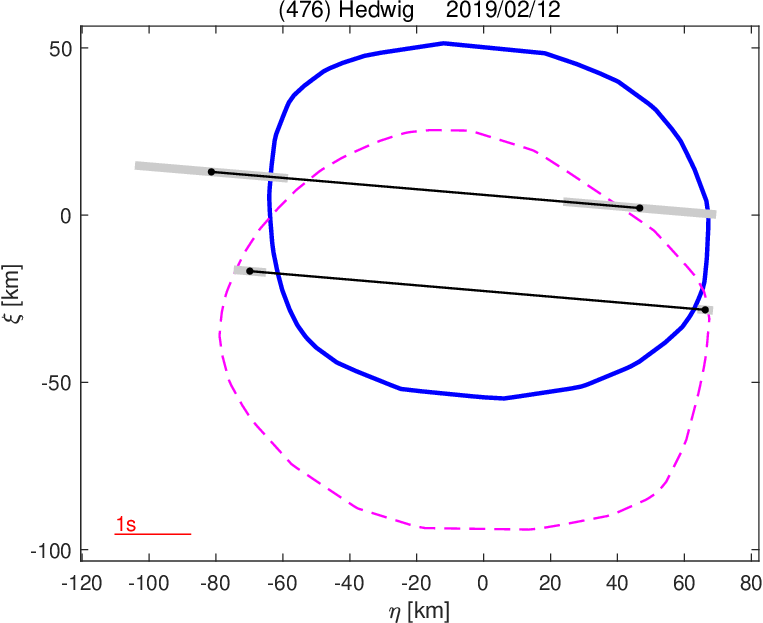}
\includegraphics[width=0.33\textwidth]{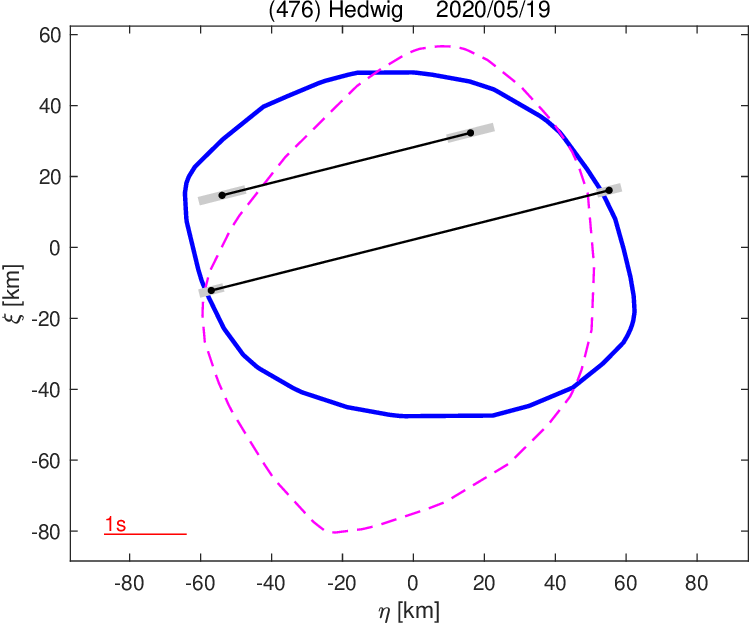}
\caption{(476) Hedwig model fit to occultation chords. Pole 1 is preferred (solid contour).}
\label{476occ}
\end{figure*}

\begin{figure*}
\includegraphics[width=0.33\textwidth]{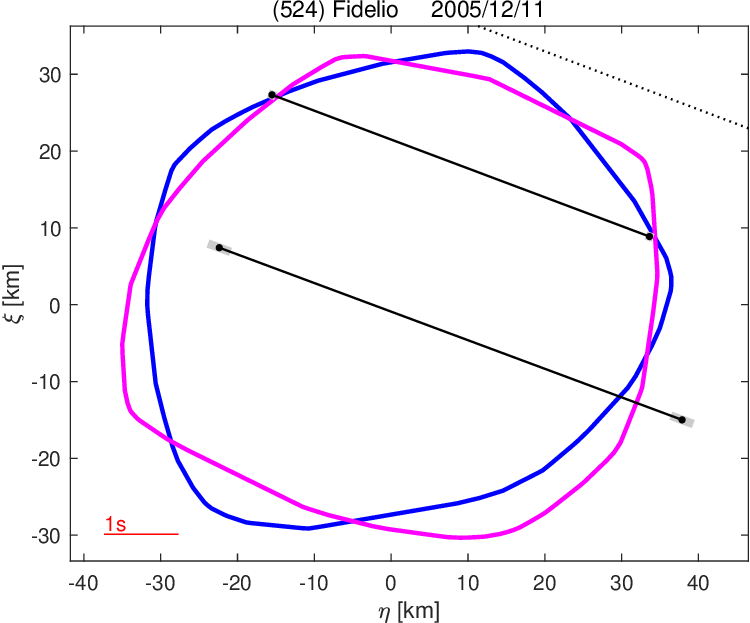}
\includegraphics[width=0.33\textwidth]{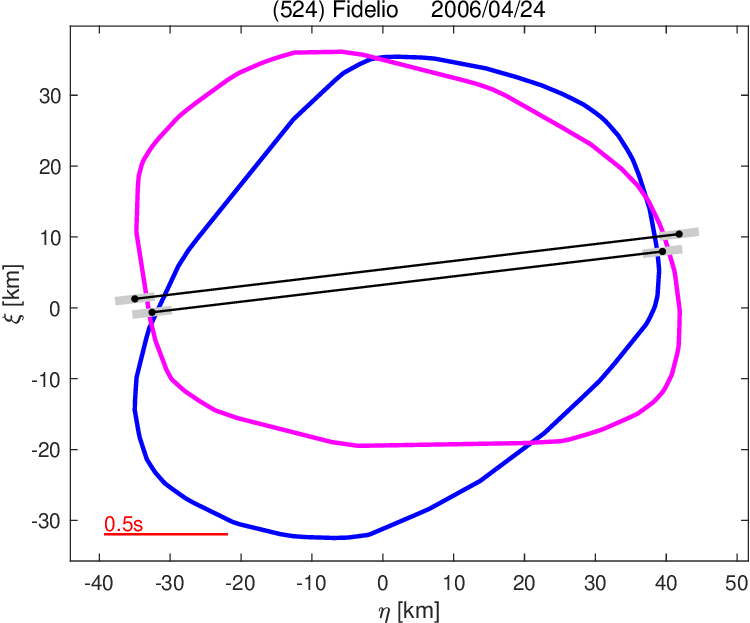}
\includegraphics[width=0.33\textwidth]{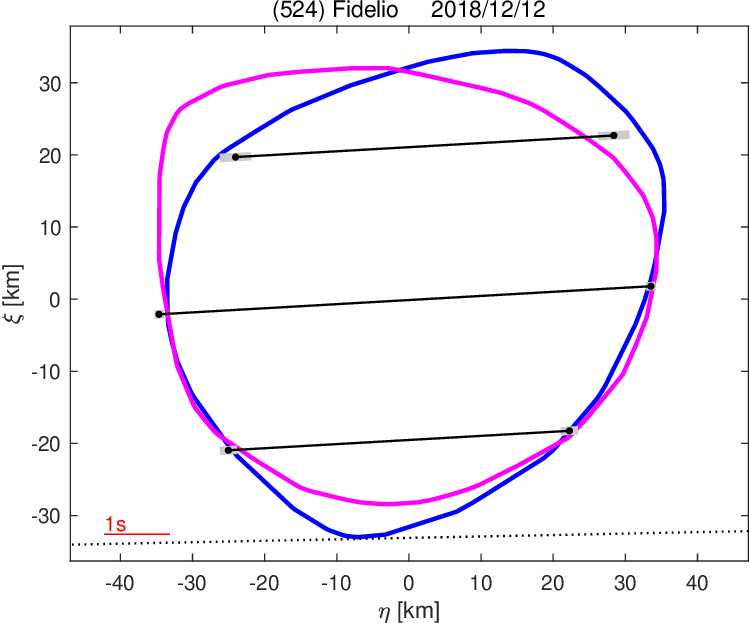}
\includegraphics[width=0.33\textwidth]{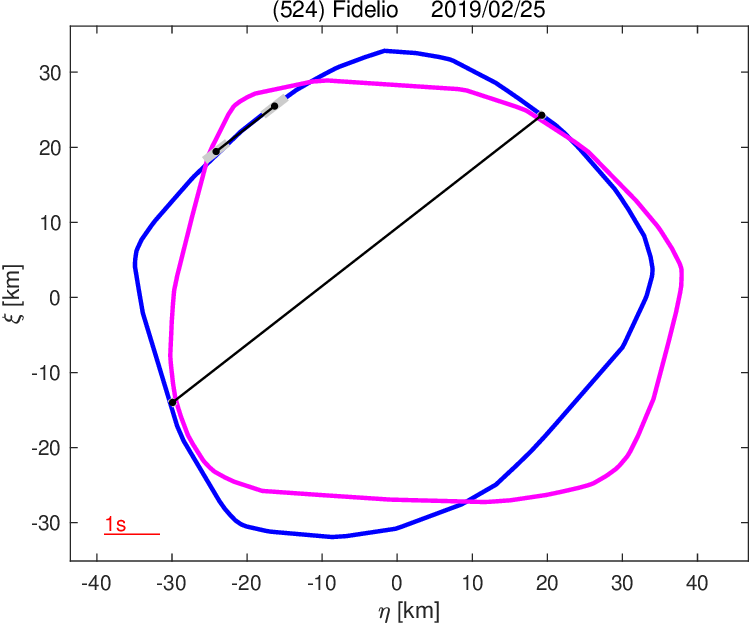}
\includegraphics[width=0.33\textwidth]{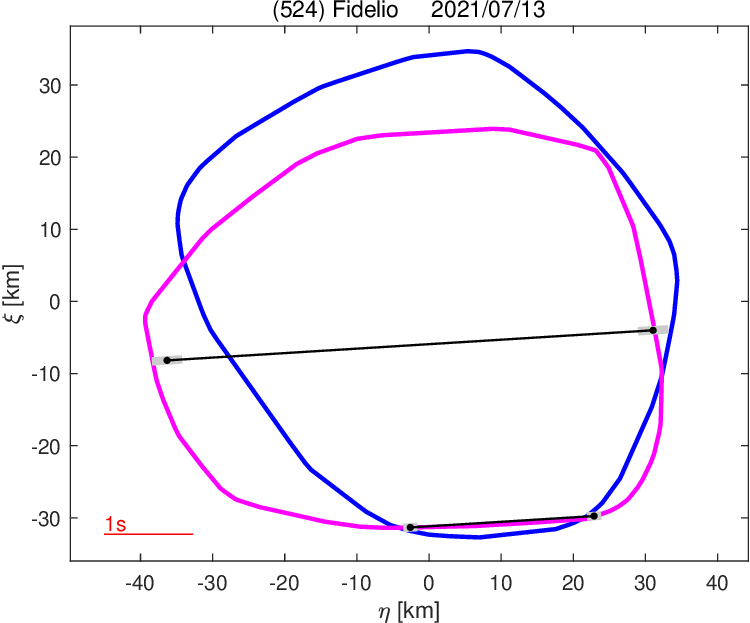}
\includegraphics[width=0.33\textwidth]{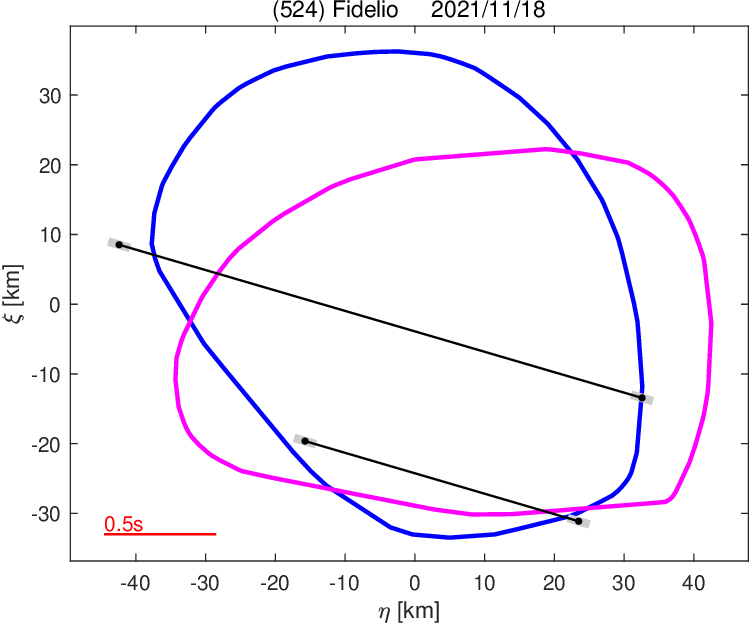}
\caption{(524) Fidelio model fit to occultation chords. Pole 2 is preferred (blue contour).}
\label{524occ}
\end{figure*}

\begin{figure*}
\includegraphics[width=0.33\textwidth]{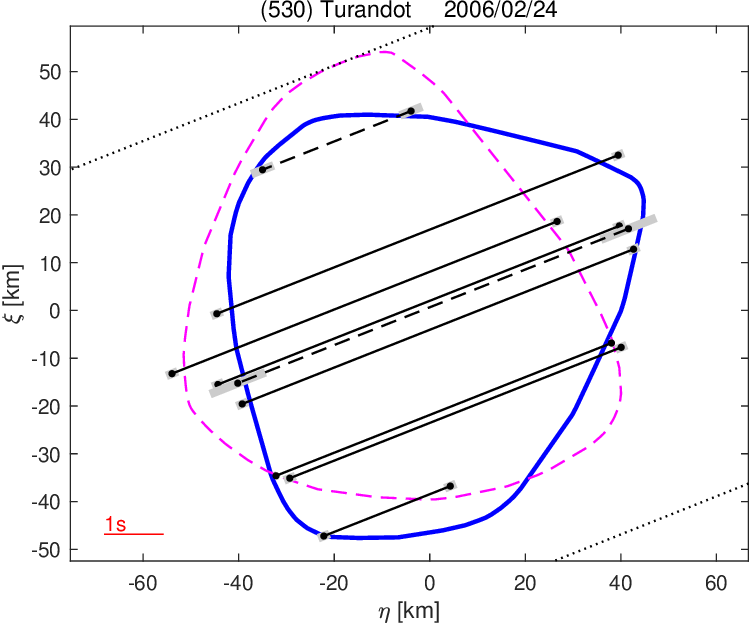}
\includegraphics[width=0.33\textwidth]{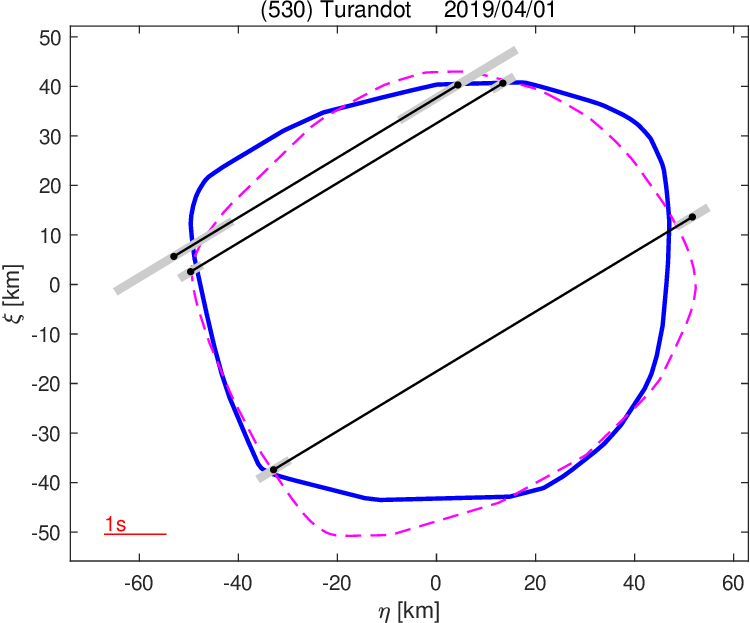}
\includegraphics[width=0.33\textwidth]{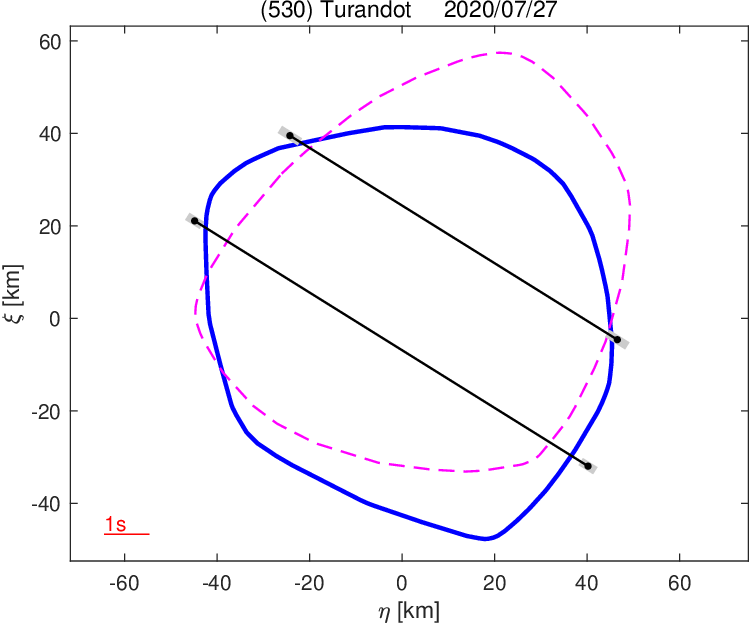}
\caption{(530) Turandot model fit to occultation chords. Pole 2 is preferred (solid contour).}
\label{530occ}
\end{figure*}

\begin{figure*}
\begin{center}
\includegraphics[width=0.3\textwidth]{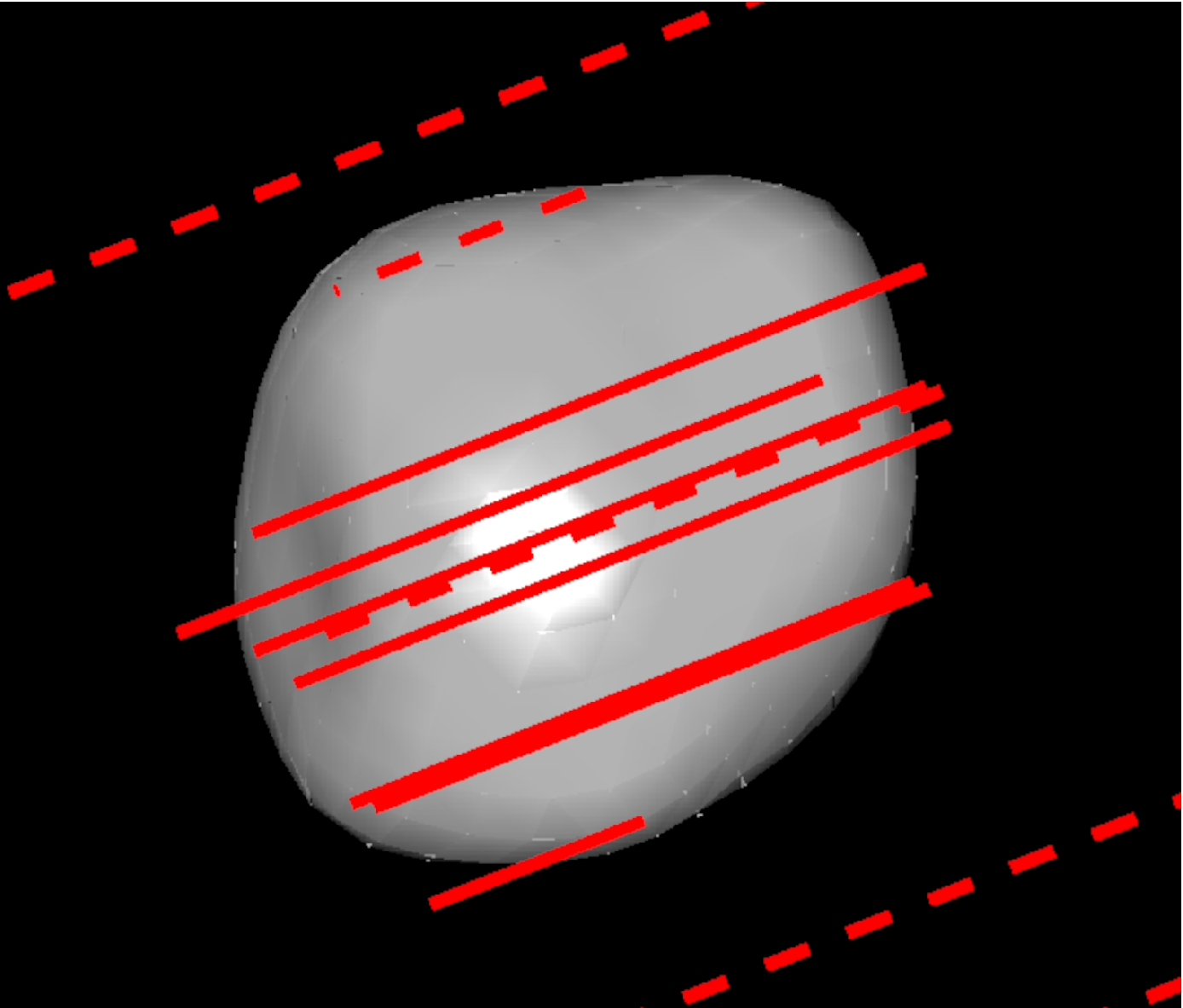}
\includegraphics[width=0.3\textwidth]{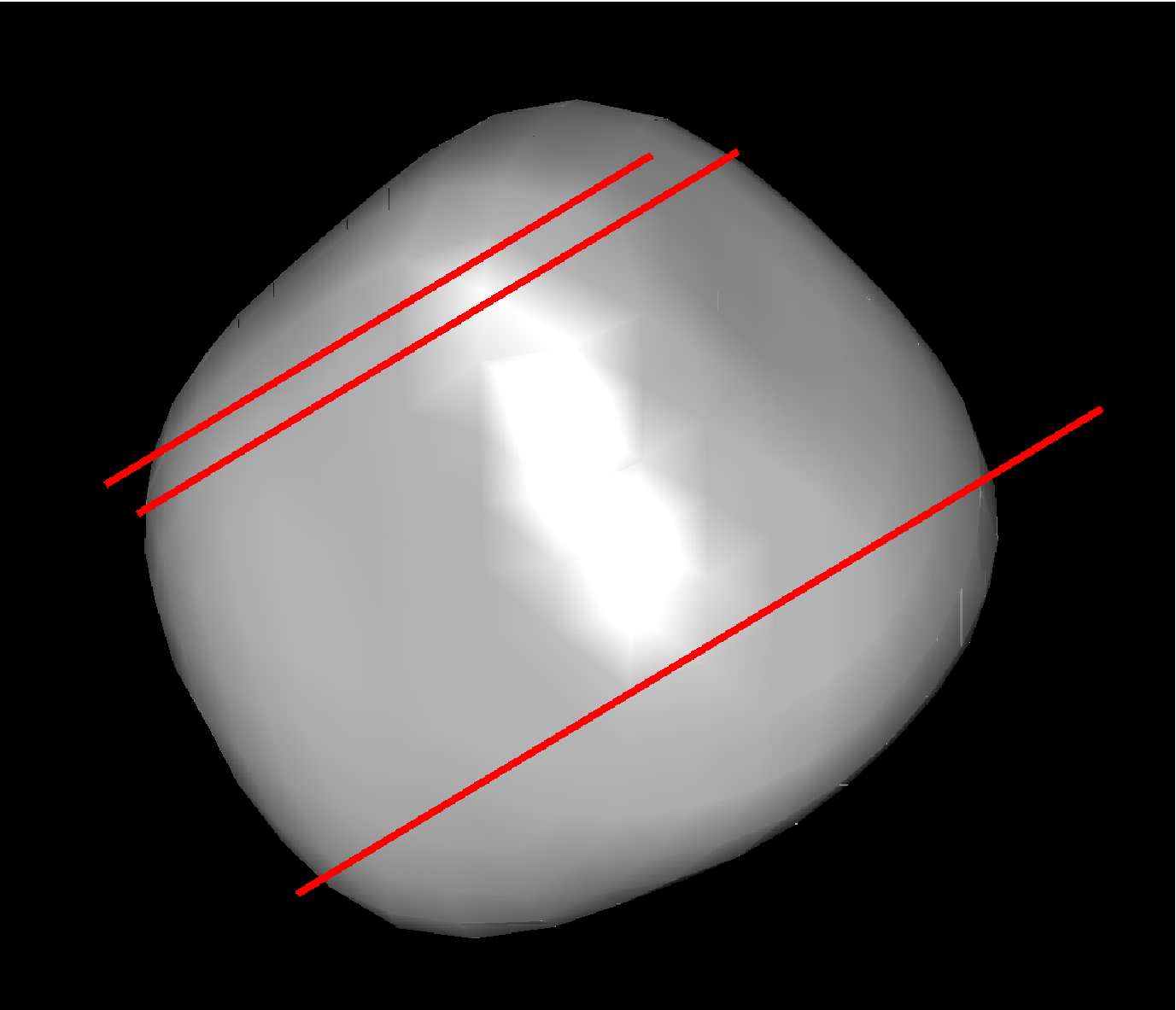}
\includegraphics[width=0.3\textwidth]{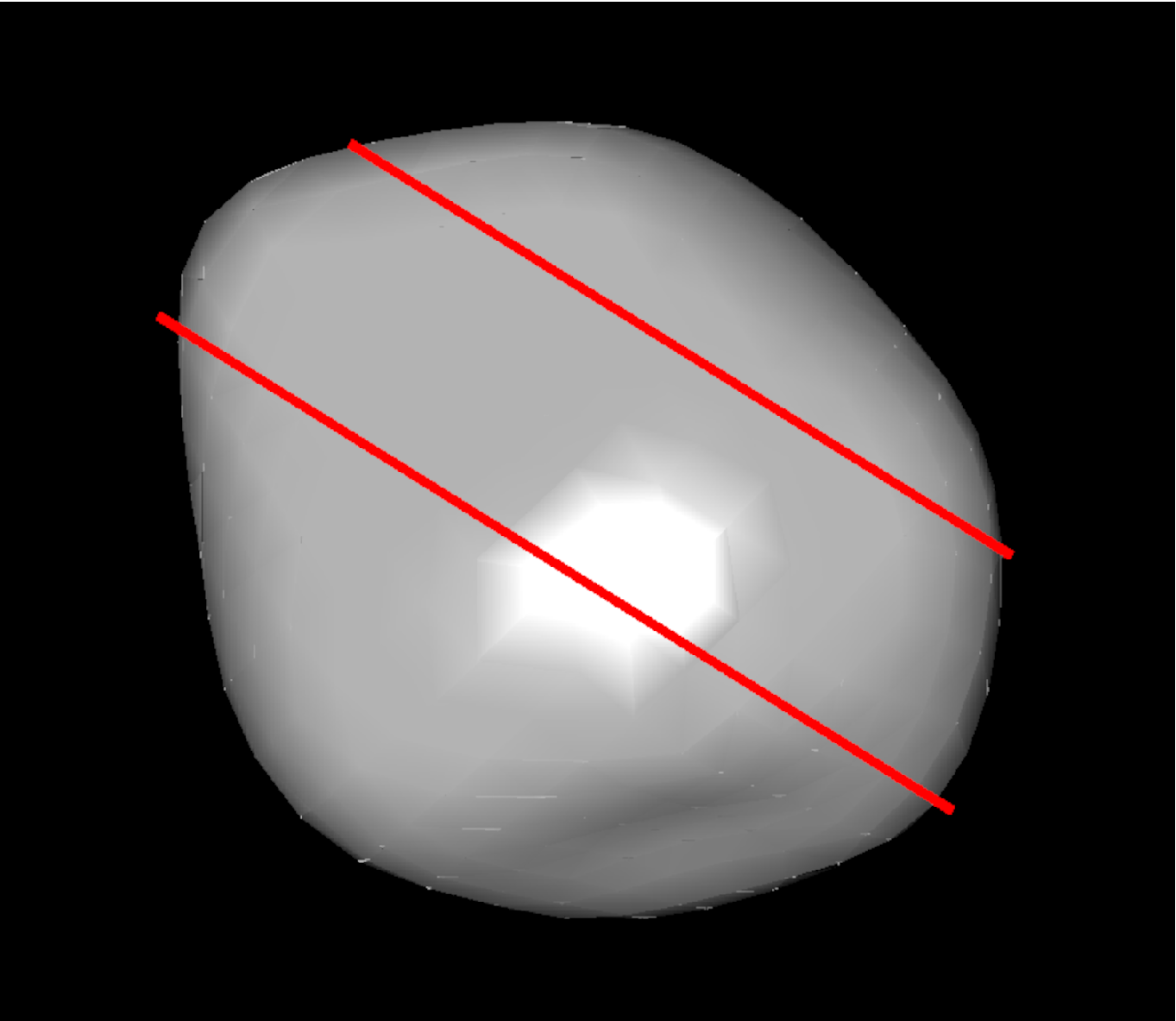}
\includegraphics[width=0.3\textwidth]{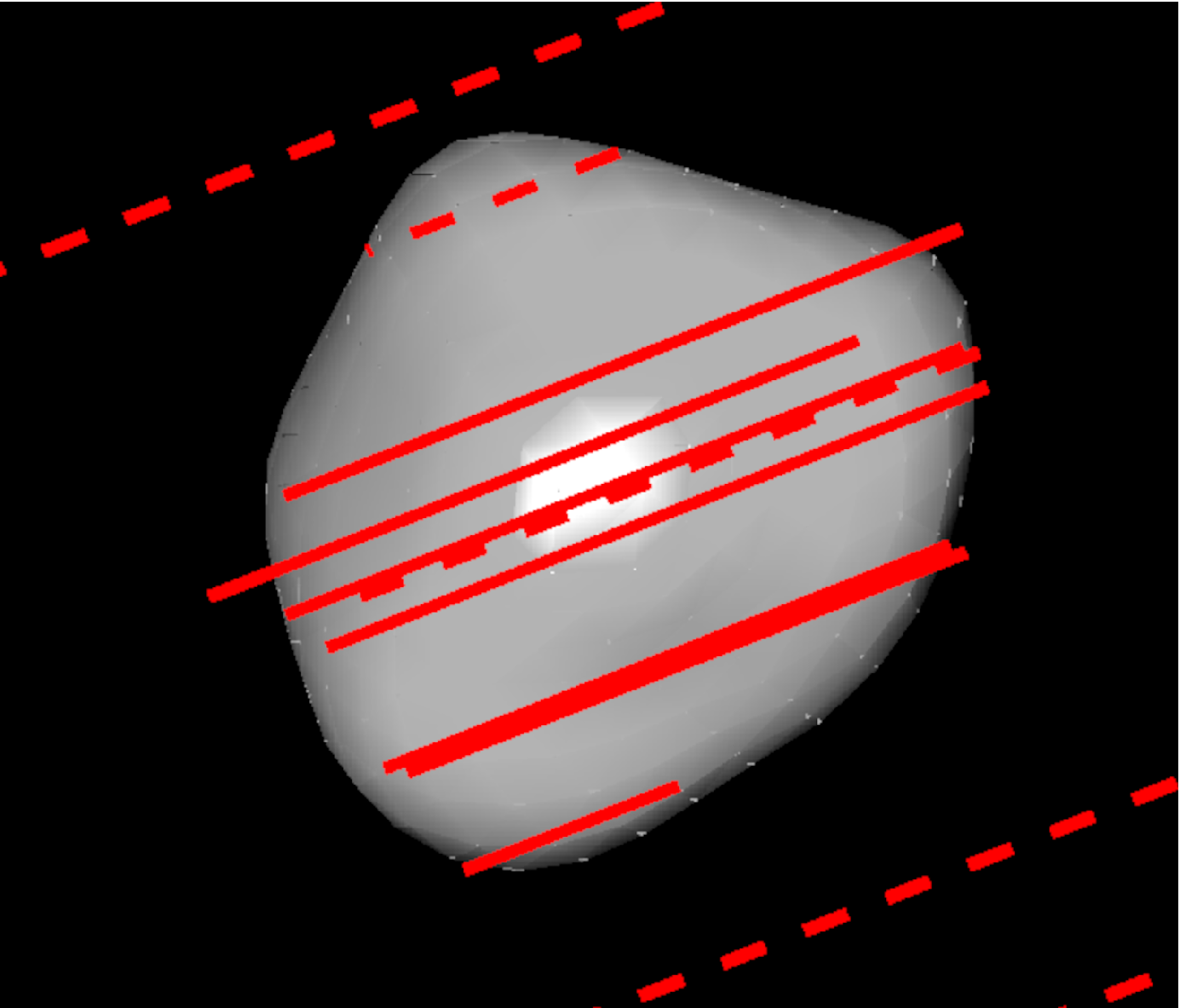}
\includegraphics[width=0.3\textwidth]{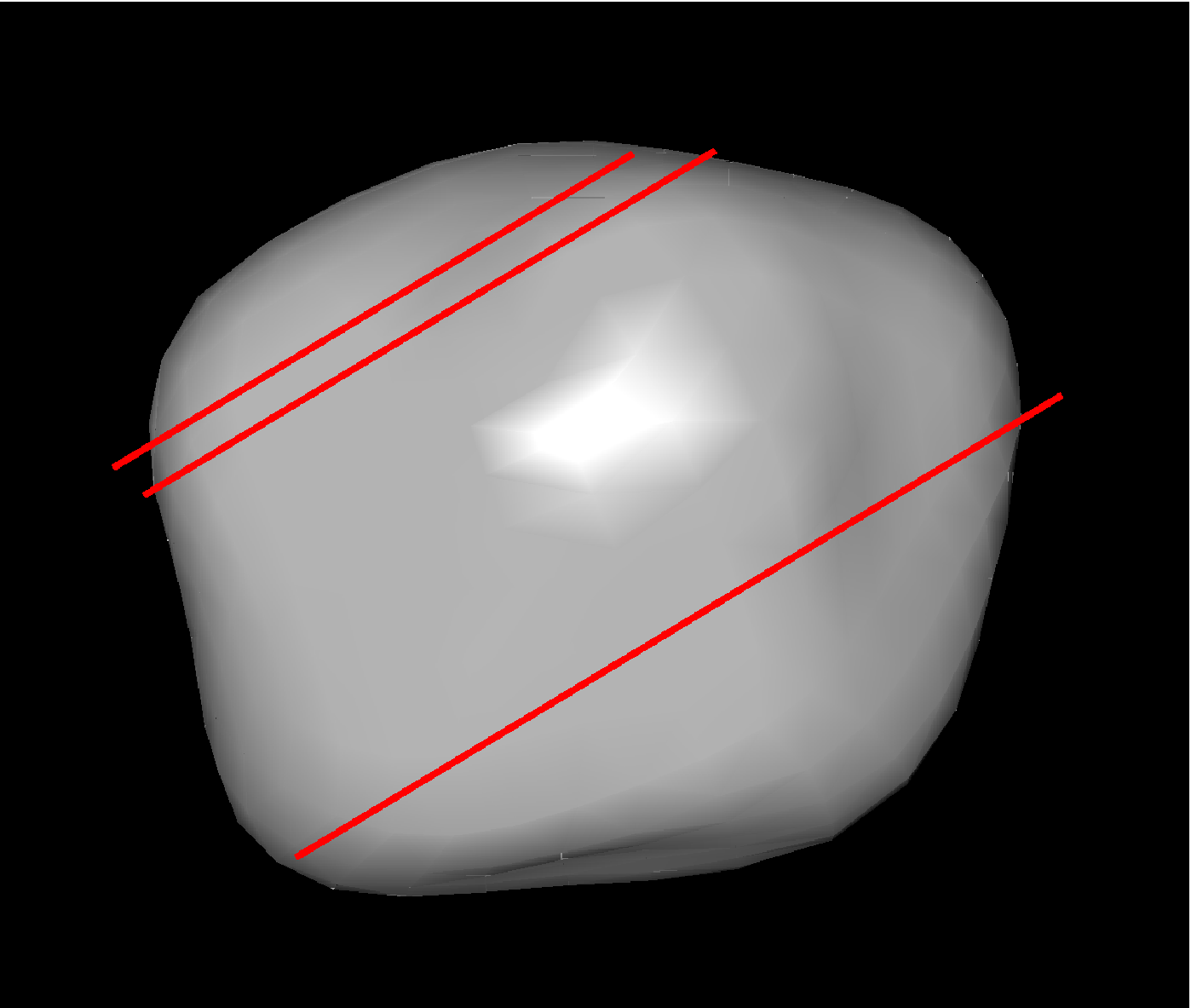}
\includegraphics[width=0.3\textwidth]{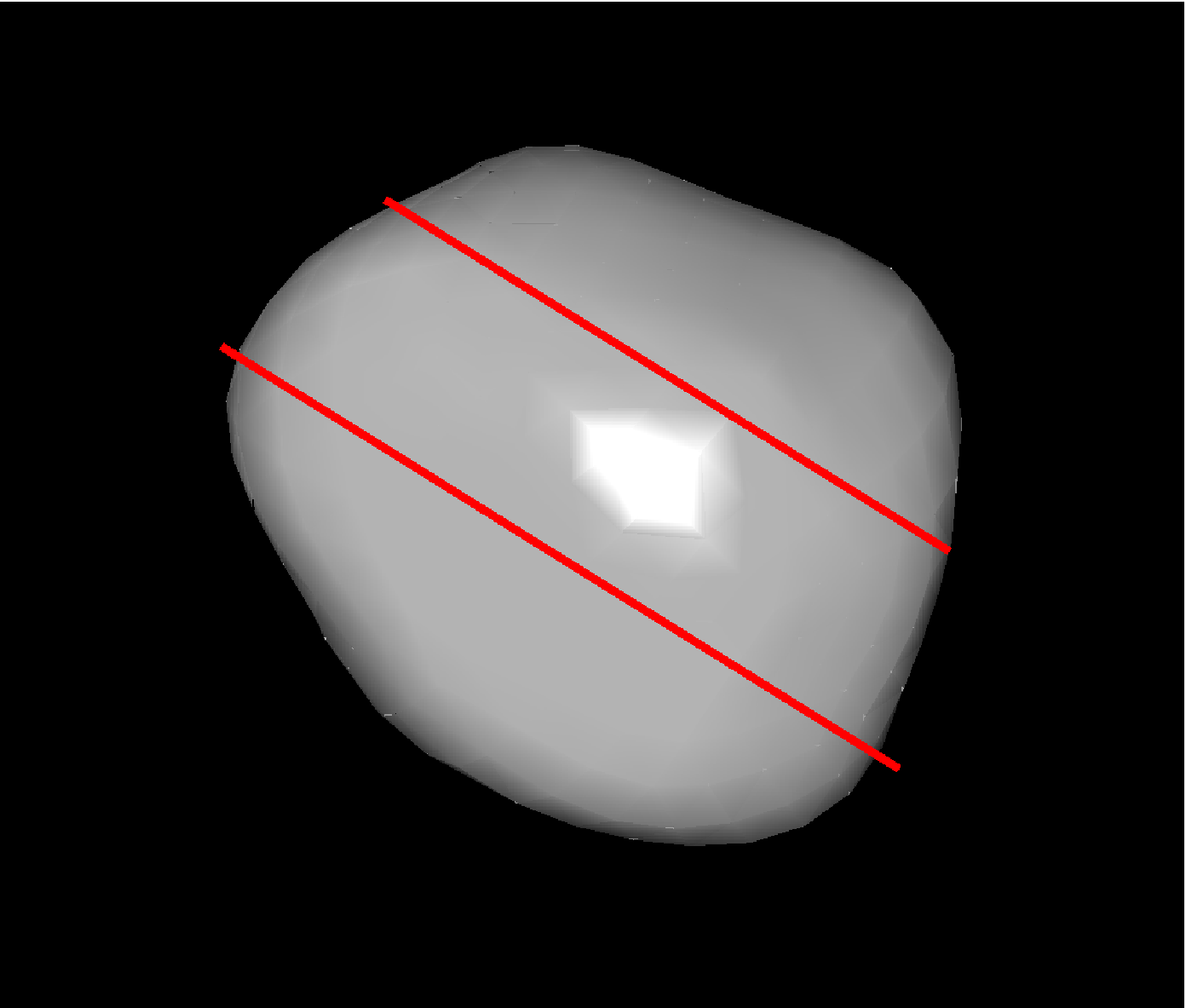}
\end{center}
\caption{(530) Turandot ADAM solution pole 1 (top row), pole 2 (bottom row), shown with occultation chords used to construct the models. 
Occultation events same as in Fig. \ref{530occ}.}
\label{530ADAM}
\end{figure*}

\begin{figure*}
\begin{center}
\includegraphics[width=0.33\textwidth]{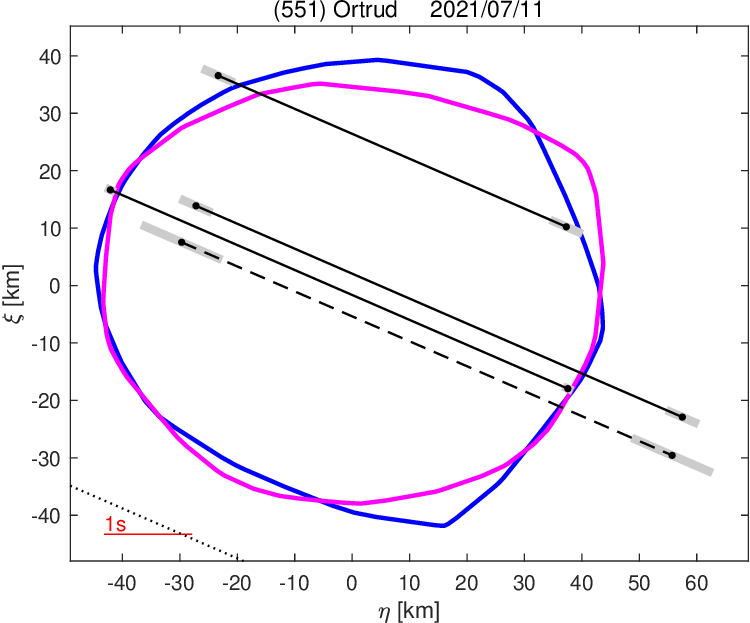}
\includegraphics[width=0.33\textwidth]{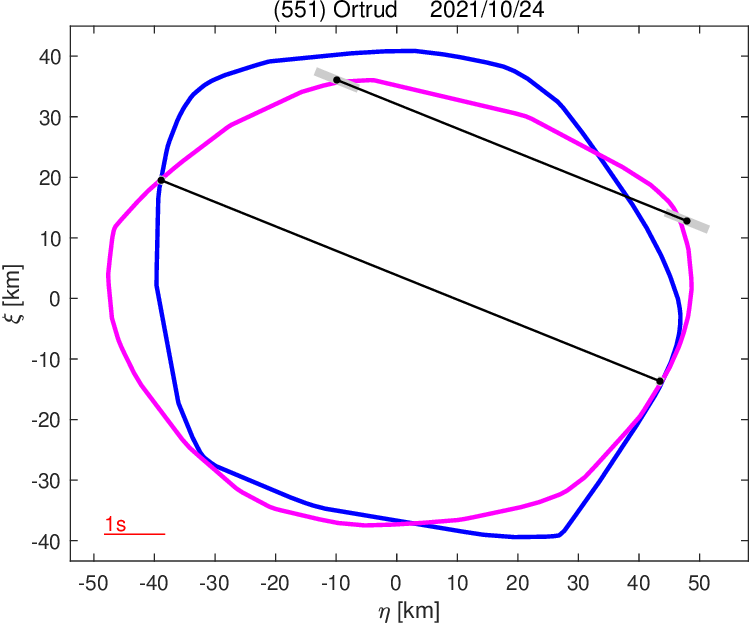}
\end{center}
\caption{(551) Ortrud model fit to occultation chords: pole 1 (blue), and pole 2 (magenta).}
\label{551occ}
\end{figure*}

\begin{figure*}
\includegraphics[width=0.33\textwidth]{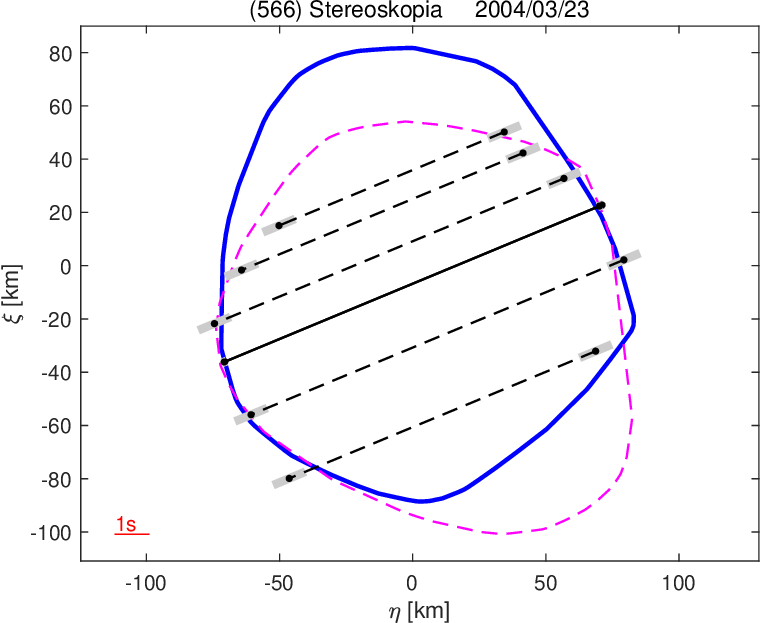}
\includegraphics[width=0.33\textwidth]{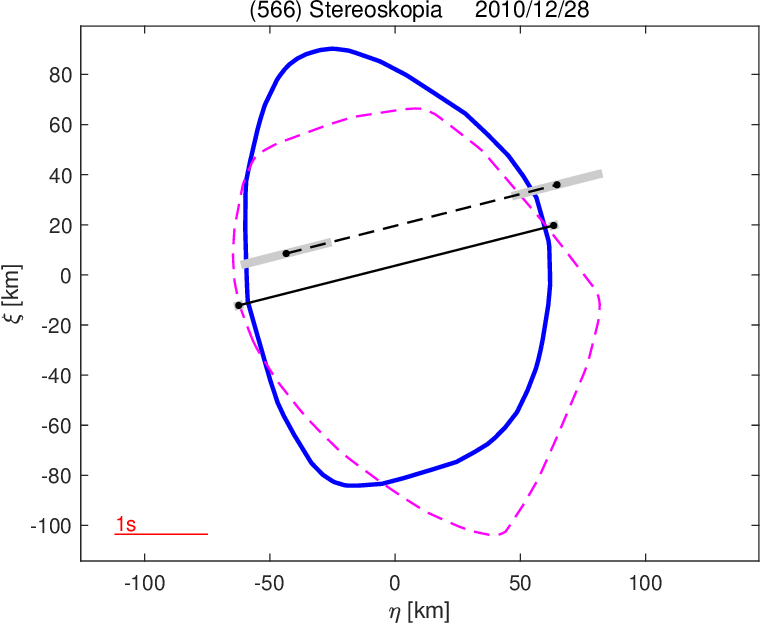}
\includegraphics[width=0.33\textwidth]{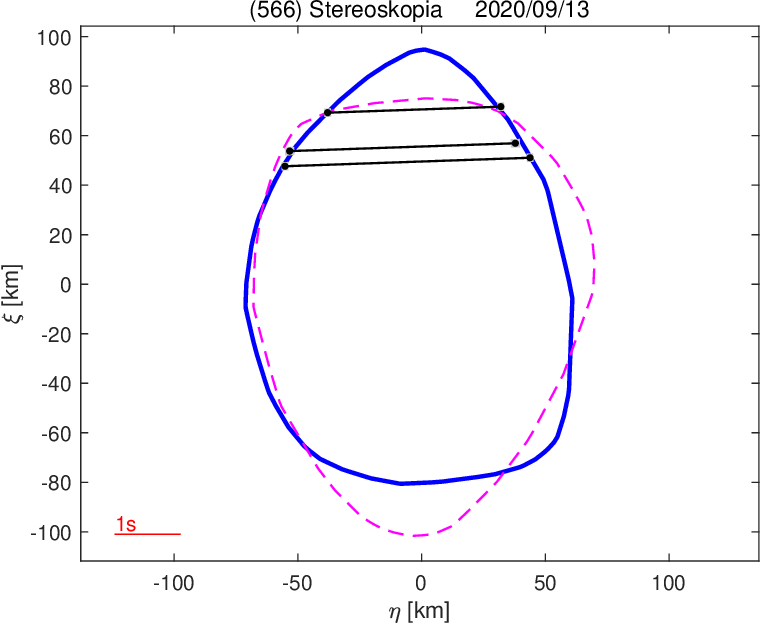}
\includegraphics[width=0.33\textwidth]{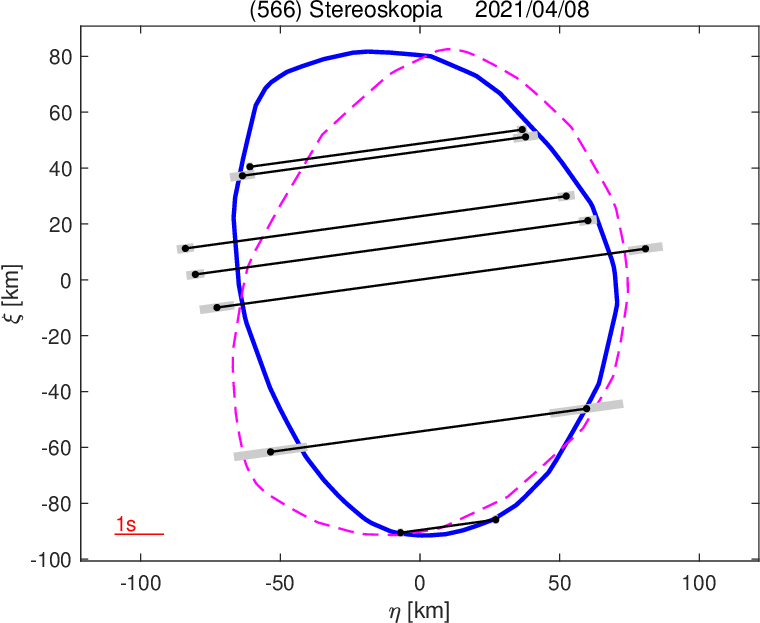}
\includegraphics[width=0.33\textwidth]{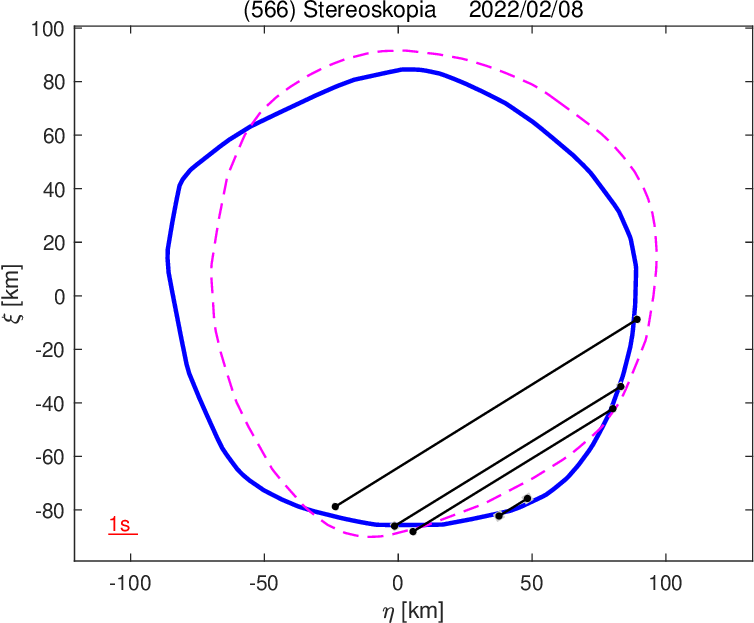}
\caption{(566) Stereoskopia model fit to occultation chords. Pole 1 is preferred (solid contour).}
\label{566occ}
\end{figure*}

\begin{figure*}
\begin{center}
\includegraphics[width=0.33\textwidth]{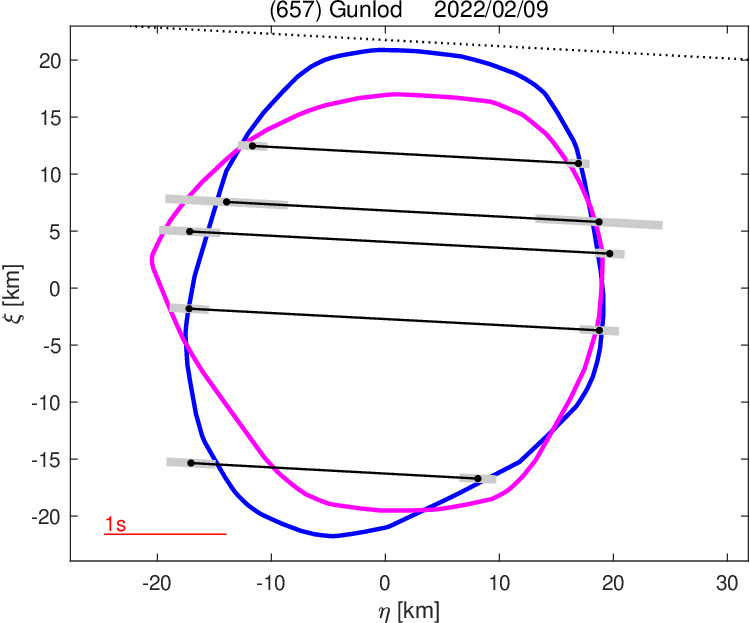}
\end{center}
%\vspace{0.5cm}
\caption{(657) Gunlod model fit to occultation chords. Pole 2 is preferred (blue contour).}
\label{657occ}
\end{figure*}

\begin{figure*}
\includegraphics[width=0.33\textwidth]{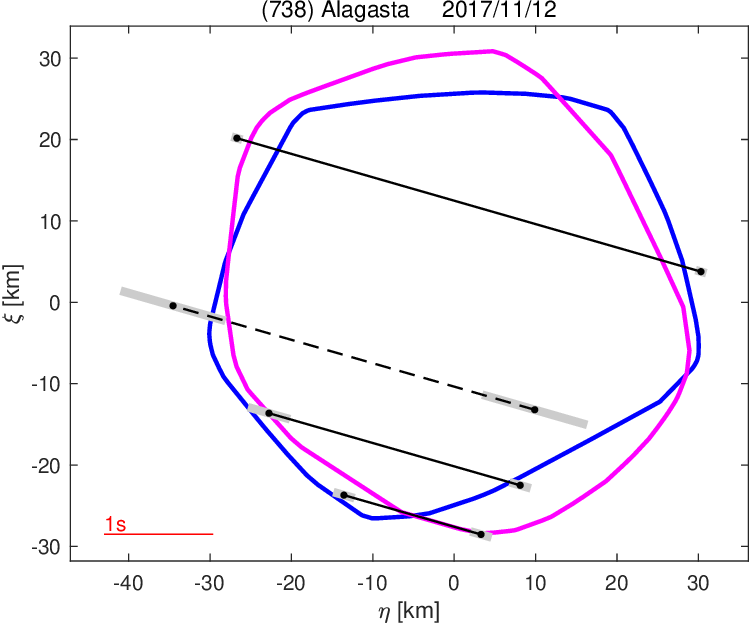}
\includegraphics[width=0.33\textwidth,clip]{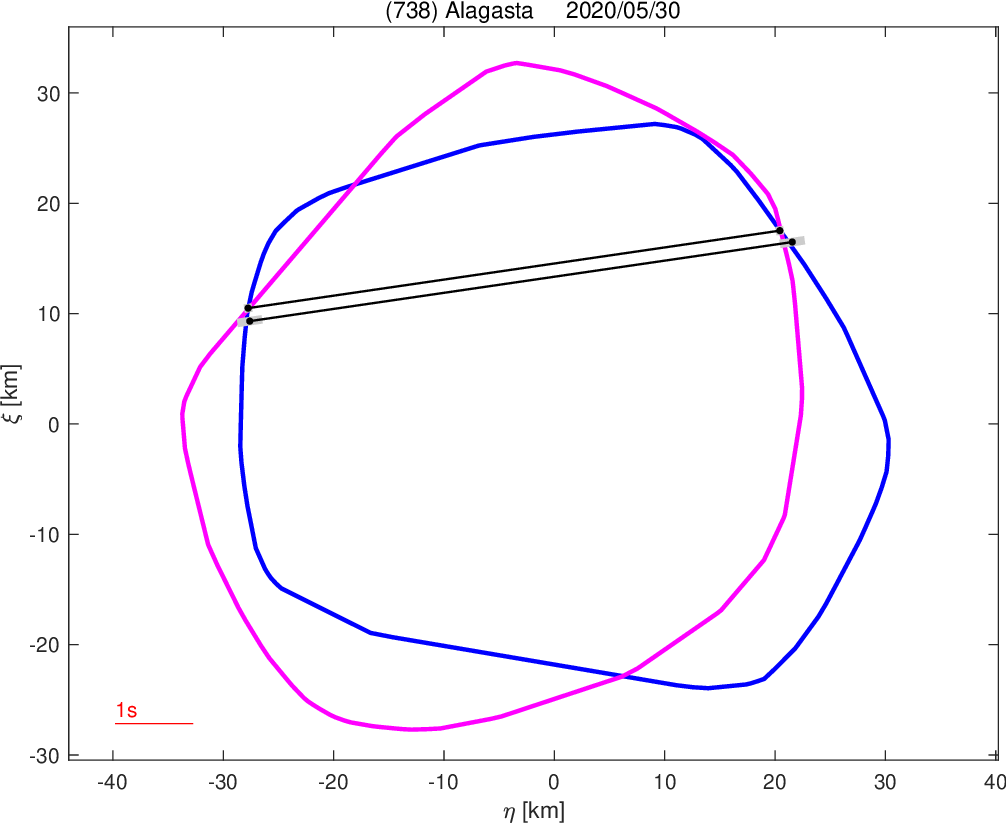}
\includegraphics[width=0.33\textwidth]{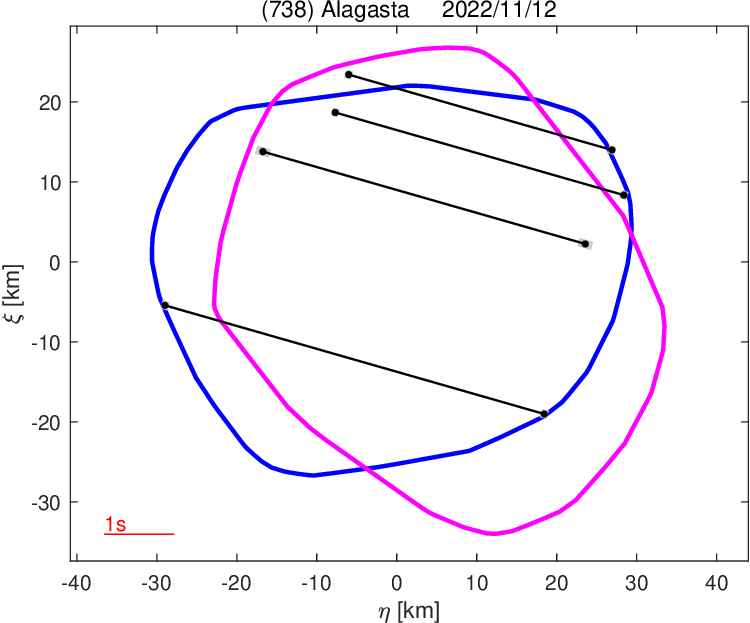}
\caption{(738) Alagasta model fit to occultation chords: pole 2 (blue), and pole 1 (magenta).}
\label{738occ}
\end{figure*}

\begin{figure*}
\begin{center}
\includegraphics[width=0.33\textwidth]{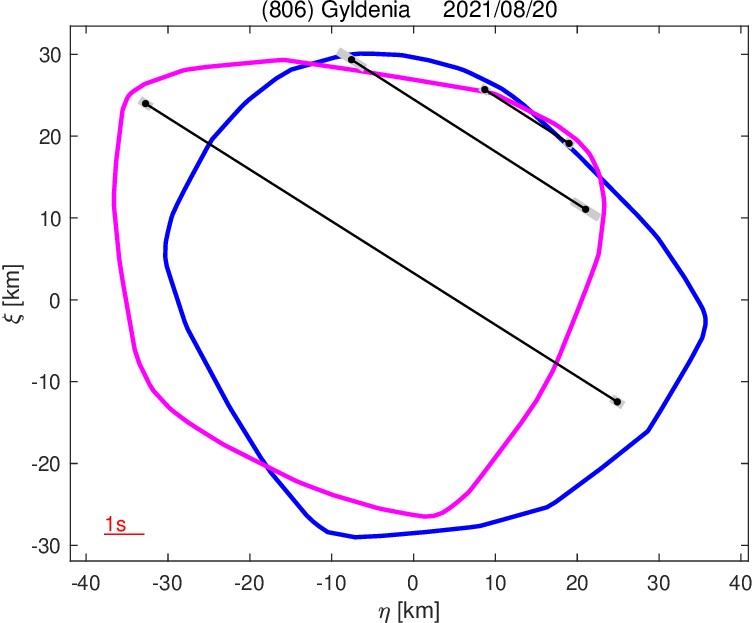}
\end{center}
\caption{(806) Gyldenia model fit to occultation chords: pole 1 (blue), and pole 2 (magenta).}
\label{806occ}
\end{figure*}

\clearpage

\section{Observing campaign details}
 This section contains summarised details of all time-series observations used for the modelling (Table \ref{obs}),
 and the list of stellar occultation event observers and sites (Table \ref{occult_obs}).

\begin{table*}[h!]
\begin{scriptsize}
\noindent
\caption{{\small Details of all photometric observations for light curves: observing dates, number of light curves, range of ecliptic longitudes of the target,
and sun-target-observer phase angles, observer name (or paper citation in case of published data), and the observing site.
 Some data come from robotic telescopes, in which case there is no observer specified.
 For data from the TESS spacecraft, N$_{lc}$ denotes the number of days of continuous observations.
 CSSS stands for Center for Solar System Studies,
 CTIO - Cerro Tololo Interamerican Observatory,
 e-EyE - Entre Encinas y Estrellas,
 ESO - European Southern Observatory,
 OASI - Observat{\'o}rio Astron{\^o}mico do Sert{\~a}o de Itaparica,
 ORM - Roque de los Muchachos Observatory,
 OT - Observatorio del Teide,
 and
 SOAO - Sobaeksan Optical Astronomy Observatory.
 }}
\begin{tabularx}{\textwidth}{ccccXX}
\hline
 Date                    & N$_{lc}$ & $\lambda$ & Phase angle & Observer & Site \\
                         &      & [deg]     &  [deg]      &          &  \\
\hline
&&&&&\\
\multicolumn{6}{l}{70 Panopaea}\\
&&&&&\\
\hline
1980 08 20 - 1980 08 21 & 2 & 329 - 330 & 10 &  \cite{Harris1989} & Mountain Observatory, CA, USA\\
1980 09 05 & 1 & 326 & 12 &  \cite{Schroll1983} & Table Mountain Observatory, CA, USA\\
2006 09 23 - 2006 12 24 & 22 & 48 - 63 & 3 - 20 & - &  Super WASP\\
\hline
\end{tabularx}
\label{obs}
%\caption{Table \ref{obs} continued.}
\end{scriptsize}
\end{table*}

\clearpage

\begin{table*}[h]
\begin{scriptsize}
\noindent
\begin{tabularx}{\textwidth}{ccccXX}
\hline
 Date                    & N$_{lc}$ & $\lambda$ & Phase angle & Observer & Site \\
                         &      & [deg]     &  [deg]      &          &  \\
\hline
&&&&&\\
\multicolumn{6}{l}{ 412 Elisabetha}\\
&&&&&\\
\hline
1990 08 22 - 1990 08 29 & 6 & 328-329 & 5-6 &  \cite{Lagerkvist1992} & ESO, La Silla, Chile\\
2012 05 10 - 2012 05 18 & 7 & 231-233 & 7 &  \cite{Stephens2012} & Rancho Cucamonga, CA USA\\
2016 02 25 - 2016 03 10 & 2 & 163-166 & 7 & K. Kami{\'n}ski &  Winer, AZ, USA\\
\hline
\end{tabularx}
%\caption{Table \ref{obs} continued.}
\end{scriptsize}
\end{table*}

\clearpage

\begin{table*}[h]
\begin{scriptsize}
\noindent
\begin{tabularx}{\textwidth}{ccccXX}
\hline
 Date                    & N$_{lc}$ & $\lambda$ & Phase angle & Observer & Site \\
                         &      & [deg]     &  [deg]      &          &  \\
\hline
                         
&&&&&\\
\multicolumn{6}{l}{ 464 Megaira}\\
&&&&&\\
\hline
1979 10 28 & 1 & 29 & 8 &  \cite{Harris1989} & Table Mountain Observatory, USA\\
2007 09 11 - 2007 09 23 & 8 & 1 - 3 & 8 - 10 &  J. Oey & Australia\\
2010 04 02 - 2010 07 01 & 24 & 182 - 189 & 4 - 18 &  \cite{Waszczak2015} &  Mt. Palomar, CA, USA\\
\hline
\end{tabularx}
%\caption{Table \ref{obs} continued.}
\end{scriptsize}
\end{table*}

\clearpage

\begin{table*}[h]
\begin{scriptsize}
\noindent
\begin{tabularx}{\textwidth}{ccccXX}
\hline
 Date                    & N$_{lc}$ & $\lambda$ & Phase angle & Observer & Site \\
                         &      & [deg]     &  [deg]      &          &  \\
\hline
&&&&&\\
\multicolumn{6}{l}{ 551 Ortrud}\\
&&&&&\\
\hline
2006 11 10 - 2006 12 08 & 8 & 46 - 51 & 1 - 11 & - &  Super WASP\\
2015 06 30 & 1 & 302 & 7 &  A. Marciniak &  OT, Tenerife, Spain\\
2015 07 07 - 2015 09 13 & 8 & 291 - 301 & 1 - 16 & - &  Montsec, Catalonia, Spain\\
\hline
\end{tabularx}
%\caption{Table \ref{obs} continued.}
\end{scriptsize}
\end{table*}

\clearpage

\begin{table*}[h]
\begin{scriptsize}
\noindent
\begin{tabularx}{\textwidth}{ccccXX}
\hline
 Date                    & N$_{lc}$ & $\lambda$ & Phase angle & Observer & Site \\
                         &      & [deg]     &  [deg]      &          &  \\
\hline
&&&&&\\
\multicolumn{6}{l}{ 806 Gyldenia}\\
&&&&&\\
\hline
2010 07 25 - 2010 12 08 & 9 & 38 - 47 & 4 - 18 &  \cite{Waszczak2015} &  Mt. Palomar, CA, USA\\
2013 03 05 - 2013 05 08 & 9 & 210 - 211 & 2 - 15 &  \cite{M2015} & Borowiec, Poland; Organ Mesa Observatory, NM, USA\\
2013 04 07 - 2013 04 18 & 9 & 212 - 214 & 2 - 6 &  \cite{Alkema2013} & - \\
2015 06 28 & 1 & 0 & 17 &  A. Marciniak &  OT, Tenerife, Spain\\
2015 09 11 - 2015 11 01 & 5 & 348 - 356 & 5 - 13 & - &  Montsec, Catalonia, Spain\\
2015 11 12 & 1 & 348 & 15 &  R. Duffard, N. Morales  &  La Sagra, Spain\\
2016 10 04 & 1 & 67 & 15 &  A. Marciniak &  ORM, La Palma, Spain\\
2016 10 07 - 2016 10 09 & 3 & 67 & 14 &  R. Szak{\'a}ts & Piszk{\'e}stet\H{o}, Hungary\\
2016 10 11 - 2016 10 30 & 3 & 65 - 67 & 8 - 13 &  R. Duffard, N. Morales &  LaSagra, Spain\\
2016 10 29 & 1 & 65 & 8 &  A. Marciniak &  CTIO, Chile\\
2018 02 13 & 1 & 143 & 7 &  M.-J. Kim, D.-H. Kim &  SOAO, South Korea\\
2018 02 13 - 2018 03 19 & 4 & 138 - 143 & 7 - 14 &  J. Skrzypek, R. Hirsch &  Borowiec, Poland\\
2018 03 17 & 1 & 138 & 13 & K. Kami{\'n}ski &  Winer, AZ, USA\\
2019 04 30 - 2019 05 21 & 4 & 230 - 234 & 3 - 5 & K. Kami{\'n}ski &  Winer, AZ, USA\\
2019 05 06 - 2019 05 07 & 2 & 233 & 2 &  D.-H. Kim, M.-J. Kim &  SOAO, South Korea\\
2020 09 17 - 2020 09 20 & 5 & 299 & 14 - 15 &  F. Monteiro, W. Mesquita, W. Pereira, M. Evangelista, E. Rond{\'o}n, J. Michimani, D. Lazzaro, T. Rodrigues  &  OASI, Itacuruba, Brasil\\
2021 09 08  - 2021 11 25 & 4 & 4 - 15 & 8 - 15 &  M. Dr{\'o}{\.z}d{\.z} &  Suhora, Poland\\
\hline
\end{tabularx}
\end{scriptsize}
\end{table*}

%=================================================
\begin{table}[h!]
\begin{tabular}{cc}
\hline
&\\
\multicolumn{2}{c}{(70) Panopaea, 2006-12-14}\\
&\\
\hline
S. Herchak & Mesa, AZ \\
B. Roberts & Shady Grove, MS \\
B. Richardson & Hewitt, TX \\
J. Barton & Robinson, TX \\
R. Nugent & Centerville, TX\\
D. Craig Smith & Ozona, TX\\
M. McCants & Oakalla, TX \\
D. Dunham & McComb, MS \\
W. Aulenbacher &  L. Bunshannon, TX \\
D. Nye & Marfa, TX \\
P. Maley & Madisonville, TX \\
D. Rask & Huntsville, TX \\
D. Clark & Providence, TX \\
K. Drake & Willis, TX \\
P. Nolan & Mountain Home, TX \\
A. Sanchez-Ibarra & Sonora, MX \\
B. Cudnik & Houston, TX \\
R. Frakenberger & San Antonio, TX \\
\hline
&\\
\multicolumn{2}{c}{(70) Panopaea, 2014-10-04}\\
&\\
\hline
J. Broughton & Eagleby, AU \\
G. Bolt & Perth, AUS \\
\hline
&\\
\multicolumn{2}{c}{(70) Panopaea, 2014-04-25}\\
&\\
\hline
A. Olsen & Urbana, IL \\
P. Maley & Marana, AZ \\
J. Moore & Blackwell, OK \\
\hline
&\\
\multicolumn{2}{c}{(275) Sapientia, 1985-04-15}\\
&\\
\hline
A. Olsen & Urbana, IL \\
P. Maley & Marana, AZ \\
J. Moore & Blackwell, OK \\
\hline
&\\
\multicolumn{2}{c}{(275) Sapientia, 2003-02-08}\\
&\\
\hline
R. Venable & Pelham, USA \\
D. Rowley & North Henderson, NC \\
\hline
&\\
\multicolumn{2}{c}{(275) Sapientia, 2003-04-12}\\
&\\
\hline
J. Nosaka & Masuda: Shimane, JP \\
M. Uchiyama & Owase: Mie, JP\\
I. Yamanihi, Naoko Kaifu et al & JP \\
M. Yamanishi & Yuasa: Wakayama, JP \\
\hline
&\\
\multicolumn{2}{c}{(275) Sapientia, 2014-09-02}\\
&\\
\hline
J. Rovira & ES \\
J. Juan & ES\\
E. Frappa, A. Klotz & FR \\
C. Perell{\'o}, A. Selva & ES \\
P. Baruffetti & IT \\
M. Bachini & IT \\
\hline
\end{tabular}
\caption{List of stellar occultation observers and locations of their observing sites. Source: Occult programme.}
\label{occult_obs}
\end{table}

\begin{table}[h!]
\begin{tabular}{cc}
\hline
&\\
\multicolumn{2}{c}{(275) Sapientia, 2015-09-30}\\
&\\
\hline
C. Hills & GB \\
S. Clarke & GB \\
J. Warell & SE \\
C. Hooker & GB \\
S. Kidd & GB \\
T. Haymes & GB \\
A. Jones & GB \\
M. Charron & GB \\
D. Arditti & GB \\
H. Paulus & DE \\
T. Law & GB \\
H. Denzau & DE \\
P. Denyer & GB \\
P. Carson & GB \\
M. Jennings & GB \\
E. Edens & NL \\
\hline
&\\
\multicolumn{2}{c}{(275) Sapientia, 2016-12-03}\\
&\\
\hline
O. Kl\"{o}s & DE \\
V. P\v{r}ib{\'a}\v{n} & CZ \\
K. Hal{\'i}\v{r} & CZ \\
M. Rottenborn & CZ \\
J. Pol{\'a}k & Cz \\
\hline
&\\
\multicolumn{2}{c}{(275) Sapientia, 2018-08-06}\\
&\\
\hline
G. Vaudescal, A. Cazaux & FR \\
P. Lindner & DE \\
C. Schnabel & ES \\
B. Kattentidt & DE \\
G. Dangl & AT \\
J. Pol{\'a}k & CZ \\
\hline
&\\
\multicolumn{2}{c}{(275) Sapientia, 2020-11-14}\\
&\\
\hline
G. Vaudescal & FR \\
E. Frappa, A. Klotz & FR \\
M. Conjat & FR \\
A. Manna & CH \\
S. Sposetti & CH \\
G. Casalnuovo & IT \\
A. Kreutzer, R. Sch\"{a}fer & AT \\
M. Billiani & FR \\
\hline
&\\
\multicolumn{2}{c}{(275) Sapientia, 2021-04-26}\\
&\\
\hline
T. George & Scottsdale, AZ \\
P. Stuart & Clear Lake Shores, TX \\
S. Brazill & Georgetown, TX \\
\hline
&\\
\multicolumn{2}{c}{(286) Iclea, 2008-06-15}\\
&\\
\hline
D. Lowe & Mt. Mee, AU \\
J. Bradshaw, A. Beck & Samford, AU \\
P. Anderson & The Gap, Qld, AU \\
D. Breadsell & Toowoomba, AU \\
\hline
&\\
\multicolumn{2}{c}{(326) Tamara, 2016-09-23}\\
&\\
\hline
J. Broughton & Reedy Creek, AU \\
J. Broughton & Brunswick Heads, AU \\
J. Broughton & Ballina, AU \\
J. Broughton & Broadwater, AU \\
\hline
\end{tabular}
%\caption{List of stellar occultation observers and locations of their observing sites.}
%\label{occult_obs}
\end{table}

\begin{table}[h!]
\begin{tabular}{cc}
\hline
&\\
\multicolumn{2}{c}{(326) Tamara, 2017-09-15}\\
&\\
\hline
D. Breit & Morgan Hill, CA \\
R. Nolthenius & Paicine, CA \\
J. Bardecker & Gardnerville, NV \\
\hline
 &\\
\multicolumn{2}{c}{(326) Tamara, 2017-12-28}\\
&\\
\hline
H. De Groot & NL \\
E. Conseil & FR \\
T. Jan{\'i}k & CZ \\
R. Boninsegna & BE \\
\hline
&\\
\multicolumn{2}{c}{(326) Tamara, 2018-05-12}\\
&\\
\hline
B. Leonard & Sherman, TX \\
M. Smith & Sherman, TX \\
K. Cobble & Princeton, TX \\
\hline
&\\
\multicolumn{2}{c}{(326) Tamara, 2018-12-21}\\
&\\
\hline
A. Manna & CH \\
S. Sposetti & CH \\
\hline
&\\
\multicolumn{2}{c}{(412) Elisabetha, 2011-01-23}\\
&\\
\hline
M. Ida & Higashiomi, Shiga, JP \\
M. Ishida & Koka, Shiga, JP\\
A. Asai & Inabe, Mie, JP\\
H. Watanabe et al & Inabe, mie, JP\\
M. Owada & Hamamatsu, Shizuoka, JP\\
\hline
&\\
\multicolumn{2}{c}{(412) Elisabetha, 2016-03-17}\\
&\\
\hline
P. Delincak & SK \\
B. G\"{a}hrken & DE \\
M. Korec & SK \\
\hline
&\\
\multicolumn{2}{c}{(412) Elisabetha, 2017-03-19}\\
&\\
\hline
A. Ossola & CH \\
A. Manna & CH \\
\hline
&\\
\multicolumn{2}{c}{(412) Elisabetha, 2019-12-04}\\
&\\
\hline
P. Baruffetti & IT \\
S. Donati & IT \\
\hline
&\\
\multicolumn{2}{c}{(426) Hippo, 2012-01-13}\\
&\\
\hline
D. Ewald & DE \\
G. Wortmann, K. Walzel & DE\\
W. Rothe & DE \\
O. Canales, F. Campos & ES \\
E. Frappa, M. Lavayssi{\`e}re & FR \\
P. Lindner & DE \\
C. Schnabel, J. Juan & ES \\
J. Rovira & ES \\
A. Selva & ES \\
\hline
&\\
\multicolumn{2}{c}{(426) Hippo, 2015-06-20}\\
&\\
\hline
R. Baldridge & Los Altos Hills, CA \\
T. Swift & Davis, CA\\
J. Bardecker & Gardnerville, NV \\
C. McPartlin & Santa Barbara, CA\\
\hline
\end{tabular}
%\caption{List of stellar occultation observers and locations of their observing sites.}
%\label{occult_obs}
\end{table}

\begin{table}[h!]
\begin{tabular}{cc}
\hline
&\\
\multicolumn{2}{c}{(426) Hippo, 2018-05-21}\\
&\\
\hline
J. Broughton & Tumbulgum, AU \\
J. Broughton & Brunswick Heads, AU \\
J. Broughton & Ballina, AU\\
\hline
&\\
\multicolumn{2}{c}{(426) Hippo, 2021-12-27}\\
&\\
\hline
H. Yoshihara & Soja, Okayama, JP \\
M. Yamashita & Ikeda, Osaka, JP \\
T. Goto & Yamada Higashi, JP \\
M. Higuchi & Shigokamachi, JP\\
M. Nishimura & Kitakuzuhacho, JP\\
H. Kasebe & Osaka, JP \\
A. Asai & Kuwana, Mie, JP\\
A. Hashimoto & Chichibu, Saitama, JP\\
H. Watanabe, H. Watanabe & Tsu, Mie, JP \\
M. Ishida & Tsu, Mie, JP\\
R. Aikawa & Sakasdo, Saitama, JP \\
M. Ida & Tsu, Mie, JP\\
R. Kukita & Nabekura, JP\\
T. Horaguchi & Tsukuba, JP \\
K. Fukui & Nagakunidai, JP\\
K. Kouno & Miyazaki, JP\\
K. Terakubo & Kokubunji, Tokyo, JP \\
K. Kitazaki & Musashino, Tokyo, JP\\
S. Uchiyama & Kashiwa, Chiba, JP \\
T. Hirose & Ohtaku, Tokyo, JP\\
M. Owada & Hamamatsu, Shizuoka, JP\\
\hline
&\\
\multicolumn{2}{c}{(426) Hippo, 2021-12-31}\\
&\\
\hline
A. Selva & ES \\
J. Juan & ES \\
C. Schnabel & ES \\
\hline
&\\
\multicolumn{2}{c}{(426) Hippo, 2022-03-23}\\
&\\
\hline
J. M{\'a}nek & CZ \\
J. Kub{\'a}nek & CZ \\
K. Hal{\'i}\v{r} & CZ \\
M. Rottenborn & CZ \\
J. Pol{\'a}k & CZ \\
E. Kowald & AT \\
B. Kattentidt & DE \\
\hline
&\\
\multicolumn{2}{c}{(439) Ohio, 2017-05-27}\\
&\\
\hline
D. Gault & Hawkesbury Heights, AU \\
T. Barry & Penrith, AU \\
D. Herald & Murrumbateman:nsw, AU \\
J. Newman & Flynn:act, AU \\
\hline
&\\
\multicolumn{2}{c}{(439) Ohio, 2022-03-11}\\
&\\
\hline
P. Zelen{\'y} & CZ \\
P. Delincak & SK \\
M. Urbanik & SK \\
M. Zawilski & PL \\
L. Benedyktowicz & PL \\
M. Borkowski & PL \\
M. Filipek & PL \\
M. Harman & SK \\
D. B{\l}a{\.z}ewicz & PL \\
\hline
\end{tabular}
%\caption{List of stellar occultation observers and locations of their observing sites.}
%\label{occult_obs}
\end{table}

\begin{table}[h!]
\begin{tabular}{cc}
\hline
&\\
\multicolumn{2}{c}{(439) Ohio, 2022-03-24}\\
&\\
\hline
O. Schreurs & BE \\
B. Goffin & BE \\
R. Boninsegna, P. Assoignon & BE \\
K. Guhl & DE \\
O. Hofschulz & DE \\
S. Andersson & DE \\
J. Delpau & FR \\
M. Borkowski & PL \\
A. Marciniak & PL \\
\hline
&\\
\multicolumn{2}{c}{(464) Megaira, 2017-12-06}\\
&\\
\hline
G. Lyzenga & Altadena, CA  \\
J. Moore & Tulsa, OK \\
C. Ellington & Edmond, OK \\
P. Maley & Congress, AZ \\
\hline
&\\
\multicolumn{2}{c}{(464) Megaira, 2017-12-18}\\
&\\
\hline
H. Tomioka & Hitachi, Ibaraki, JP \\
H. Watanabe & Tarui, Gifu, JP \\
S. Uchiyama & Hamamatsu, Shizuoka, JP \\
A. Asai & Inabe, Mie, JP\\
H. Watanabe & Inabe, Mie, JP\\
M. Owada & Hamamatsu, Shizuoka, JP \\
M. Ishida & Moriyama, Shiga, JP\\
Y. Ikari & Moriyama, Shiga, JP\\
T. Ito & Suzuka, Mie, JP\\
H. Yamamura & Tsu, Mie, JP \\
M. Ida & Tsu, Mie, JP\\
\hline
&\\
\multicolumn{2}{c}{(476) Hedwig, 2000-11-07}\\
&\\
\hline
N. Quinn & GB \\
J. Moellmann & DE \\
D. Ewald & DE \\
N. W\"{u}nsche & DE \\
M. Dentel & DE \\
S. Andersson, M. Haupt & DE \\
P. Enskonatus & DE \\
J. A. Berdejo & ES \\
J. Grados & ES \\
\hline
&\\
\multicolumn{2}{c}{(476) Hedwig, 2017-02-24}\\
&\\
\hline
B. Loader & Nelson, NZL \\
G. McKay & Blenheim \\
\hline
&\\
\multicolumn{2}{c}{(476) Hedwig, 2018-07-26}\\
&\\
\hline
P. Delincak & SK \\
J. Kub{\'a}nek & HU \\
M. Jarmoc & PL \\
\hline
&\\
\multicolumn{2}{c}{(476) Hedwig, 2018-11-13}\\
&\\
\hline
C. Perell{\'o}, A. Selva & ES \\
C. Perell{\'o}, A. Selva & ES \\
J. Rovira & ES \\
J. Rovira & ES \\
S. Meister & CH \\
A. Ossola & CH \\
S. Sposetti & CH \\
J. Pol{\'a}k & CZ \\
M. Rottenborn & CZ \\
C. Weber & DE \\
C. Weber & DE \\
\hline
\end{tabular}
%\caption{List of stellar occultation observers and locations of their observing sites.}
%\label{occult_obs}
\end{table}

\begin{table}[h!]
\begin{tabular}{cc}
\hline
&\\
\multicolumn{2}{c}{(476) Hedwig, 2019-02-12}\\
&\\
\hline
E. Frappa & FR \\
M. Conjat & FR \\
\hline
&\\
\multicolumn{2}{c}{(476) Hedwig, 2020-05-19}\\
&\\
\hline
P. Le Cam & FR \\
E. Frappa & FR \\
E. Frappa, A. Klotz & FR \\
\hline
&\\
\multicolumn{2}{c}{(524) Fidelio, 2005-12-11}\\
&\\
\hline
R. Gon\c{c}alves & PT \\
O. Canales, J. L. Marco & ES \\
R. Poncy & FR \\
P. Bernascolle & FR \\
\hline
&\\
\multicolumn{2}{c}{(524) Fidelio, 2006-04-24}\\
&\\
\hline
H. Denzau & DE \\
W. Rothe & DE \\
T. Jan{\'i}k & CZ \\
V. P\v{r}ib{\'a}\v{n} & CZ \\
J. Vilagi, L. Kornos & SK \\
G. Dangl & AT \\
E. Frappa, A. Klotz & Tarot, Calern, FR \\
J. Lecacheux & FR \\
\hline
&\\
\multicolumn{2}{c}{(524) Fidelio, 2018-12-12}\\
&\\
\hline
E. Frappa, A. Klotz & Grasse, FR\\
R. Venable & Lovejoy, GA \\
R. Venable & Griffin, GA \\
R. Venable & Barnesville, GA \\
V. Tsamis, A. Tserionis et al. & Ellinogermaniki, GR\\
\hline
&\\
\multicolumn{2}{c}{(524) Fidelio, 2019-02-25}\\
&\\
\hline
R. Reaves & Parker, AZ \\
P. Maley & Dateland, AZ \\
\hline
&\\
\multicolumn{2}{c}{(524) Fidelio, 2021-07-13}\\
&\\
\hline
J. Newman & Flynn, Act, AU \\
W. Hanna & Yass, Nswn AU \\
\hline
&\\
\multicolumn{2}{c}{(524) Fidelio, 2021-11-18}\\
&\\
\hline
R. Venable & Chester, GA \\
R. Venable & Tarversvillle, GA \\
\hline
&\\
\multicolumn{2}{c}{(530) Turandot, 2006-02-24}\\
&\\
\hline
S. Jamieson & Eagle, WI \\
S. Ballaron & Lake Villa, IL \\
C. Bueter & Gas City, IN \\
D. Drake & Indian Head Park, IL \\
D. Dunham & Gaston, IN \\
G. Samolyk & Shirland, IL \\
B. Oldham & Pelham, NC \\
D. Oesper & Dodgeville, WI \\
S. Messner & Harvest Moon Obs, MN \\
R. Huziak & Saskatoon:sk, CAN \\
P. Maley et al & Cincinnati, OH \\
D. Dunham & Fisher, IN \\
J. Armor & Villa Hilles, KY \\
M. Hoskinson & Vilna Alberta, CAN \\
G. Stone & Lorborn:sk, CAN \\
D. Carton & Dark Sky Obs, CAN \\
O. Piechowski & Versailles, KY \\
D. Breit & Martinez, CA \\
\hline
\end{tabular}
%\caption{List of stellar occultation observers and locations of their observing sites.}
%\label{occult_obs}
\end{table}

\begin{table}[h!]
\begin{tabular}{cc}
\hline
&\\
\multicolumn{2}{c}{(530) Turandot, 2019-04-01}\\
&\\
\hline
P. Maley & Carefree, AZ \\
T. George & Scottsdale, AZ \\
W. Thomas & Florence, AZ \\
\hline
&\\
\multicolumn{2}{c}{(530) Turandot, 2020-07-27}\\
&\\
\hline
S. Preston & Medina, WA \\
D. Gamble & Summerland, BC \\
\hline
&\\
\multicolumn{2}{c}{(551) Ortrud, 2021-07-11}\\
&\\
\hline
A. Asai & Inabe, Mie, JP \\
H. Yamamura & Suzuka, Mie, JP \\
K. Isobe & Uda, Mie, JP \\
A. Hashimoto & Chichibu, Saitama, JP \\
M. Owada & Hammatsu, Shizuoka, JP \\
\hline
&\\
\multicolumn{2}{c}{(551) Ortrud, 2021-10-24}\\
&\\
\hline
J. Barton & Robinson, TX \\
D. Eisfeldt & Waco, TX \\
\hline
&\\
\multicolumn{2}{c}{(566) Stereoskopia, 2004-03-23}\\
&\\
\hline
T. Jan{\'i}k & CZ \\
D. Naillon & FR \\
H. McGee & GB \\
A. Christou, D. Asher & GB \\
J. Lecacheux & FR \\
T. Midavaine, O. Dechambre et al & FR \\
B. Christophe, J. M. Vugnon & FR \\
L. Parmeggiani & FR \\
G. Kirby & GB\\
\hline
&\\
\multicolumn{2}{c}{(566) Stereoskopia, 2010-12-28}\\
&\\
\hline
H. Tomioka & Hitachie, Ibaraki, JP \\
R. Aikawa & Sakado, Saitama, JP \\
K. Kitazaki & Musashino, Tokyo, JP\\
M. Owada & Hamamatsu, Shizuoka, JP \\
H. Watanabe & Inabe, Mie, JP\\
\hline
&\\
\multicolumn{2}{c}{(566) Stereoskopia, 2020-09-13}\\
&\\
\hline
H. Watanabe & Tarui: Gifu, JP\\
H. Yamamura & Wanouchi: Gifu, JP \\
A. Asai & Inabe: Mie, JP\\
\hline
&\\
\multicolumn{2}{c}{(566) Stereoskopia, 2021-04-08}\\
&\\
\hline
R. Boninsegna & BE \\
K. L. Bath & DE \\
S. Meister & CH \\
M. Simon & DE \\
W. Hasubick & DE \\
G. Krannich & DE \\
B. G\"{a}hrken & DE \\
B. Kattentidt & DE \\
\hline
\end{tabular}
%\caption{List of stellar occultation observers and locations of their observing sites.}
%\label{occult_obs}
\end{table}

\begin{table}[h!]
\begin{tabular}{cc}
\hline
&\\
\multicolumn{2}{c}{(566) Stereoskopia, 2022-02-08}\\
&\\
\hline
P. Ceravolo & Osoyoos, CAN \\
V. Nikitin & Boulder, CO \\
S. Messner & Jefferson, TX \\
M. Skrutskie & Nederland, CO \\
\hline
&\\
\multicolumn{2}{c}{(657) Gunlod, 2022-02-09}\\
&\\
\hline
J. J. Castellani & FR \\
P. Baruffetti & IT \\
M. Conjat & FR \\
M. Conjat & FR \\
G. Vaudescal & FR \\
Y. Argentin & FR \\
P. Fini & IT \\
\hline
&\\
\multicolumn{2}{c}{(738) Alagasta, 2017-11-12}\\
&\\
\hline
M. Kashiwagura & \={Oe}, Yamagata, JP \\
H. Togashi & \={Oe}, Yamagata, JP \\
I. Otsuki & Date, Fukushima, JP \\
K. Hosoi & Miharu, Fukushima, JP \\
A. Hashimoto & Ono, Fukushima, JP \\
H. Yamamura & Hodatsushimizu, Ishikawa, JP \\
M. Ida & Kahoku, Ishikawa, JP \\
M. Ishida & Kanazawa, Ishikawa, JP \\
H. Tomioka & Hitachi, Ibaraki, JP \\
A. Yaeza & Hitachi, Ibaraki, JP\\
A. Asai & Inabe, Mie, JP \\
H. Watanabe & Inabe, Mie, JP \\
Y. Ikari & Moriyama, Shiga, JP\\
\hline
&\\
\multicolumn{2}{c}{(738) Alagasta, 2020-05-30}\\
&\\
\hline
J. Bourgeois & BE \\
F. Van Den Abbeel & BE \\
O. Schreurs, E. Fernandez & BE \\
J. M. Winkel & NL \\
J. Pol{\'a}k & CZ \\
M. Rottenborn & CZ \\
J. Kub{\'a}nek & CZ \\
\hline
&\\
\multicolumn{2}{c}{(738) Alagasta, 2022-11-12}\\
&\\
\hline
K. Kitazaki & Musashino, Tokyo, JP \\
Ha. Watanabe, Hi. Watanabe & Tamaki, JP \\
M. Owada & Hamamatsu, Shizuoka, JP \\
H. Matsushita & Shima, Mie, JP\\
N. Manago & Kamitonda, Wakayama, JP \\
\hline
&\\
\multicolumn{2}{c}{(806) Gyldenia, 2021-08-20}\\
&\\
\hline
W. Thomas & Florence, AZ \\
P. Maley & Toltec, AZ \\
S. Messner & Hudson, WI \\
\hline
\end{tabular}
%\caption{List of stellar occultation observers and locations of their observing sites.}
%\label{occult_obs}
\end{table}

\clearpage

\section{Light curves}
 Composite light curves presenting new data for target asteroids
 (Figures \ref{70composit2017} - \ref{806composit2021}). 
%Composits 1
    \begin{table*}[ht]
    \centering
\vspace{0.5cm}
\begin{tabularx}{\linewidth}{XX}
\includegraphics[width=0.46\textwidth,clip]{70composit2017.eps} 
\captionof{figure}{Composite light curve of (70) Panopaea from the years 2016-2017.}
\label{70composit2017}
&
\includegraphics[width=0.46\textwidth,clip]{70composit2018a.eps} 
\captionof{figure}{Composite light curve of (70) Panopaea from the year 2018.}
\label{70composit2018a}
\\
\includegraphics[width=0.46\textwidth,clip]{70composit2018b.eps} 
\captionof{figure}{Composite light curve of (70) Panopaea from the year 2018 observed by TESS spacecraft.}
\label{70composit2018b}
&
\includegraphics[width=0.46\textwidth,clip]{70composit2019.eps} 
\captionof{figure}{Composite light curve of (70) Panopaea from the year 2019.}
\label{70composit2019}
\\
\includegraphics[width=0.46\textwidth,clip]{275composit2017.eps} 
\captionof{figure}{Composite light curve of (275) Sapientia from the years 2016-2017.}
\label{275composit2017}
&
\includegraphics[width=0.46\textwidth,clip]{275composit2018.eps} 
\captionof{figure}{Composite light curve of (275) Sapientia from the year 2018.}
\label{275composit2018}
\end{tabularx}
    \end{table*}

%Composits 2
    \begin{table*}[ht]
    \centering
\vspace{0.5cm}
\begin{tabularx}{\linewidth}{XX}
\includegraphics[width=0.46\textwidth,clip]{286composit2007.eps} 
\captionof{figure}{Composite light curve of (286) Iclea from the year 2007.}
\label{286composit2007}
&
\includegraphics[width=0.46\textwidth,clip]{286composit2013.eps} 
\captionof{figure}{Composite light curve of (286) Iclea from the year 2013.}
\label{286composit2013}
\\
\includegraphics[width=0.46\textwidth,clip]{286composit2014.eps} 
\captionof{figure}{Composite light curve of (286) Iclea from the year 2014.}
\label{286composit2014}
&
\includegraphics[width=0.46\textwidth,clip]{286composit2015.eps} 
\captionof{figure}{Composite light curve of (286) Iclea from the year 2015.}
\label{286composit2015}
\\
\includegraphics[width=0.46\textwidth,clip]{286composit2017.eps} 
\captionof{figure}{Composite light curve of (286) Iclea from the years 2016-2017.}
\label{286composit2017}
&
\includegraphics[width=0.46\textwidth,clip]{286composit2018.eps} 
\captionof{figure}{Composite light curve of (286) Iclea from the years 2017-2018.}
\label{286composit2018}
\end{tabularx}
    \end{table*}

%Composits 3
    \begin{table*}[ht]
    \centering
\vspace{0.5cm}
\begin{tabularx}{\linewidth}{XX}
\includegraphics[width=0.46\textwidth,clip]{286composit2019.eps} 
\captionof{figure}{Composite light curve of (286) Iclea from the year 2019.}
\label{286composit2019}
&
\includegraphics[width=0.46\textwidth,clip]{326composit2012.eps} 
\captionof{figure}{Composite light curve of (326) Tamara from the year 2012.}
\label{326composit2012}
\\
\includegraphics[width=0.46\textwidth,clip]{326composit2015.eps} 
\captionof{figure}{Composite light curve of (326) Tamara from the year 2015.}
\label{326composit2015}
&
\includegraphics[width=0.46\textwidth,clip]{326composit2016a.eps} 
\captionof{figure}{Composite light curve of (326) Tamara from March 2016.}
\label{326composit2016a}
\\
\includegraphics[width=0.46\textwidth,clip]{326composit2016b.eps} 
\captionof{figure}{Composite light curve of (326) Tamara August-September 2016.}
\label{326composit2016b}
&
\includegraphics[width=0.46\textwidth,clip]{326composit2016c.eps} 
\captionof{figure}{Composite light curve of (326) Tamara from October 2016. 
Data in 2016 were gathered in greatly varied phase angles and ecliptic longitudes, 
preventing the creation of one composite light curve (see Fig. \ref{326composit2016a}, 
and \ref{326composit2016b}).}
\label{326composit2016c}
\end{tabularx}
    \end{table*}

%Composits 4
    \begin{table*}[ht]
    \centering
\vspace{0.5cm}
\begin{tabularx}{\linewidth}{XX}
\includegraphics[width=0.46\textwidth,clip]{326composit2017.eps} 
\captionof{figure}{Composite light curve of (326) Tamara from the year 2017.}
\label{326composit2017}
&
\includegraphics[width=0.46\textwidth,clip]{326composit2019.eps} 
\captionof{figure}{Composite light curve of (326) Tamara from the year 2019.}
\label{326composit2019}
\\
\includegraphics[width=0.46\textwidth,clip]{412composit2016.eps} 
\captionof{figure}{Composite light curve of (412) Elisabetha from the year 2016.}
\label{412composit2016}
&
\includegraphics[width=0.46\textwidth,clip]{412composit2017.eps} 
\captionof{figure}{Composite light curve of (412) Elisabetha from the year 2017.}
\label{412composit2017}
\\
\includegraphics[width=0.46\textwidth,clip]{412composit2018.eps} 
\captionof{figure}{Composite light curve of (412) Elisabetha from the year 2018.}
\label{412composit2018}
&
\includegraphics[width=0.46\textwidth,clip]{412composit2020.eps} 
\captionof{figure}{Composite light curve of (412) Elisabetha from the years 2019-2020.}
\label{412composit2020}
\end{tabularx}
    \end{table*}

%Composits 5
    \begin{table*}[ht]
    \centering
\vspace{0.5cm}
\begin{tabularx}{\linewidth}{XX}
\includegraphics[width=0.46\textwidth,clip]{412composit2021.eps} 
\captionof{figure}{Composite light curve of (412) Elisabetha from the year 2021.}
\label{412composit2021}
&
\includegraphics[width=0.46\textwidth,clip]{426composit2015.eps} 
\captionof{figure}{Composite light curve of (426) Hippo from the year 2015.}
\label{426composit2015}
\\
\includegraphics[width=0.46\textwidth,clip]{426composit2016.eps} 
\captionof{figure}{Composite light curve of (426) Hippo from the years 2016-2017.}
\label{426composit2016}
&
\includegraphics[width=0.46\textwidth,clip]{426composit2019.eps} 
\captionof{figure}{Composite light curve of (426) Hippo from the year 2019 observed by TESS spacecraft.}
\label{426composit2019}
\\
\includegraphics[width=0.46\textwidth,clip]{426composit2021.eps} 
\captionof{figure}{Composite light curve of (426) Hippo from the year 2021.}
\label{426composit2021}
&
\includegraphics[width=0.46\textwidth,clip]{439composit2015.eps} 
\captionof{figure}{Composite light curve of (439) Ohio from the years 2015-2016.}
\label{439composit2015}
\end{tabularx}
    \end{table*}

%Composits 6
    \begin{table*}[ht]
    \centering
\vspace{0.5cm}
\begin{tabularx}{\linewidth}{XX}
\includegraphics[width=0.46\textwidth,clip]{439composit2018.eps} 
\captionof{figure}{Composite light curve of (439) Ohio from the year 2018.}
\label{439composit2018}
&
\includegraphics[width=0.46\textwidth,clip]{439composit2019.eps} 
\captionof{figure}{Composite light curve of (439) Ohio from the year 2019.}
\label{439composit2019}
\\
\includegraphics[width=0.46\textwidth,clip]{439composit2020.eps} 
\captionof{figure}{Composite light curve of (439) Ohio from the year 2020.}
\label{439composit2020}
&
\includegraphics[width=0.46\textwidth,clip]{464composit2015.eps} 
\captionof{figure}{Composite light curve of (464) Megaira from the year 2015.}
\label{464composit2015}
\\
\includegraphics[width=0.46\textwidth,clip]{464composit2016.eps} 
\captionof{figure}{Composite light curve of (464) Megaira from the year 2016.}
\label{464composit2016}
&
\includegraphics[width=0.46\textwidth,clip]{464composit2017.eps} 
\captionof{figure}{Composite light curve of (464) Megaira from the year 2017.}
\label{464composit2017}
\end{tabularx}
    \end{table*}

%Composits 7
    \begin{table*}[ht]
    \centering
\vspace{0.5cm}
\begin{tabularx}{\linewidth}{XX}
\includegraphics[width=0.46\textwidth,clip]{464composit2019.eps} 
\captionof{figure}{Composite light curve of (464) Megaira from the years 2018-2019.}
\label{464composit2019}
&
\includegraphics[width=0.46\textwidth,clip]{464composit2020.eps} 
\captionof{figure}{Composite light curve of (464) Megaira from the year 2020.}
\label{464composit2020}
\\
\includegraphics[width=0.46\textwidth,clip]{476composit2013.eps} 
\captionof{figure}{Composite light curve of (476) Hedwig from the year 2013.}
\label{476composit2013}
&
\includegraphics[width=0.46\textwidth,clip]{476composit2014.eps} 
\captionof{figure}{Composite light curve of (476) Hedwig from the years 2014-2015.}
\label{476composit2014}
\\
\includegraphics[width=0.46\textwidth,clip]{476composit2016.eps} 
\captionof{figure}{Composite light curve of (476) Hedwig from the year 2016.}
\label{476composit2016}
&
\includegraphics[width=0.46\textwidth,clip]{476composit2017.eps} 
\captionof{figure}{Composite light curve of (476) Hedwig from the year 2017.}
\label{476composit2017}
\end{tabularx}
    \end{table*}

%Composits 8
    \begin{table*}[ht]
    \centering
\vspace{0.5cm}
\begin{tabularx}{\linewidth}{XX}
\includegraphics[width=0.46\textwidth,clip]{476composit2018.eps} 
\captionof{figure}{Composite light curve of (476) Hedwig from the year 2018.}
\label{476composit2018}
&
\includegraphics[width=0.46\textwidth,clip]{524composit2013.eps} 
\captionof{figure}{Composite light curve of (524) Fidelio from the year 2013.}
\label{524composit2013}
\\
\includegraphics[width=0.46\textwidth,clip]{524composit2015.eps} 
\captionof{figure}{Composite light curve of (524) Fidelio from the years 2014-2015.}
\label{524composit2015}
&
\includegraphics[width=0.46\textwidth,clip]{524composit2016.eps} 
\captionof{figure}{Composite light curve of (524) Fidelio from the year 2016.}
\label{524composit2016}
\\
\includegraphics[width=0.46\textwidth,clip]{524composit2017.eps} 
\captionof{figure}{Composite light curve of (524) Fidelio from the year 2017.}
\label{524composit2017}
&
\includegraphics[width=0.46\textwidth,clip]{524composit2018.eps} 
\captionof{figure}{Composite light curve of (524) Fidelio from the years 2018-2019.}
\label{524composit2018}
\end{tabularx}
    \end{table*}

%Composits 9
    \begin{table*}[ht]
    \centering
\vspace{0.5cm}
\begin{tabularx}{\linewidth}{XX}
\includegraphics[width=0.46\textwidth,clip]{530composit2015.eps} 
\captionof{figure}{Composite light curve of (530) Turandot from the years 2015-2016.}
\label{530composit2015}
&
\includegraphics[width=0.46\textwidth,clip]{530composit2016.eps} 
\captionof{figure}{Composite light curve of (530) Turandot from the year 2016.}
\label{530composit2016}
\\
\includegraphics[width=0.46\textwidth,clip]{530composit2018.eps} 
\captionof{figure}{Composite light curve of (530) Turandot from the year 2018.}
\label{530composit2018}
&
\includegraphics[width=0.46\textwidth,clip]{530composit2019.eps} 
\captionof{figure}{Composite light curve of (530) Turandot from the year 2019.}
\label{530composit2019}
\\
\includegraphics[width=0.46\textwidth,clip]{551composit2015.eps} 
\captionof{figure}{Composite light curve of (551) Ortrud from the year 2015.}
\label{551composit2015}
&
\includegraphics[width=0.46\textwidth,clip]{551composit2018.eps} 
\captionof{figure}{Composite light curve of (551) Ortrud from the year 2018.}
\label{551composit2018}
\end{tabularx}
    \end{table*}

%Composits 10
    \begin{table*}[ht]
    \centering
\vspace{0.5cm}
\begin{tabularx}{\linewidth}{XX}
\includegraphics[width=0.46\textwidth,clip]{551composit2019.eps} 
\captionof{figure}{Composite light curve of (551) Ortrud from the year 2019.}
\label{551composit2019}
&
\includegraphics[width=0.46\textwidth,clip]{551composit2020.eps} 
\captionof{figure}{Composite light curve of (551) Ortrud from the year 2020.}
\label{551composit2020}
\\
\includegraphics[width=0.46\textwidth,clip]{551composit2021.eps} 
\captionof{figure}{Composite light curve of (551) Ortrud from the year 2021, 
observed mainly from TESS spacecraft.}
\label{551composit2021}
&
\includegraphics[width=0.46\textwidth,clip]{566composit2017.eps} 
\captionof{figure}{Composite light curve of (566) Stereoskopia from the year 2017, 
observed by Kepler mission.}
\label{566composit2017}
\\
\includegraphics[width=0.46\textwidth,clip]{566composit2019.eps} 
\captionof{figure}{Composite light curve of (566) Stereoskopia from the years 2019-2020.}
\label{566composit2019}
&
\includegraphics[width=0.46\textwidth,clip]{566composit2020.eps} 
\captionof{figure}{Composite light curve of (566) Stereoskopia from the years 2020-2021.}
\label{566composit2020}
\end{tabularx}
    \end{table*}

%Composits 11
    \begin{table*}[ht]
    \centering
\vspace{0.5cm}
\begin{tabularx}{\linewidth}{XX}
\includegraphics[width=0.46\textwidth,clip]{566composit2022.eps} 
\captionof{figure}{Composite light curve of (566) Stereoskopia from the year 2022.}
\label{566composit2022}
&
\includegraphics[width=0.46\textwidth,clip]{657composit2015.eps} 
\captionof{figure}{Composite light curve of (657) Gunlod from the year 2015.}
\label{657composit2015}
\\
\includegraphics[width=0.46\textwidth,clip]{657composit2016.eps} 
\captionof{figure}{Composite light curve of (657) Gunlod from the year 2016, 
observed mainly from TESS spacecraft.}
\label{657composit2016}
&
\includegraphics[width=0.46\textwidth,clip]{657composit2018.eps} 
\captionof{figure}{Composite light curve of (657) Gunlod from the year 2018, 
observed by Kepler mission.}
\label{657composit2018}
\\
\includegraphics[width=0.46\textwidth,clip]{657composit2020.eps} 
\captionof{figure}{Composite light curve of (657) Gunlod from the year 2020.}
\label{657composit2020}
&
\includegraphics[width=0.46\textwidth,clip]{657composit2021a.eps} 
\captionof{figure}{Composite light curve of (657) Gunlod from the year 2021 
from TESS spacecraft.}
\label{657composit2021a}
\end{tabularx}
    \end{table*}

%Composits 12
    \begin{table*}[ht]
    \centering
\vspace{0.5cm}
\begin{tabularx}{\linewidth}{XX}
\includegraphics[width=0.46\textwidth,clip]{657composit2021b.eps} 
\captionof{figure}{Composite light curve of (657) Gunlod from the years 2021-2022.
Given the different character of the light curve caused by slightly different viewing 
geometry, it could not be folded with the light curve from Fig. \ref{657composit2021a}.}
\label{657composit2021b}
&
\includegraphics[width=0.46\textwidth,clip]{738composit2015.eps} 
\captionof{figure}{Composite light curve of (738) Alagasta from the year 2015.}
\label{738composit2015}
\\
\includegraphics[width=0.46\textwidth,clip]{738composit2016.eps} 
\captionof{figure}{Composite light curve of (738) Alagasta from the year 2016.}
\label{738composit2016}
&
\includegraphics[width=0.46\textwidth,clip]{738composit2017.eps} 
\captionof{figure}{Composite light curve of (738) Alagasta from the year 2017.}
\label{738composit2017}
\\
\includegraphics[width=0.46\textwidth,clip]{738composit2019.eps} 
\captionof{figure}{Composite light curve of (738) Alagasta from the years 2018-2019.}
\label{738composit2019}
&
\includegraphics[width=0.46\textwidth,clip]{738composit2020.eps} 
\captionof{figure}{Composite light curve of (738) Alagasta from the year 2020.}
\label{738composit2020}
\end{tabularx}
    \end{table*}

%Composits 13
    \begin{table*}[ht]
    \centering
\vspace{0.5cm}
\begin{tabularx}{\linewidth}{XX}
\includegraphics[width=0.46\textwidth,clip]{806composit2015.eps} 
\captionof{figure}{Composite light curve of (806) Gyldenia from the year 2015.}
\label{806composit2015}
&
\includegraphics[width=0.46\textwidth,clip]{806composit2016.eps} 
\captionof{figure}{Composite light curve of (806) Gyldenia from the year 2016.}
\label{806composit2016}
\\
\includegraphics[width=0.46\textwidth,clip]{806composit2018.eps} 
\captionof{figure}{Composite light curve of (806) Gyldenia from the year 2018.}
\label{806composit2018}
&
\includegraphics[width=0.46\textwidth,clip]{806composit2020.eps} 
\captionof{figure}{Composite light curve of (806) Gyldenia from the year 2020.}
\label{806composit2020}
\\
\includegraphics[width=0.46\textwidth,clip]{806composit2021.eps} 
\captionof{figure}{Composite light curve of (806) Gyldenia from the year 2021.}
\label{806composit2021}
\end{tabularx}
    \end{table*}

\end{appendix} 
 
\end{document}